\documentclass[journal]{IEEEtran}
\usepackage{cite,amsmath,amssymb,amsthm,bm,threeparttable,booktabs,multirow,array,graphics,scalerel,times,bbm}
\usepackage[table]{xcolor}
\usepackage[T1]{fontenc}
\usepackage[hidelinks]{hyperref}
\usepackage[caption=false, font=footnotesize,subrefformat=parens]{subfig}

\usepackage[ruled,norelsize]{algorithm2e}
\SetAlCapFnt{\small}
\SetAlCapNameFnt{\small}
\SetAlFnt{\small}

\let\oldarraystretch\arraystretch
\newlength{\oldtabcolsep}
\begin{document}
\title{
	Automatic Generation of Topology Diagrams for Strongly-Meshed Power Transmission Systems
}
\author{
	Jingyu~Wang,~\IEEEmembership{Member,~IEEE}, %
	Jinfu~Chen,~\IEEEmembership{Senior~Member,~IEEE}, %
	Dongyuan~Shi,~\IEEEmembership{Senior~Member,~IEEE}, %
	and Xianzhong Duan,~\IEEEmembership{Senior~Member,~IEEE}%
	\thanks{
		The work was supported by the State Grid Corporation of China Science and Technology Project under Grant 5100-202199558A-0-5-ZN. \emph{(Corresponding author: Dongyuan Shi)}
	}%
	\thanks{
		J. Wang, J. Chen, and D. Shi are with the State Key Laboratory of Advanced Electromagnetic Engineering and Technology, School of Electrical and Electronic Engineering, Huazhong University of Science and Technology, Wuhan, China (e-mails: \href{jywang@hust.edu.cn}{jywang@hust.edu.cn}, \href{chenjinfu@hust.edu.cn}{chenjinfu@hust.edu.cn}, \href{dongyuanshi@hust.edu.cn}{dongyuanshi@hust.edu.cn}).
	}%
	\thanks{
		X. Duan is with the College of Electrical and Information Engineering, Hunan University, Changsha, China (e-mail: \href{duanxz@hnu.edu.cn}{duanxz@hnu.edu.cn}).
	}
}
\maketitle

\begin{abstract}
	Topology diagrams are widely seen in power system applications, but their automatic generation is often easier said than done. When facing power transmission systems with strongly-meshed structures, existing approaches can hardly produce topology diagrams catering to the aesthetics of readers. This paper proposes an integrated framework for generating aesthetically-pleasing topology diagrams for power transmission systems. Input with a rough layout, the framework first conducts visibility region analysis to reduce line crossings and then solves a mixed-integer linear programming problem to optimize the arrangement of nodes. Given that the complexity of both modules is pretty high, simplification heuristics are also proposed to enhance the efficiency of the framework. Case studies on several power transmission systems containing up to 2,046 nodes demonstrate the capability of the proposed framework in generating topology diagrams conforming to aesthetic criteria in the power system community. Compared with the widespread force-directed algorithm, the proposed framework can preserve the relative positions of nodes in the original layout to a great extent, which significantly contributes to the identification of electrical elements on the diagrams. Meanwhile, the time consumption is acceptable for practical applications.
\end{abstract}


\begin{table*}[!t]
	\centering
	\caption{Aesthetic Criteria of Topology Diagrams}
	\begin{tabular}{cllc}
		\toprule
		\textbf{Abbr.} & \multicolumn{1}{c}{\textbf{Description}} & \multicolumn{1}{c}{\textbf{Metric}} & \textbf{Priority} \\
		\midrule
		EX    & Avoid unnecessary line crossings & $m_{\text{EX}} = -C[\Gamma(G)]$ & High \\
		EL    & Avoid too small edge lengths & $m_{\text{EL}} = \min(\mathcal{L}) / \mathrm{avr}(\mathcal{L})$ & High \\
		ND    & Avoid too small node-to-node distances & $m_{\text{ND}} = \min(\mathcal{D}) / \mathrm{avr}(\mathcal{D})$ & High \\
		IA    & Avoid too small angles between incident edges & $m_{\text{IA}} = \min(\mathcal{R}) / \mathrm{avr}(\mathcal{R})$ & High \\
		RP    & Maintain relative positions of nodes & $m_{\text{RP}} = 1 - \mathrm{avr}(\mathcal{Q}) / 180^{\circ}$ & Low \\
		OR    & Improve line orthogonality & $m_{\text{OR}} = 1-\mathrm{avr}(\mathcal{O})$ & Low \\
		EV    & Improve spatial evenness of the diagram & $m_{\text{EV}} = -\sigma^2(\tilde{\mathcal{D}}_v^M)$ & Low \\
		\bottomrule
	\end{tabular}%
	\label{tab:aesthetic_criteria}%
\end{table*}%

\section{Introduction}
\label{sec:introduction}

\IEEEPARstart{T}{opology} diagrams can be obtained by drawing substations or buses as nodes and their interconnections as edges. They are widely used in power system applications to depict the overall structure of power grids and visualize the spatial correlations among electrical data. Currently, most software applications ask users to draw topology diagrams manually. Although it grants users complete control over the layout, it shifts the difficulty of designing accurate and aesthetically-pleasing topology diagrams to users in the meantime. Users need to pay undivided attention to the correctness of node connectivity and repeatedly tune the positions of graphic elements for better aesthetic quality, both of which require great patience and extensive labor. Additionally, topology diagrams cannot remain static during the lengthy service cycle of power systems, as the continuous extension and rewiring may negatively impact their aesthetic quality. Therefore, research on the automatic generation of topology diagrams is of significant value.

Some studies suggest visualizing power systems directly using the latitude and longitude provided by the Geographic Information System (GIS) \cite{Ten_Wuergler_2008,Wong_Schneider_2009,Overbye_Wert_2021}. This strategy can preserve the spatial relationship between power systems and other geographic landmarks. However, the downside is that urban areas that are usually of more concern may look overcrowded as power grids in those areas are often very dense. Compared with precise coordinates of substations and buses, whether their relative positions are consistent with the mental map of readers matters more for understanding topology diagrams. In \cite{Li_Feng_2008,Wu_Lin_2018,Shang_Hu_2019}, topology diagrams of power distribution systems are generated with GIS coordinates serving as a reference. The yielded diagrams achieve good readability as these approaches not only consider aesthetic criteria but also maintain the spatial relationship between feeders. However, these approaches do not support generating topology diagrams for power transmission systems with strongly-meshed structures. In \cite{Birchfield_Overbye_2018}, Birthfield \textit{et al.} develop a framework to draw topology diagrams for power transmission systems, which shares many of the goals of this paper. It employs a force-directed approach and a greedy approach to optimize the size and location of substations while keeping their original geographic context. It also contains a line routing algorithm based on Delaunay triangulation to prevent lines from overlapping with substations. Nonetheless, \cite{Birchfield_Overbye_2018} does not explicitly reduce line crossings and has no restriction on the line orientation, which might lead to decreased legibility of yielded diagrams. Besides, it is difficult for users to extend \cite{Birchfield_Overbye_2018} to consider other aesthetic criteria for meeting specific visualization demands.

More researchers ignore the geographic property of power systems and investigate topology diagram generation methods merely from aesthetic viewpoints. The earliest exploration \cite{Canales_Garibay_1979} dates to the late 1970s, which generates topology diagrams for small power systems by solving linear programming problems. Since then, many rule-based, physics-based, and optimization-based methods have been proposed. Rule-based methods plot diagrams based on fixed cartographic specifications \cite{Raman_Khincha_1986,Nagendra_Deekshit_2004,Peng_Wang_2016,Ding_Meng_2016,Hussain_Aslam_2018,Cuffe_Keane_2017,KOVACEV2023118733}. Due to their limited adaptability, they are mainly used in distribution grids with radial structures. Physics-based methods model power grids as physical systems and compute layouts based on physical laws. The force-directed algorithm \cite{Eades_1984,Walshaw_2000,Teja_Yemula_2014,Mota_Mota_2011} is a representative of this class. It assigns forces to nodes and edges and harvests their positions after driving them into a state of equilibrium. It is arguably the most popular approach today because it can generate topology diagrams efficiently, without restrictions on the structure of power systems. However, since the final positions of graphic elements are entirely dominated by kinematic laws, the yielded layouts will likely not conform to the aesthetic consensus of the power system community. Optimization-based methods formulate aesthetic criteria into optimization problems. With human experience integrated, these methods should be capable of generating topology diagrams for complex power grids. However, previous studies based on branch-and-bound optimization \cite{Kovacev_Lendak_2013}, genetic algorithms \cite{Zhou_Sun_2016}, and particle swarm optimization \cite{Lendak_Erdeljan_2010,Lin_Xing_2010} are mostly intended for power distribution systems.

Some progress has also been made in general-purpose graph drawing algorithms in the past few years. In \cite{Jacomy_Venturini_2014}, Gephi's team introduces ForceAtlas2, an improved force-directed graph drawing algorithm. By integrating techniques like the Barnes Hut simulation, degree-dependent repulsive force, and local and global adaptive temperatures, ForceAtlas2 achieves better quality and speed than similar algorithms. In \cite{Zheng_Pawar_2019}, an improvement of multidimensional scaling-based force-directed algorithms is proposed, which uses stochastic gradient descent to minimize the energy function. Results show that this improvement can reach lower stress levels faster and more consistently than previous methods. In \cite{Kruiger_Rauber_2017}, the authors propose a graph layout method, tsNET, based on the t-distributed Stochastic Neighbor Embedding (t-SNE) dimensionality reduction technique. The application of deep learning techniques in graph drawing is first explored in \cite{Wang_Jin_2020}, where a graph Long Short-Term Memory (LSTM) model is trained with set of layout examples to generate layouts in a similar style for new graphs. In \cite{Ahmed_Luca_2022}, a layout approach is proposed to simultaneously optimize multiple readability criteria that can be described by a differentiable function. Compared with existing algorithms that optimize one aesthetic criteria at the expense of others, this approach reaches a better comprehensive graph drawing quality.

Topology diagrams with fewer line crossings and neatly arranged graphic elements are often considered elegant. Recent progress in computational geometry can help in these aspects. In \cite{Radermacher_Reichard_2019}, line crossings in rectilinear graphs are reduced by greedily moving terminal nodes of intersecting edges to their optimal positions determined by visibility region analysis. In \cite{Nickel_Nollenburg_2020}, a Mixed-Integer Linear Programming (MILP) model is devised to place metro stations to make the yielded metro map clear and nice-looking. Despite that both techniques perform well in their respective application scenarios, they cannot be trivially applied to generate topology diagrams for power transmission systems. On the one hand, the fundamental algorithm of \cite{Radermacher_Reichard_2019}, i.e., the one searching for the optimal position of a node, has a superquadratic complexity, making the overall crossing reduction approach time-consuming even for moderate-scale power systems. On the other hand, power transmission systems have more complex structures than metro networks. Directly using the MILP model in \cite{Nickel_Nollenburg_2020} may result in low convergence speed and undesirable output layouts.

This paper proposes an integrated framework to generate aesthetically-pleasing topology diagrams for strongly-meshed power transmission systems. Compared with the force-directed algorithm, the proposed framework can keep the spatial relationships of nodes roughly identical to the given initial layout, making it easier for readers to recognize electrical elements. It advances the state-of-the-art in the following aspects:

\begin{enumerate}
	\item A set of topology diagram assessment metrics reflecting the aesthetic convention of the power system community are formulated, which can be used to compare the quality of different topology diagrams.
	\item The computational geometry-based crossing-reduction method in \cite{Radermacher_Reichard_2019} is considerably enhanced in efficiency by lowering the number of node moving and only considering subgraphs surrounding the intersecting edges.
	\item The MILP modeling method in \cite{Nickel_Nollenburg_2020} is modified according to the cartographic convention of power grid topology diagrams, with special considerations taken to improve convergence speed.
	\item Users are allowed to formulate new aesthetic criteria as optimization objectives and constraints to extend the proposed framework. To the best of authors' knowledge, it is the first to leave room for users to customize the topology layout of transmission systems over 1,000 nodes according to specific visualization demands.
\end{enumerate}

The rest of the paper is organized as follows. Section \ref{sec:grid_topology_aesthetics} summarizes typical aesthetic criteria of power grid topology diagrams. Section \ref{sec:edge_crossing_reduction} and Section \ref{sec:optimal_layout_generation} elaborate the crossing reduction and layout planning methods used in the proposed framework, respectively. Section \ref{sec:case_studies} validates the performance of the framework through case studies. Section \ref{sec:conclusion} concludes the paper and envisions the future work.

\textbf{Notations:} In this paper, a power transmission system is represented by an undirected graph $G=(\mathcal{V}, \mathcal{E})$, with $\mathcal{V}$ and $\mathcal{E}$ denoting the sets of nodes and edges in $G$, respectively. The set of neighbors of a node $v$ is represented as $\mathcal{N}_v$. The set of edges incident to $v$ is denoted as $\mathcal{E}_v$, with the set of edges irrelevant to $v$ written as $\mathcal{E}'_v = \mathcal{E} \backslash \mathcal{E}_v$. The layout of $G$ is expressed by a set of points $\Gamma(G)=\{p_v \mid v \in \mathcal{V}\}$, where $p_v=(x_v, y_v)$ denotes $v$'s coordinates. Except as otherwise noted, an edge $(i,j)$ in $\Gamma(G)$ can be uniquely determined by the coordinates of its endpoints, i.e., $(p_i, p_j)$. Parallel lines are merged into one abstract line during crossing reduction and layout planning. The initial, crossing-reduced, and optimized layouts are respectively represented as $\Gamma^0$, $\bar{\Gamma}$, and $\hat{\Gamma}$. Decision variables are expressed in the sans-serif font, e.g., $(\mathsf{x}_v, \mathsf{y}_v)$, to discriminate from known values.

\section{Aesthetics in Topology Diagrams}
\label{sec:grid_topology_aesthetics}

Beauty is in the eye of the beholder. Each application may have unique preferences in the style of power grid topology diagrams. Nonetheless, some representative aesthetic criteria are commonly followed in most scenarios to improve the legibility and tidiness of the yielded topology diagrams. Coupled with the suggestions in \cite{Wong_Schneider_2009,Purchase_2002,Bennett_Ryall_2007,Cuffe_Keane_2016} and the authors' experience, seven common aesthetic criteria are listed in Table \ref{tab:aesthetic_criteria}. Also, a set of metrics $m_x$ ($x \in \{\text{EX, EL, ND, IA, RP, OR, EV}\}$) are defined to quantitatively describe the obedience degree of a topology diagram $\Gamma(G)$ to the above criteria. These metrics can be used to evaluate and compare different topology diagram generation algorithms, with higher values regarded as better in aesthetics. 

The metric $m_{\text{EX}}$ is defined as the opposite number of the total count of line crossings, i.e., $-C[\Gamma(G)]$, to reflect the penalty of line intersections on the diagram quality. Metrics $m_{\text{EL}}$, $m_{\text{ND}}$, and $m_{\text{IA}}$ are defined in a similar form, i.e., dividing the minimum element of a set by the average of all elements in the set. For $m_{\text{EL}}$, the set $\mathcal{L}=\{\ell_e \mid e \in \mathcal{E}\}$ is used, which contains the lengths of all edges. For $m_{\text{ND}}$, the set $\mathcal{D}=\{d^N_v \mid v \in \mathcal{V}\}$ is used, with $d^N_v$ denoting the distance from $v$ to its closest neighbor in $\mathcal{N}_v$. For $m_{\text{IA}}$, the set $\mathcal{R}=\{d^A_v \mid v \in \mathcal{V}\}$ is used, where $d^A_v=a^{\min}_v / a^{*}_v$ is the relative angular resolution of $v$. Here, $a^{\min}_v$ is the minimum angle between adjacent edges incident to $v$, and $a^{*}_v = 360^{\circ} / n_v$, with $n_v$ the degree of $v$. 

The metric $m_{\text{RP}}$ is defined as $1 - \mathrm{avr}(\mathcal{Q}) / 180^{\circ}$, where $\mathcal{Q} = \{\Delta\theta_{ij} \mid (i,j) \in \mathcal{E}\}$ contains the angle $\Delta\theta_{ij}$ between vectors $(p_i^0, p_j^0)$ and $(p_i,p_j)$ for each edge $(i, j)$. $p_i^0$ and $p_j^0$ stand for the coordinates of nodes $i$ and $j$ in the initial layout $\Gamma^0(G)$, while $p_i$ and $p_j$ represent the corresponding coordinates in the output layout $\Gamma(G)$. The metric $m_\text{OR}$ is defined as $1-\mathrm{avr}(\mathcal{O})$, where $\mathcal{O}=\{\delta_e \mid e\in \mathcal{E}\}$ contains the edge deviation factor $\delta_e = \min\{\theta_e, |90^{\circ} - \theta_e|, 180^{\circ} - \theta_e\} / 45^{\circ}$ of all edges. The edge deviation factor is proposed in \cite{Purchase_2002}, which represents the deviation from $\theta_e$, the included angle between $e$ and the $x$-axis, to the closest orthogonal angle. The metric $m_{\text{EV}}$ is defined as the opposite value of the variance of $\tilde{\mathcal{D}}_v^M$, where $\tilde{\mathcal{D}}_v^M$ is the Min-Max normalization of $\mathcal{D}_v^M=\{d_v^i \mid 1 \leq i \leq M, v \in \mathcal{V}\}$. Here, $d_v^i$ denotes the distance between $v$ and its $i$-th nearest neighbor. $M$ is a parameter determining how many neighbors are considered, which is suggested to be 10\% of total nodes.

Note that the criteria in Table \ref{tab:aesthetic_criteria} may conflict with one another. For instance, reducing line crossings based on \cite{Radermacher_Reichard_2019} involves moving nodes to their optimal locations, making the relative position of nodes likely changed. Thus, insisting on EX may lead to violation of RP. Another example is that pursuing line orthogonality may forcibly align nodes along horizontal or vertical lines and thus may also change the relative position of nodes, causing potential conflicts between OR and RP. These conflicts can be untied by assigning different priority levels. Criteria for ensuring basic legibility of topology diagrams, including EX, EL, ND, and IA, should be given higher priority. On the contrary, others can have relatively lower priority levels as they mainly aim to facilitate information acquisition and enhance visual comfort. In Section \ref{sec:optimal_layout_generation}, aesthetic criteria of high priority will be modeled as MILP constraints, with those of low priority jointly composing the optimization objective. The tradeoff between different low-priority rules can be realized by placing different weights on their corresponding cost terms in the objective.

\section{Line Crossing Reduction}
\label{sec:edge_crossing_reduction}

\begin{figure}[!tb]
	\centerline{\includegraphics[width=0.36\textwidth]{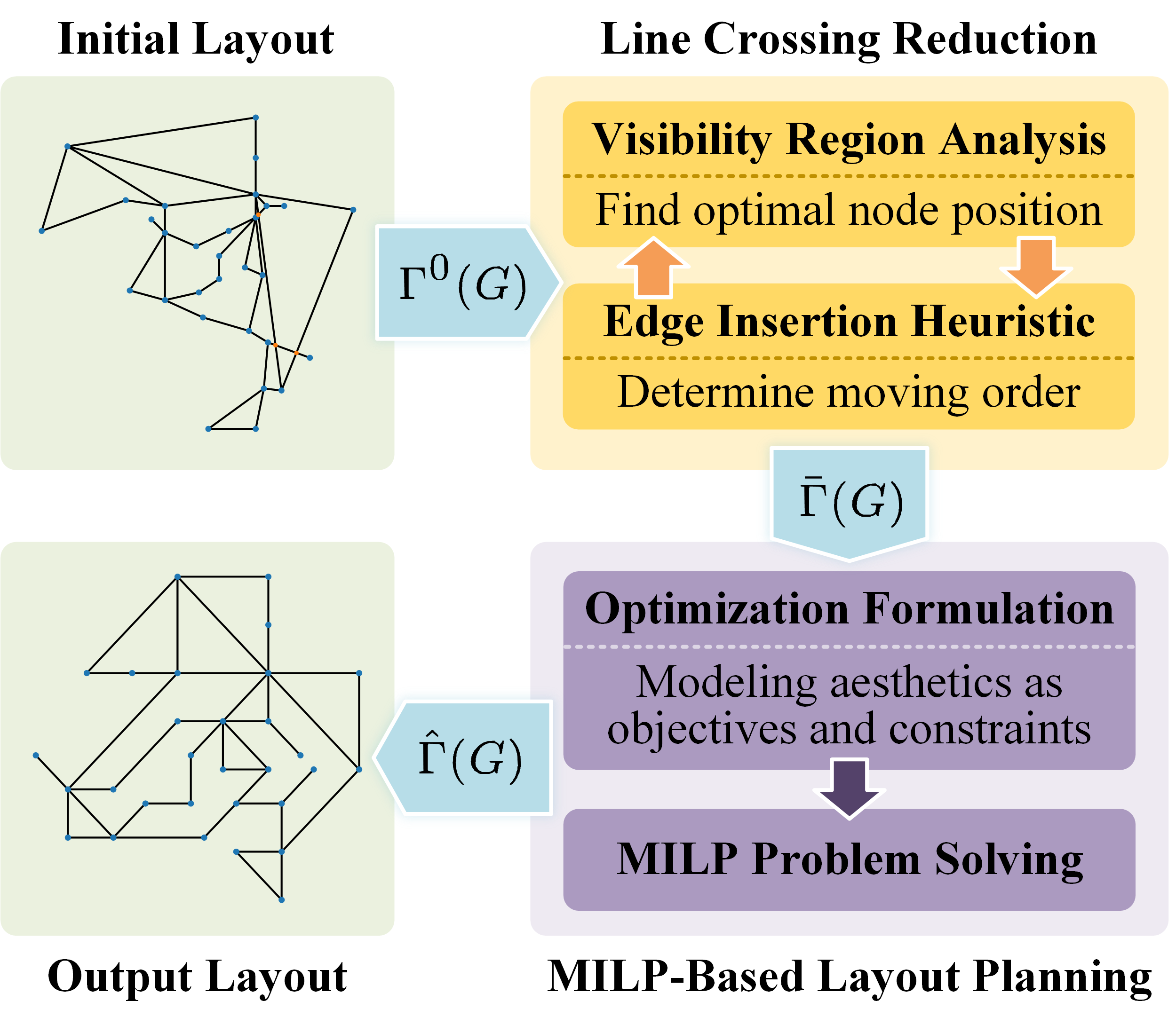}}
	\caption{The architecture of the topology diagram generation framework.}
	\label{fig:overall_framework}
\end{figure}

As depicted in Fig. \ref{fig:overall_framework}, the proposed topology diagram generation framework is composed of two primary modules, including a line crossing reduction module based on computational geometry and a node layout planning module based on MILP optimization. The framework takes a poor initial layout $\Gamma^0(G)$ as the input. The layout may contain many line crossings, which should be mitigated as much as possible to improve readability. This section starts with an introduction of the crossing reduction method in \cite{Radermacher_Reichard_2019}, followed by the presentation of two simplification heuristics to decrease the time overhead. The node layout planning module will be discussed in the next section.

\subsection{Crossing Reduction by Node Moving}
\label{sub:crossing_reduction_by_moving_nodes}

\begin{algorithm}[t]
\caption{Optimal Node Moving} \label{alg:optimal_node_moving}
	\SetAlgoLined
	\DontPrintSemicolon
	\KwIn{$G=(\mathcal{V},\mathcal{E})$, initial layout $\Gamma^0(G)$, node $v$ to move}
	\KwOut{optimal position $p_v^*$}
	\vspace{2mm}
	Initialize a canvas $\mathcal{C}_v \leftarrow \varnothing$\;
	\ForEach{node $u\in \mathcal{N}_v$}{
		\ForEach{edge $e \in \mathcal{E}_v' \cap \mathcal{E}_u'$}{
			Draw visibility region boundary $\mathcal{B}[u;e]$ on $\mathcal{C}_v$}}
	Arrangement $\mathcal{A} \leftarrow {}$divide $\mathcal{C}_v$ along boundaries into faces\;
	\ForEach{face $f \in \mathcal{A}$}{
		\ForEach{face $g \in \mathcal{A}\backslash{\{f\}}$}{
			$e_{fg} \leftarrow{}$demarcation edge between $f$ and $g$\;
			\If{$f$, $g$ are adjacent}{
				\eIf{$e_{fg}$ is on an edge $e \in \mathcal{E}_v' \cap \mathcal{E}_u'$}{
					Increment $\Delta_{fg} \leftarrow N_v^f-N_v^g$\;}{
					// $e_{fg}$ is on a ray $R_{uz} \in \mathcal{B}[u; (z, \cdot)]$\;
					Increment $\Delta_{fg} \leftarrow N_z^g-N_z^f$\;}
				}}}
	Compute the dual graph $G_d=(\mathcal{V}_d, \mathcal{E}_d)$ of $\mathcal{A}$\;
	Initialize a queue $\mathcal{Q} \leftarrow \varnothing$, $\Delta f^0 \leftarrow 0$\;
	Enqueue the initial face $f^0$ where $v$ lies into $\mathcal{Q}$\;
	\While{$\mathcal{Q}$ is not empty}{
		$f \leftarrow{}$ Dequeue $\mathcal{Q}$\;
		\ForEach{child node $g$ of $f$}{
			Increment $\Delta f_c \leftarrow \Delta f + \Delta_{fg}$\;
			Enqueue $g$ into $\mathcal{Q}$\;
		}
	}
	Find the optimal face $f_v^*$ by sorting $\mathcal{V}_d$ as descending $\Delta f$\;
	Optimal point $p_v^* \leftarrow{}$ a random point in $f_v^*$\;
	\KwRet{$p_v^*$}\;
\end{algorithm}

\begin{figure}[!tb]
	\centerline{\includegraphics{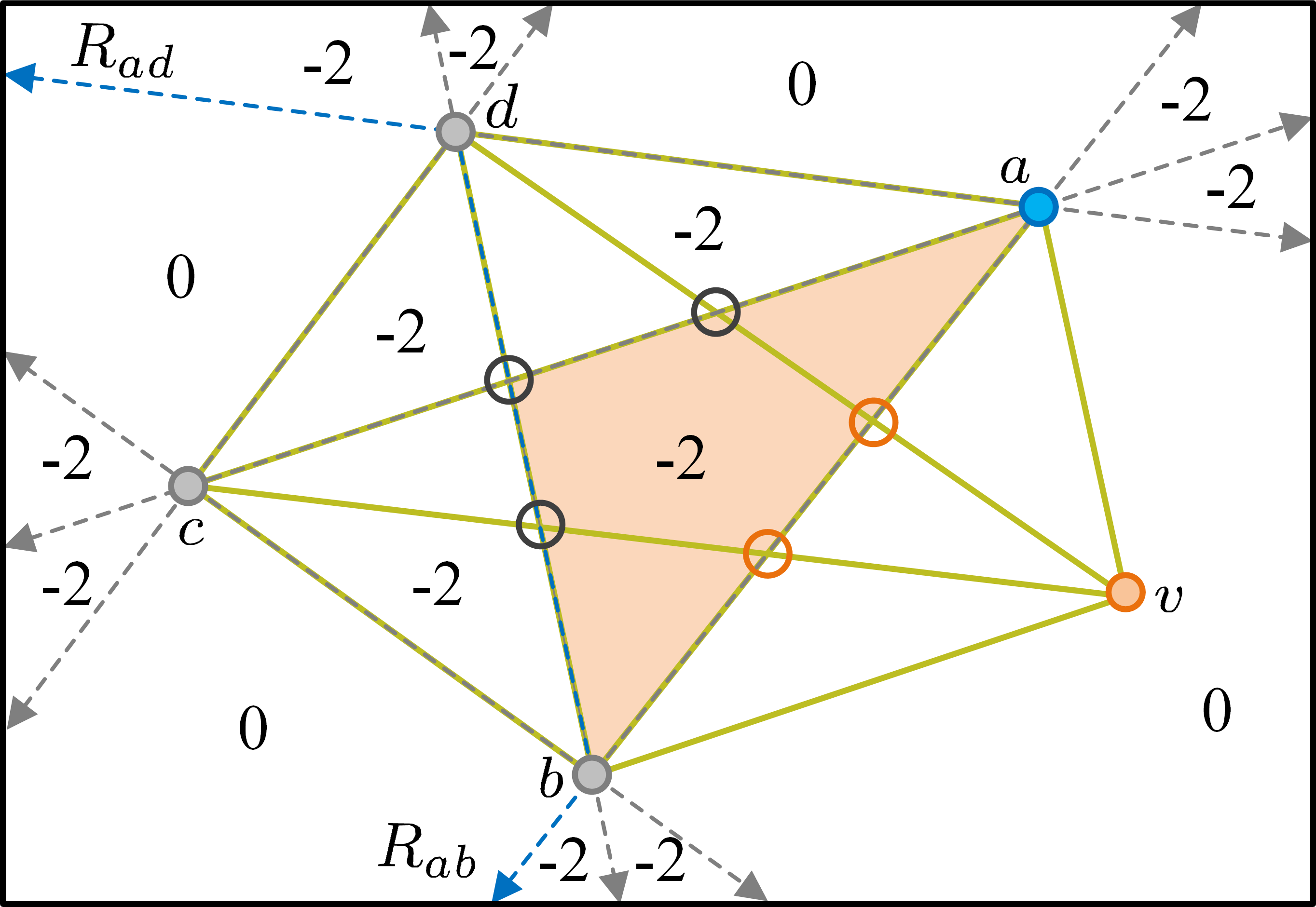}}
	\caption{The arrangement for finding the best position to move $v$. Line crossings of $\Gamma^0(K_5)$ are marked with circles, among which the orange ones can be eliminated by moving $v$ into the orange region. The visibility region boundary $\mathcal{B}[a;(b,d)]$ is marked in blue.}
	\label{fig:visibility_region_analysis}
\end{figure}

The core idea of the crossing reduction method in \cite{Radermacher_Reichard_2019} is to greedily reposition nodes on the topology diagram so that the total number of line crossings can approach the global minimum. Visibility region analysis is used to find the optimal position to place each node, as shown in Algorithm \ref{alg:optimal_node_moving}. Take the complete graph $K_5$ consisting of five nodes $\mathcal{V}=\{v,a,b,c,d\}$ as an example. As shown in Fig. \ref{fig:visibility_region_analysis}, the initial layout $\Gamma^0(K_5)$ consists of an outer pentagon and an inner pentagram, whose total crossings $C[\Gamma^0(K_5)]=5$. To find the position $p^{*}_v$ such that moving $v$ to $p^{*}_v$ yields the fewest crossings, the visibility region of each node $u \in \mathcal{N}_v$ against each edge $e \in \mathcal{E}'_v \cap \mathcal{E}'_u$ should be computed. To be specific, the visibility region of the node $a$ against the edge $(b, d)$ is bounded by $(b, d)$ itself and the two rays $R_{ab}$ and $R_{ad}$ that start from $b$ and $d$ and diverge along vectors $(a, b)$ and $(a, d)$, respectively. As the name indicates, if $v$ locates in the visibility region, edges $(v, a)$ and $(b, d)$ will not intersect. The boundary of the visibility region is denoted as $\mathcal{B}[a;(b,d)]$. By drawing $\mathcal{B}[u;e]$ for all possible $(u, e)$ pairs on the same canvas, an arrangement $\mathcal{A}$ that divides the canvas into a set of \textit{faces} can be obtained. Note that the number of line crossings may change only when $v$ is moved from one face to another. Suppose that $f$ and $g$ are adjacent faces and that $e_{fg}$ is their demarcation line. The increment of line crossings $\Delta_{fg}$ on moving $v$ from $f$ to $g$ depends on the type of $e_{fg}$:
\begin{enumerate}
	\item If $e_{fg}$ is on an edge $e \in \mathcal{E}'_v \cap \mathcal{E}'_u$, $\Delta_{fg} = N_v^f - N_v^g$. $N_v^f$ and $N_v^g$ are the numbers of $v$'s neighbors on the half planes $H_f$ and $H_g$ containing $f$ and $g$, respectively.
	\item If $e_{fg}$ is on a ray $R_{uz} \in \mathcal{B}[u; (z, \cdot)]$, $\Delta_{fg} = N_z^g - N_z^f$. $N_z^f$ and $N_z^g$ are the numbers of $z$'s neighbors on the half planes $H_f$ and $H_g$ containing $f$ and $g$, respectively.
\end{enumerate}

Since moving $v$ within a face does not change the number of crossings, searching for $p^{*}_v$ is equivalent to searching for the optimal face $f^{*}_v$. Suppose that $v$ initially lies in the face $f^0$. The increment of line crossings $\Delta_{f}$ after moving $v$ to any face $f\in\mathcal{A}$ can be computed by conducting breadth-first traversal of all nodes on the dual graph of $\mathcal{A}$ starting from $f^0$. The face holding the minimum $\Delta_{f}$, e.g., the one colored in orange in Fig. \ref{fig:visibility_region_analysis}, is the desired optimal face $f^{*}_v$.

\begin{algorithm}[t]
\caption{Edge Insertion Heuristic} \label{alg:edge_insertion_heuristic}
	\SetAlgoLined
	\DontPrintSemicolon
	\KwIn{$G=(\mathcal{V},\mathcal{E})$, initial layout $\Gamma^0(G)$}
	\KwOut{crossing-reduced layout $\bar{\Gamma}(G)$}
	\vspace{2mm}
	Compute the set $\mathcal{P}$ of intersecting edge pairs in $\Gamma^0(G)$ using the Bentley-Ottmann algorithm\;
	Initialize a set $\mathcal{E}_r \leftarrow \varnothing$, $\Gamma \leftarrow \Gamma^0$\;
	\ForEach{edge pair $(e_i, e_j) \in \mathcal{P}$}{
		$\Gamma \leftarrow{}$ Remove $e_i$ from $\Gamma$\;
		$\mathcal{E}_r \leftarrow \mathcal{E}_r \cup \{e_i\}$\;
	}
	\ForEach{edge $e = (u, v)\in \mathcal{E}_r$}{
		$\Gamma \leftarrow{}$ Add $e$ back to $\Gamma$ by connecting $u$ and $v$\; 
		Initialize a set $\mathcal{V}_e^C \leftarrow \{u, v\}$\;
		\ForEach{edge $e'=(u',v')$ intersecting with $e$}{
			$\mathcal{V}_e^C \leftarrow \mathcal{V}_e^C \cup \{u', v'\}$}
		\ForEach{node $v \in \mathcal{V}_e^C$}{
			Compute $c_v \leftarrow \sum_{\varepsilon\in\mathcal{E}_v}C[\Gamma;\varepsilon]^2$\;}
		$\mathcal{V}_e^C \leftarrow{}$ Sort $\mathcal{V}_e^C$ according to the descending order of $c_v$\;
		\For{$i=1$ \KwTo $|\mathcal{V}_e^C|$}{
			Node $v \leftarrow i$-th element of $\mathcal{V}_e^C$\;
			Move $v$ to $p_v^*$ computed by Algorithm \ref{alg:optimal_node_moving}\;
		}
	}
	$\bar{\Gamma}(G)\leftarrow \Gamma$\;
	\KwRet{$\bar{\Gamma}(G)$}\;
\end{algorithm}

Visibility region analysis only ensures the local optimality of moving a single node. Questions of ``Which nodes need to be moved?'' and ``In what order should these nodes be moved?'' remain to be answered to reduce line crossings globally. After comparing several heuristics, \cite{Radermacher_Reichard_2019} concludes that the edge insertion heuristic tends to produce the best result in most cases. As shown in Algorithm \ref{alg:edge_insertion_heuristic}, the heuristic includes the following steps:
\begin{enumerate}
	\item Find out all intersecting edge pairs and remove an edge from each pair to obtain a planar layout.
	\item Iteratively insert the removed edges, one at a time, back to the planar layout. On inserting $e$, the terminal nodes of $e$ itself and any other edge intersecting with $e$ are selected as the nodes to be moved and are put into $\mathcal{V}^C_e$.
	\item Compute $c_v=\sum_{\varepsilon \in \mathcal{E}_v} C[\Gamma; \varepsilon]^2$ for each $v \in \mathcal{V}^C_e$, where $C[\Gamma; \varepsilon]$ represents the number of line crossings lying on an incident edge $\varepsilon$ of $v$.
	\item Move nodes in $\mathcal{V}^C_e$ to their respective optimal positions in the descending order of $c_v$.
\end{enumerate}

\subsection{Heuristics for Lowering Time Overhead}
\label{sub:heuristics_for_lowering_overheads}

Power transmission systems usually have strongly-meshed structures, making their topology diagrams likely to contain many line crossings. Unfortunately, directly applying \cite{Radermacher_Reichard_2019} to reduce line crossings for power transmission systems often results in unacceptable time overhead. That is because the elementary algorithm computing the optimal node position has a superquadratic complexity level. Moreover, eliminating an intersection requires calling the elementary algorithm for many times. To lower the time overhead, two heuristics are proposed below to simplify the original crossing reduction method.

\begin{figure}[!tb]
    \centering
    \subfloat[\label{fig:bfs_tree}]{\includegraphics{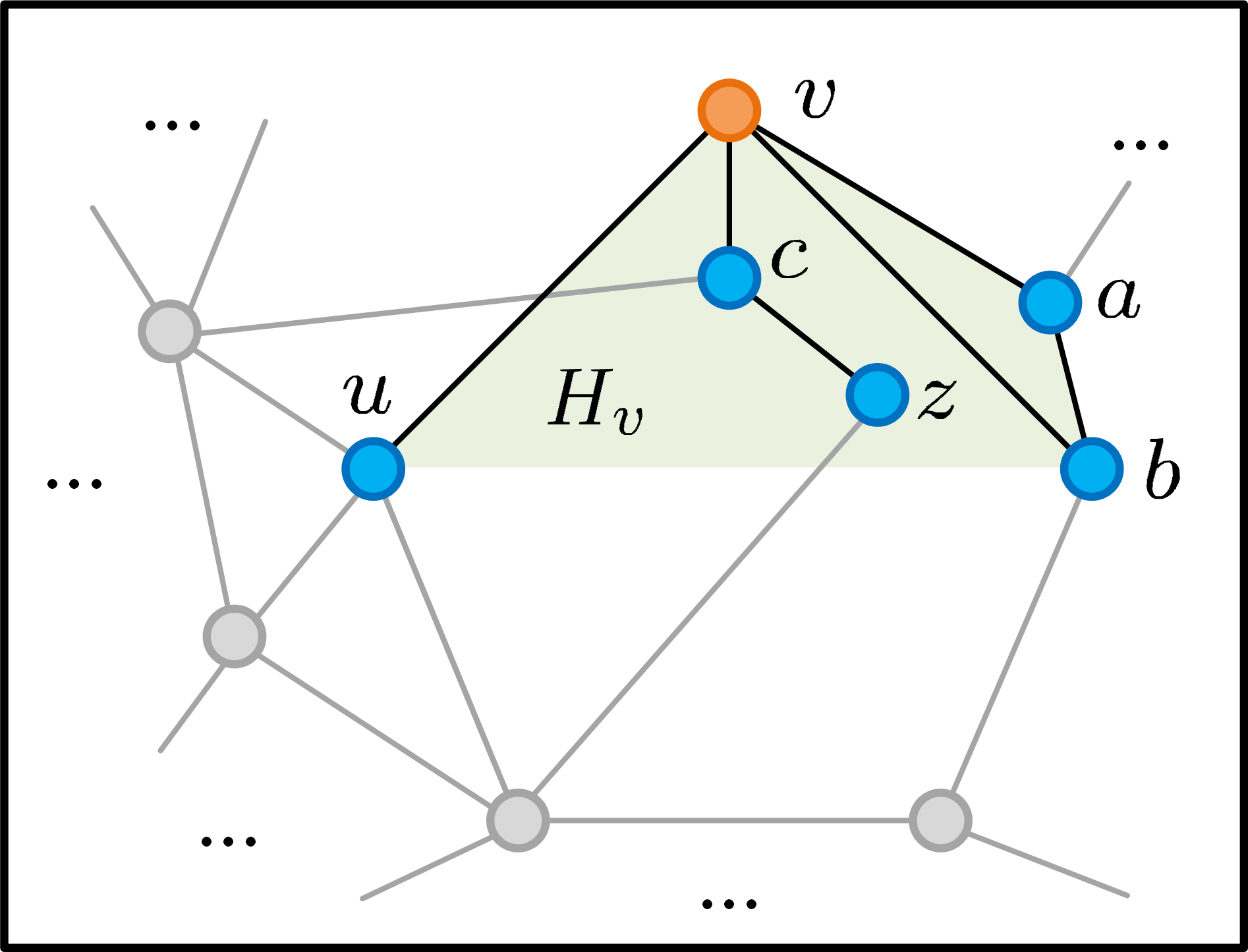}}
    \hfil
    \subfloat[\label{fig:partial_arrangement}]{\includegraphics{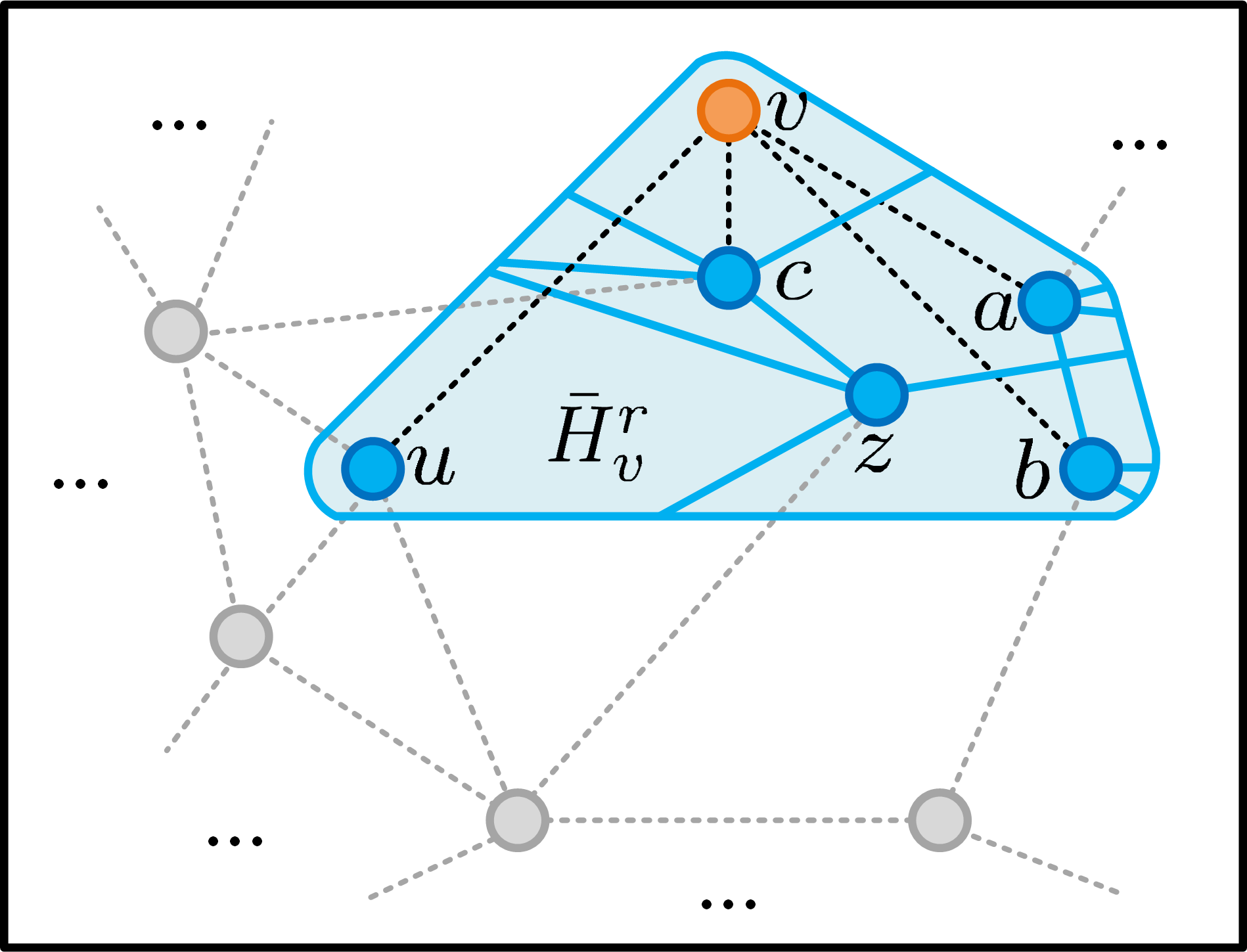}}
    \caption{The heuristic H1 restricts the space in which the optimal node position is searched. (a) The subgraph $G_v$ on $d_v^{\mathcal{T}} = 1$, with the convex hull $H_v$ colored in green. (b) The arrangement $\mathcal{A}_v$ built for $G_v$, with $\bar{H}_v^r$ shaded in blue.}
    \label{fig:princile_of_h2}
\end{figure}

\begin{algorithm}[t]
\caption{Simplification Heuristic H1} \label{alg:simplification_heuristic_h1}
	\SetAlgoLined
	\DontPrintSemicolon
	\KwIn{$G=(\mathcal{V},\mathcal{E})$, node $v$ to move, BFS-tree depth $d_v^\mathcal{T}$, expansion radius $r$}
	\KwOut{simplified arrangement $\mathcal{A}_v$}
	\vspace{2mm}
	$\mathcal{T}_v \leftarrow{}$BFS-tree rooted at $v$ with $d_v^T$ as the depth\;
	$H_v\leftarrow{}$convex hull of nodes in $\mathcal{T}_v$\;
	$G_v=(\hat{\mathcal{V}},\hat{\mathcal{E}}) \leftarrow{}$subgraph of $G$ with nodes not covered by $H_v$ removed\;
	Initialize a bounded canvas $\bar{H}_v^r \leftarrow{}$expand $H_v$ by $r$\;
	\ForEach{node $u\in \hat{\mathcal{V}}$}{
		\ForEach{edge $e \in \hat{\mathcal{E}}_v' \cap \hat{\mathcal{E}}_u'$}{
			Draw visibility region boundary $\mathcal{B}[u;e]$ on $\bar{H}_v^r$}}
	Arrangement $\mathcal{A}_v \leftarrow {}$divide $\bar{H}_v^r$ along boundaries into faces\;
	\KwRet{$\mathcal{A}_v$}\;
\end{algorithm}

The first heuristic H1 shown in Algorithm \ref{alg:simplification_heuristic_h1} restricts the space in which the optimal face $f^{*}_v$ of the node $v$ is searched. It is based on an observation that $f^{*}_v$ is unlikely to be geometrically far away from $v$. Instead of building the arrangement $\mathcal{A}$ for the whole diagram, H1 only computes an arrangement $\mathcal{A}_v$ for a subgraph $G_v$ around $v$. $G_v$ can be obtain by removing the nodes and edges not covered by the convex hull $H_v$ of a subset of nodes $\mathcal{T}_v$, with $\mathcal{T}_v$ a Breadth-First-Search (BFS) tree rooted at $v$ and having a tunable depth $d_v^{\mathcal{T}}$. As shown in Fig. \subref*{fig:bfs_tree}, given $d_v^{\mathcal{T}} = 1$, $\mathcal{T}_v=\{v,u,a,b,c\}$. Although $z \notin \mathcal{T}_v$, it lies within $H_v$, so it will be included in $G_v$. After $G_v$ is determined, $\mathcal{A}_v$ can be constructed by drawing the boundaries of visibility regions of $v$'s neighbors. In this process, unlike \cite{Radermacher_Reichard_2019} that allows rays to extend to the canvas border, H1 requires them to stop at the boundary of $\bar{H}_v^r$, which is obtained by expanding $H_v$ by a tunable radius $r$. With the extension of rays limited, $f^{*}_v$ can only be found in the blue region in Fig. \subref*{fig:partial_arrangement}, which avoids inducing new crossings to the unconcerned part of $\Gamma^0(G)$.

\begin{algorithm}[t]
\caption{Simplification Heuristic H2} \label{alg:simplification_heuristic_h2}
	\SetAlgoLined
	\DontPrintSemicolon
	\KwIn{planar layout $\Gamma^0$, removed edges $\mathcal{E}_r$}
	\KwOut{crossing-reduced layout $\bar{\Gamma}(G)$}
	\vspace{2mm}
	Initialize $\Gamma \leftarrow \Gamma^0$\;
	\ForEach{edge $e=(u,v) \in \mathcal{E}_r$}{
		Number of crossings $N_{-e} \leftarrow C[\Gamma]$\;
		$\Gamma' \leftarrow{}$add $e$ to $\Gamma$ by connecting $u$ and $v$\;
		$\Gamma' \leftarrow{}$move $u$ and $v$ to their optimal positions\;
		Number of crossings $N_{+e} \leftarrow C[\Gamma']$\;
		\eIf{$N_{+e} < N_{-e}$}{$\Gamma \leftarrow \Gamma'$\;}{
			$\Gamma' \leftarrow{}$add an inflection node $\hat{v}_e$ to $\Gamma$, on $e$\;
			$\Gamma' \leftarrow{}$move $\hat{v}_e$ to its optimal position\;
			Number of crossings $N_{+e} \leftarrow C[\Gamma']$\;
			\If{$N_{+e} < N_{-e}$}{$\Gamma \leftarrow \Gamma'$\;}}}
	$\bar{\Gamma}(G)\leftarrow \Gamma$\;
	\KwRet{$\bar{\Gamma}(G)$}\;
\end{algorithm}

The other heuristic H2 shown in Algorithm \ref{alg:simplification_heuristic_h2} reduces the number of nodes to be moved in each round of edge insertion. On inserting an edge $e$, H2 first tries to move the terminal nodes of $e$ to their optimal positions. If the total number of line crossings decreases, the current round is finished and the next edge is being considered; otherwise, H2 adds an inflection point $\hat{v}_e$ to $e$ and moves $\hat{v}_e$ to the best position. The rationality behind this is that adding inflection points does not impact the topological connectivity of edges but does increase the degree of freedom for crossing reduction. Consider the edge $(a, b)$ in Fig. \ref{fig:visibility_region_analysis}. If a virtual node is added to $(a,b)$ and dragged to the right of $v$, the two crossings originally on $(a, b)$ can be removed. Note that the added inflection points are kept into the final output only if moving item effectively reduces the total number of crossings. Instead of moving all nodes in $\mathcal{V}^C_e$ in each round of edge insertion, H2 only moves two or three nodes and thus enhances the efficiency of the method in \cite{Radermacher_Reichard_2019}.

\section{MILP-Based Layout Planning}
\label{sec:optimal_layout_generation}

Crossing reduction is necessary but not sufficient for drawing clear and beautiful topology diagrams, as it only improves the conformance to the EX criterion. In this section, an MILP model is constructed to optimize the crossing-reduced layout $\bar{\Gamma}(G)$ for better compliance with other criteria.

\subsection{Modeling of Constraints}
\label{sub:modeling_of_constraints}

\begin{figure}[!tb]
	\centerline{\includegraphics{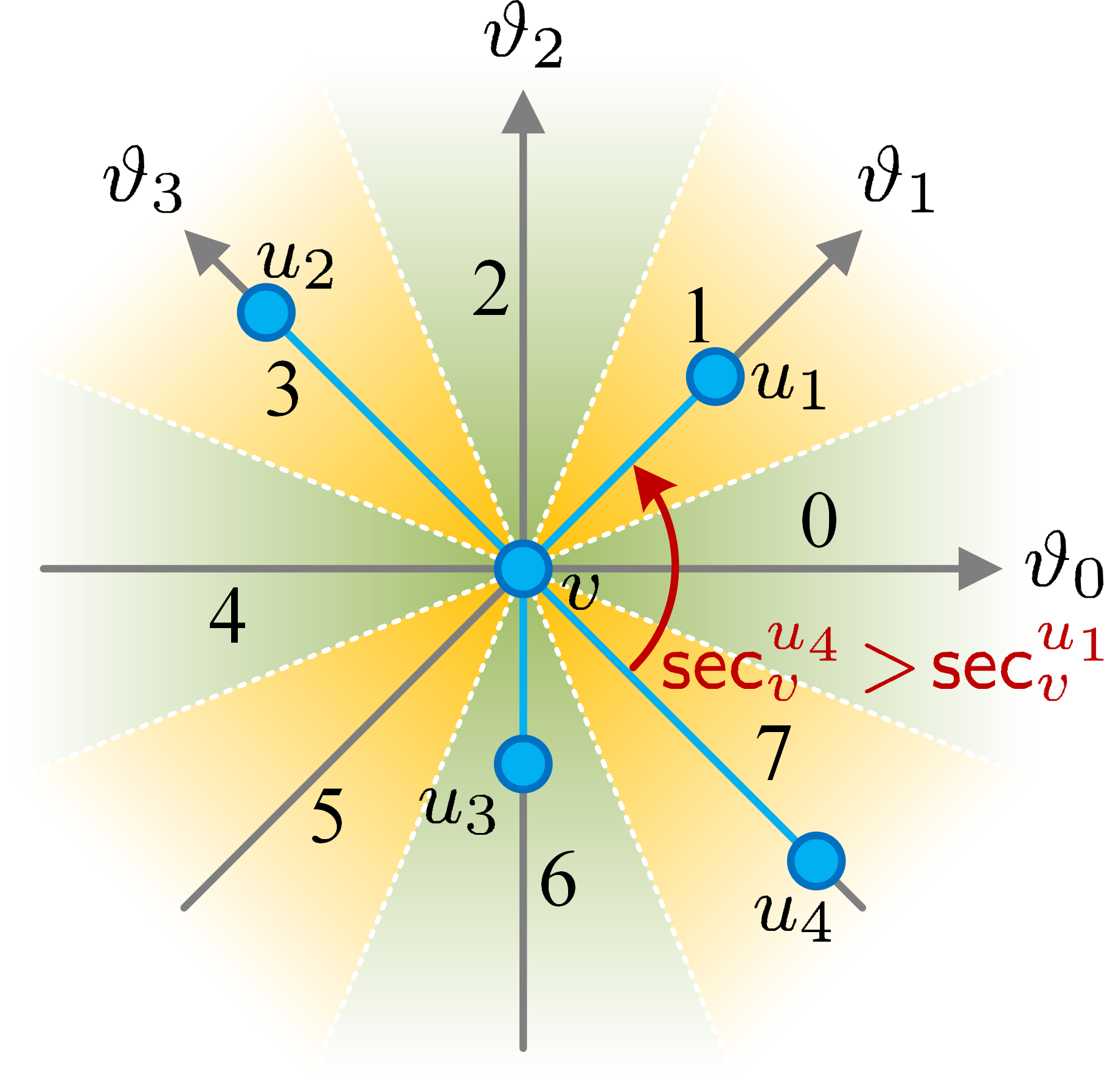}}
	\caption{A $K$-linear ($K=4$) coordinate system with sectors indexed.}
	\label{fig:k_linear_coordinate_system}
\end{figure}

The constraints of an optimization problem are propositions that all feasible solutions must obey. Four groups of constraints are included in the MILP model to make the yielded diagrams follow the high-priority aesthetic criteria listed in Table \ref{tab:aesthetic_criteria}.

\subsubsection{Coordinate system}
\label{ssub:coordinate_system}
The first group of constraints sets up a $K$-linear coordinate system like the one illustrated in Fig. \ref{fig:k_linear_coordinate_system}. The $K$ axes are defined to be along a set of equiangular directions $\mathcal{C}_K = \{\vartheta_i = i\pi/K \mid i = 0,1,\dots,K-1 \}$. The plane is divided into $2K$ sectors, each of which is bisected by the positive or negative part of an axis. The sector index $j$ and the axis index $i$ follow $i = j \mod K$. Suppose that $(\mathsf{x}_v, \mathsf{y}_v)$ represent the Euclidean coordinates of a node $v$, and that $\mathsf{z}^i_v$ is its coordinate on the $i$-th axis of $\mathcal{C}_K$. The relationship between $\mathsf{z}^i_v$ and $(\mathsf{x}_v, \mathsf{y}_v)$ is constrained by
\begin{equation}
	\mathsf{z}^i_v = \cos\vartheta_i \cdot \mathsf{x}_v + \sin\vartheta_i \cdot \mathsf{y}_v
\end{equation}

Denote the coordinate system rotating $\mathcal{C}_K$ by a counterclockwise $\pi/2$ as $\bar{\mathcal{C}}_K$. The coordinate of $v$ on the $i$-th axis of $\bar{\mathcal{C}}_K$ is constrained by
\begin{equation}
	\bar{\mathsf{z}}^i_v = -\sin\vartheta_i \cdot \mathsf{x}_v + \cos\vartheta_i \cdot \mathsf{y}_v
	\label{eq:orthogonal_coordinate}
\end{equation}

During layout planning, nodes are not allowed to be placed arbitrarily, but rather in a way that ensures each edge to be parallel with an axis of $\mathcal{C}_K$. If $(u,v)$ is parallel with $\vartheta_i$, $u$ and $v$ should have the same coordinate on the $i$-th axis of $\bar{\mathcal{C}}_K$. The relationship between $\bar{\mathsf{z}}^i_u$ and $\bar{\mathsf{z}}^i_v$ can be constrained by
\begin{equation}
	\begin{gathered}
		\bar{\mathsf{z}}_u^i - \bar{\mathsf{z}}_v^i \leq M(1 - \mathsf{in}^{u_v}_j) \\
		\bar{\mathsf{z}}_v^i - \bar{\mathsf{z}}_u^i \leq M(1 - \mathsf{in}^{u_v}_j)
	\end{gathered}
	\label{eq:alignment_on_orthogonal}
\end{equation}
where $M$ is a large positive constant and $\mathsf{in}^{u_v}_j$ is a binary variable indicating whether $u$ lies in the $j$-th sector if $v$ locates at the origin. If $\mathsf{in}^{u_v}_j = 0$, constraints in \eqref{eq:alignment_on_orthogonal} can be trivially satisfied as long as $M$ is sufficiently large. However, if $\mathsf{in}^{u_v}_j = 1$, constraints in \eqref{eq:alignment_on_orthogonal} can only be satisfied if $\bar{\mathsf{z}}_u^i = \bar{\mathsf{z}}_v^i$. Thus, the above constraints can ensure the edge $(u, v)$ to be parallel with $\vartheta_i$ if $u$ lies in the sector $j$ with respect to $v$.

The purpose of using the $K$-linear coordinate system is twofold. On the one hand, forcing edges to be parallel with axes of $\mathcal{C}_K$ simplifies the layout planning problem, allowing linear programming to model complex aesthetic rules. On the other hand, the minimum included angle between consecutive incident edges in the output diagram is equal to $\pi / K$. The IA criterion in Table \ref{tab:aesthetic_criteria} can be met as long as $K$ is not too large.

\subsubsection{Relative position}
\label{ssub:combinatorial_embedding}
The second group of constraints keeps the relative position of nodes in the output layout $\hat{\Gamma}(G)$ roughly identical to the input layout $\bar{\Gamma}(G)$ so that the RP criterion can be satisfied. This goal is realized in three stages, two of which introduce constraints to the MILP model, with the other participating in forming the optimization objective. In the first stage, constraints are formulated to maintain the cyclic order of each node's neighbors. Suppose that $\mathsf{sec}^u_v$ represents the index of the sector where $u$ locates with respect to $v$. Maintaining the cyclic order of $u_0, u_1, \dots, u_{n_v} \in \mathcal{N}_v$ is equivalent to requiring $\mathsf{sec}^{u_0}_v, \mathsf{sec}^{u_1}_v, \dots, \mathsf{sec}^{u_{n_v}}_v$ to be strictly increasing, except the case going from the last used sector to the first one (e.g., the case marked in Fig. \ref{fig:k_linear_coordinate_system}). This requirement can be constrained by
\begin{equation}
	\begin{gathered}
		\begin{aligned}
			\mathsf{sec}_v^{u_1} + 1 &\leq \mathsf{sec}_v^{u_2} + 2K \cdot \mathsf{exc}^1_v \\
			\mathsf{sec}_v^{u_2} + 1 &\leq \mathsf{sec}_v^{u_3} + 2K \cdot \mathsf{exc}^2_v \\
			&\vdots \\
			\mathsf{sec}_v^{u_{n_v}} + 1 &\leq \mathsf{sec}_v^{u_1} + 2K \cdot \mathsf{exc}^{n_v}_v
		\end{aligned}\\
		\sum_{i=1}^{n_v} \mathsf{exc}^i_v = 1 \\
	\end{gathered}
	\label{eq:combinatorial_embedding}
\end{equation}
where $\mathsf{exc}^i_v$ is a binary variable indicating whether the sector where $u_i$ lies is the last used sector. 

Constraints in \eqref{eq:combinatorial_embedding} make $\hat{\Gamma}(G)$ and $\bar{\Gamma}(G)$ have the same combinatorial embedding, but they do not maintain edge directions. In the second stage, edge directions in $\hat{\Gamma}(G)$ are constrained to be close to those in $\bar{\Gamma}(G)$. Suppose that the index of the sector where $u$ locates with respect to $v$ in the input layout $\bar{\Gamma}(G)$ is $\sigma^u_v$. To model the similarity instead of identicality of edge directions between input and output layouts, $\mathsf{sec}^u_v$ should not be restricted to $\sigma^u_v$, but rather confined to a set $\mathcal{S}^u_v = [\sigma^u_v - s, \sigma^u_v + s]$, where $s \geq 1$ is a tunable parameter introduced to allow some flexibility. The inclusion relation between $\mathsf{sec}^u_v$ and $\mathcal{S}^u_v$ can be modeled as
\begin{equation}
	\begin{gathered}
		\sum_{j=0}^{2K-1} \mathbbm{1}[j \in \mathcal{S}_v^u] \cdot \mathsf{in}^{u_v}_j = 1 \\
		\begin{aligned}
			\mathsf{sec}^u_v = &\sum_{j=0}^{2K-1} j \cdot \mathbbm{1}[j \in \mathcal{S}_v^u] \cdot \mathsf{in}^{u_v}_j \\
			\mathsf{sec}^v_u = &~ \mathsf{sec}^u_v + K
		\end{aligned}
	\end{gathered}
	\label{eq:relative_position_constraint}
\end{equation}
where $\mathbbm{1}[\cdot]$ is the characteristic function that returns one if the condition holds and otherwise zero. 

The last stage minimizes the difference between $\mathsf{sec}^u_v$ and $\sigma^u_v$, which will be discussed in detail in Section \ref{sub:modeling_of_objectives}.

\subsubsection{Edge length}
\label{ssub:edge_length}
The third group of constraints is introduced to avoid too short edges in $\hat{\Gamma}(G)$. The idea is to keep the terminal nodes $u$ and $v$ of each edge $(u, v)$ apart for at least a certain length $\ell_{\min}$ along the direction of $(u, v)$. This group of constraints can be constructed as
\begin{equation}
	\begin{cases}
		\mathsf{z}_u^i - \mathsf{z}_v^i \geq -M(1-\mathsf{in}^{u_v}_j) + \ell_{\min} & \text{if} \quad j < K \\
		\mathsf{z}_v^i - \mathsf{z}_u^i \geq -M(1-\mathsf{in}^{u_v}_j) + \ell_{\min} & \text{if} \quad j \geq K
	\end{cases}
	\label{eq:edge_length}
\end{equation}
If $\mathsf{in}^{u_v}_j = 0$, constraints in \eqref{eq:edge_length} can be trivially satisfied. Otherwise, \eqref{eq:edge_length} is degenerated into $|\mathsf{z}_u^i - \mathsf{z}_v^i| \geq \ell_{\min}$.

\subsubsection{Planarity}
\label{ssub:node_distance}
The crossing reduction method in Section \ref{sec:edge_crossing_reduction} cannot always remove all intersections from $\Gamma^0(G)$. A dummy node is placed at each remaining crossing to ensure that the $\bar{\Gamma}(G)$ inputting into the layout planning procedure is planar. On this basis, the last group of constraints is introduced to prevent layout optimization from causing new line crossings. Note that a pair of edges $e=(u,v)$ and $e'=(u',v')$ do not intersect if and only if they are separable in at least one direction in $\mathcal{C}_K$. Hence, the following constraints should be modeled for each pair of edges that may cross with each other 
\begin{equation}
	\begin{gathered}
		\sum_{j=0}^{2K-1} \mathsf{sep}^{ee'}_j \geq 1 \\
		\begin{aligned}
			\mathsf{z}_{u'}^i - \mathsf{z}_u^i &\geq -M(1-\mathsf{sep}^{ee'}_j) + d_{\min}\\
			\mathsf{z}_{u'}^i - \mathsf{z}_v^i &\geq -M(1-\mathsf{sep}^{ee'}_j) + d_{\min}\\
			\mathsf{z}_{v'}^i - \mathsf{z}_u^i &\geq -M(1-\mathsf{sep}^{ee'}_j) + d_{\min}\\
			\mathsf{z}_{v'}^i - \mathsf{z}_v^i &\geq -M(1-\mathsf{sep}^{ee'}_j) + d_{\min}
		\end{aligned}
	\end{gathered}
\label{eq:planarity}
\end{equation}
where $\mathsf{sep}^{ee'}_j$ is a binary variable indicating whether $e$ and $e'$ can be separated along the bisecting ray of the sector $j$, and $d_{\min}$ is the minimum allowable distance between two edges. Note that bisecting rays are assumed to originate from the origin. Thus, the last four constraints in \eqref{eq:planarity} are applicable only if $j < K$. For $j \geq K$, subtraction terms in the left-hand side of the inequations need to be inverted.

In theory, if constraints in \eqref{eq:planarity} are formulated for each pair of non-adjacent edges in each face of $\bar{\Gamma}(G)$, new crossings are guaranteed not to appear. Nonetheless, this strategy may add a large number of constraints to the MILP problem, as eligible edge pairs will increase drastically with the scale of the graph. It makes the efficiency of solving the layout planning problem pretty low. However, in practice, which pairs of edges are prone to intersect during layout optimization is often subject to the structure of the input layout. Most non-adjacent edges in the same face will likely not intersect with each other even if the constraints in \eqref{eq:planarity} are not explicitly added. Thus, to enhance efficiency, the layout planning process can be started with this group of constraints completely omitted. By iteratively adding constraints corresponding to each pair of intersecting edges in last round's output, a final layout that contains no line crossing is likely to be derived in much shorter time. 

\subsection{Modeling of Optimization Objective}
\label{sub:modeling_of_objectives}

Unlike constraints, optimization objectives of MILP models are not used to set up bottom lines to screen out feasible solutions but instead to find better solutions that can improve the aesthetic quality of the output layout. It is formulated as
\begin{equation}
	\min\quad \sum_{i \in \{\text{RP,OR,EV}\}} w_i \cdot \mathsf{cost}_i
	\label{eq:objective}
\end{equation}
i.e., minimizing the weighted average of three cost terms quantifying the deviation of the output layout from the low-priority aesthetic criteria RP, OR, and NE. In \eqref{eq:objective}, $w_i$ is the user-specified weight of $\mathsf{cost}_i$, with $w_\text{RP} + w_\text{OR} + w_\text{NE} =1$. The definition of each cost term is delineated as follows.

\subsubsection{Relative position}
\label{ssub:relative_position_objective}
The first cost term in \eqref{eq:objective} works along with the second group of constraints to preserve the relative position of nodes to the maximum extent. Recall that a tunable parameter $s$ is involved in \eqref{eq:relative_position_constraint} to leave some flexibility for the MILP model to explore possibly better layouts. The goal of the cost term here is to keep the exploration conservative. It is realized by minimizing the total difference between edge directions in $\bar{\Gamma}(G)$ and $\hat{\Gamma}(G)$, i.e., $\sum_{(u,v) \in \mathcal{E}}|\sigma^u_v - \mathsf{sec}^u_v|$. By defining an integer variable $\mathsf{diff}^u_v$ and imposing
\begin{equation}
	\begin{aligned}
		\sigma^u_v - \mathsf{sec}^u_v &\leq \mathsf{diff}_v^u \\
		\mathsf{sec}^u_v - \sigma^u_v &\leq \mathsf{diff}_v^u \\
	\end{aligned}
\end{equation}
the cost term corresponding to RP can be formulated as 
\begin{equation}
	\mathsf{cost}_\text{RP} = \sum_{(u, v) \in \mathcal{E}} \mathsf{diff}_v^u
	\label{eq:cost_rp}
\end{equation}

\subsubsection{Line orthogonality}
\label{ssub:line_orthogonality}
This cost term aims to increase the line orthogonality of $\hat{\Gamma}(G)$, improving the conformance to the OR criterion. It can be implemented by defining a binary variable $\mathsf{hor}^u_v$ and an optional binary variable $\mathsf{ver}^u_v$ for each edge $(u,v) \in \mathcal{E}$ to indicate whether the edge is horizontal or vertical. Note that $\mathsf{ver}^u_v$ is needed only when $K$ is an even number; otherwise, no edges will be in the vertical direction. If $\mathsf{hor}^u_v = 1$, $\mathsf{sec}^u_v$ should be equal to zero. Instead, if $\mathsf{ver}^u_v = 1$, $\mathsf{sec}^u_v$ should be equal to $K / 2$. Thus, by setting up
\begin{equation}
	\begin{aligned}
		\mathsf{sec}^u_v \leq M(1-\mathsf{hor}^u_v) \\
		-\mathsf{sec}^u_v \leq M(1-\mathsf{hor}^u_v) \\
		\mathsf{sec}^u_v - K / 2 \leq M(1-\mathsf{ver}^u_v) \\
		-(\mathsf{sec}^u_v - K / 2) \leq M(1-\mathsf{ver}^u_v) \\
	\end{aligned}
	\label{eq:objectives_3}
\end{equation}
the cost term corresponding to OR can be defined as
\begin{equation}
	\mathsf{cost}_\text{OR}=\sum_{(u,v)\in\mathcal{E}}\left(1-\mathsf{hor}^u_v-\mathsf{ver}^u_v\right)
	\label{eq:cost_or}
\end{equation}
which returns one if the edge $(u,v)$ is neither horizontal nor vertical; otherwise, it returns zero.

\subsubsection{Edge length compactness and evenness}
\label{ssub:edge_length_evenness}
The third cost term is applied to improve the compactness and evenness of edge lengths, which is beneficial for enhancing the spatial evenness of the diagram. The Euclidean length of an edge $(u, v)$ in the $K$-linear coordinate system is defined as the maximum absolute value $|\mathsf{z}^i_u - \mathsf{z}^i_v|$ in all $\vartheta_i \in \mathcal{C}_K$, which can be modeled by introducing an integer variable $\mathsf{len}_v^u$ and setting up the following constraints
\begin{equation}
	\begin{aligned}
		\mathsf{z}_u^i - \mathsf{z}_v^i &\leq \mathsf{len}_v^u \\
		\mathsf{z}_v^i - \mathsf{z}_u^i &\leq \mathsf{len}_v^u \\
	\end{aligned}
\end{equation}
for $i=0,1,\dots,K-1$. The deviation of $\mathsf{len}_v^u$ can then be modeled by defining an integer variable $\mathsf{dev}_v^u$ and setting up
\begin{equation}
	\begin{aligned}
		\mathsf{dev}_v^u & \leq \mathsf{len}_v^u - \frac{1}{m}\sum_{(u, v) \in \mathcal{E}} \mathsf{len}_v^u\\
		-\mathsf{dev}_v^u & \leq \mathsf{len}_v^u - \frac{1}{m}\sum_{(u, v) \in \mathcal{E}} \mathsf{len}_v^u \\
	\end{aligned}
\end{equation}
where $m$ is the total number of edges in the layout. The cost term corresponding to EV can thus be defined as
\begin{equation}
	\mathsf{cost}_\text{EV} = \frac{1}{m} \sum_{(u,v) \in \mathcal{E}} \mathsf{len}_v^u + \frac{1}{m} \sum_{(u, v) \in \mathcal{E}} \mathsf{dev}_v^u
\end{equation}

It should be noted that although the proposed framework aim to enhance the seven aesthetic criteria listed in Table \ref{tab:aesthetic_criteria}, discussions in Section \ref{sec:grid_topology_aesthetics} are not exclusive. Depending on the demonstration demands of different applications, not all of the aesthetic criteria covered in this paper are necessary to be included in the MILP optimization model in practical applications. Also, users are free to model other aesthetic criteria as optimization objectives and constraints as needed to customize the proposed framework. Compared with existing topology diagram generation approaches, the proposed framework has better practicability as it leaves room for user customization.

\section{Case Studies}
\label{sec:case_studies}

\renewcommand{\arraystretch}{1.3}
\setlength{\tabcolsep}{5pt}
\begin{table*}[!t]
  \centering
  \caption{Topology Diagrams of Different Grid Models}
    \begin{tabular}{*{4}{c}@{\hskip 6mm}*{2}{m{4em}<{\centering}}}
    \toprule
    \multirow[c]{2}[2]{*}{\textbf{Models}} & \multicolumn{3}{c}{\textbf{Layouts}} & \multicolumn{2}{c}{\textbf{Parameters}} \\ \cmidrule{2-6}
    & Initial $\Gamma^0(G)$ & Crossing-reduced $\bar{\Gamma}(G)$ & Optimized $\hat{\Gamma}(G)$ & Name & Value \\ \midrule
    \multirow[c]{7}{*}{\shortstack{IEEE 30\\[1em] \\ Nodes: 30 \\ Edges: 41 \\[1em] Crossings:\\ Initial: 3\\Final: 0}} & 
    \multirow[c]{7}{*}{\includegraphics[width=1.1in]{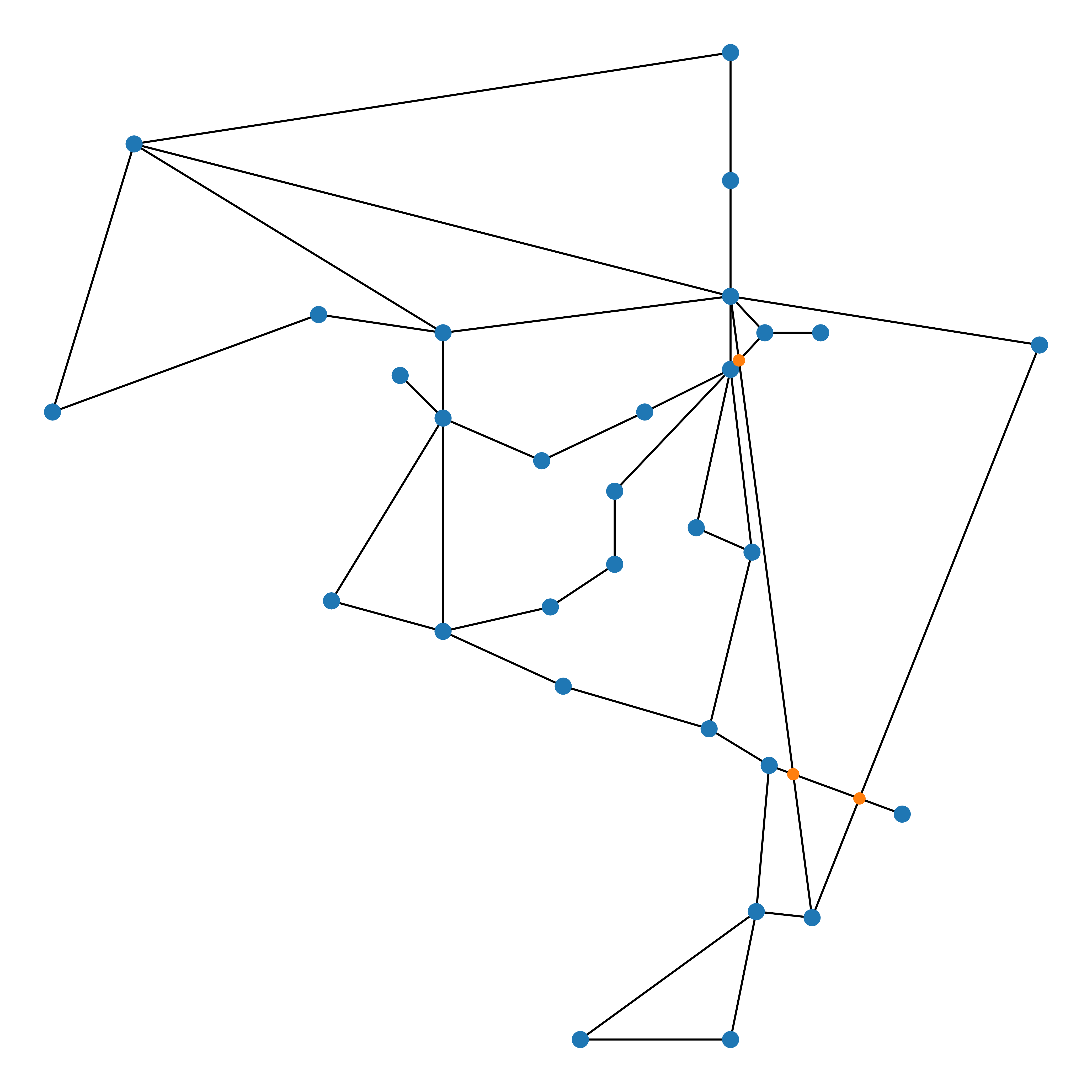}} & 
    \multirow[c]{7}{*}{\includegraphics[width=1.1in]{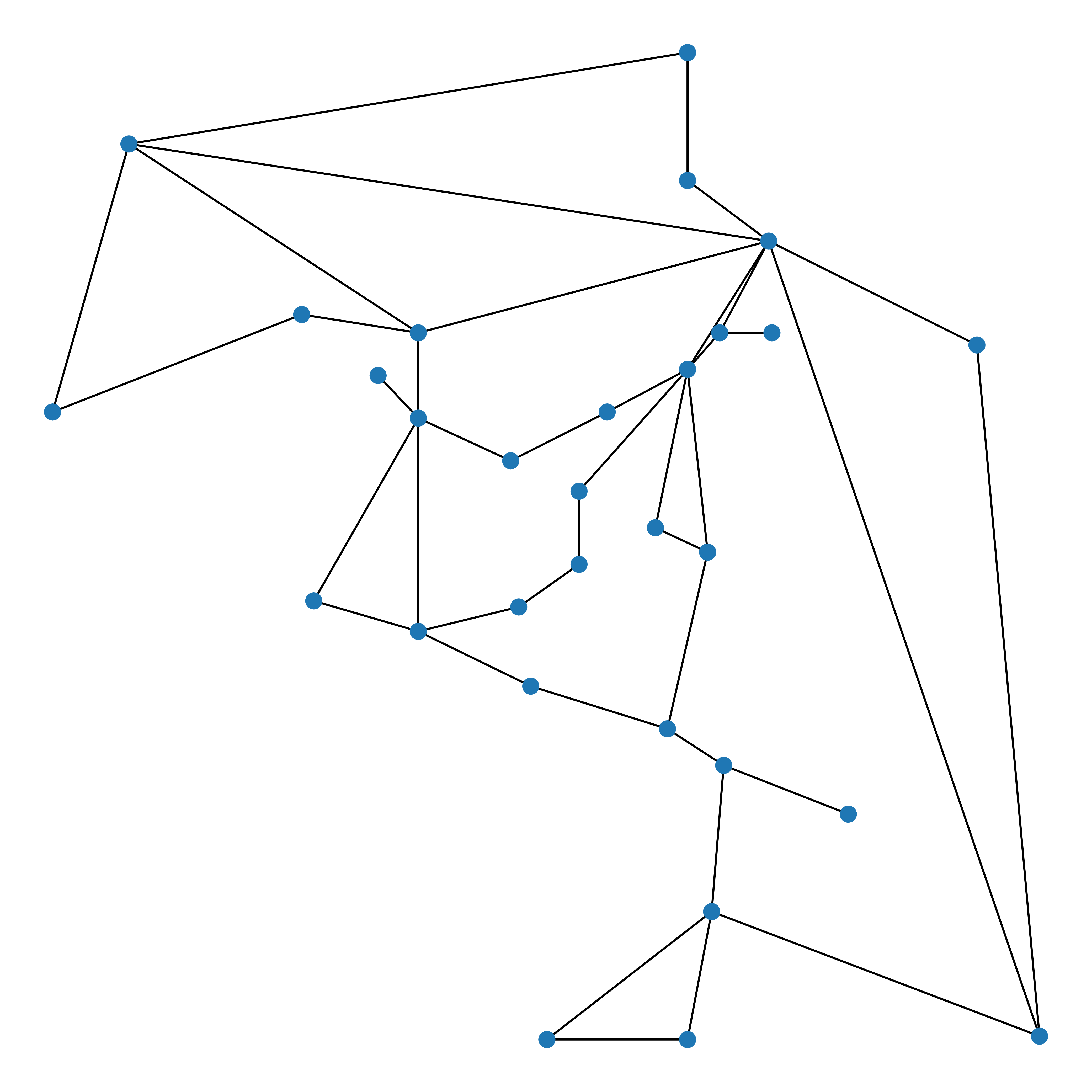}} & 
    \multirow[c]{7}{*}{\includegraphics[width=1.1in]{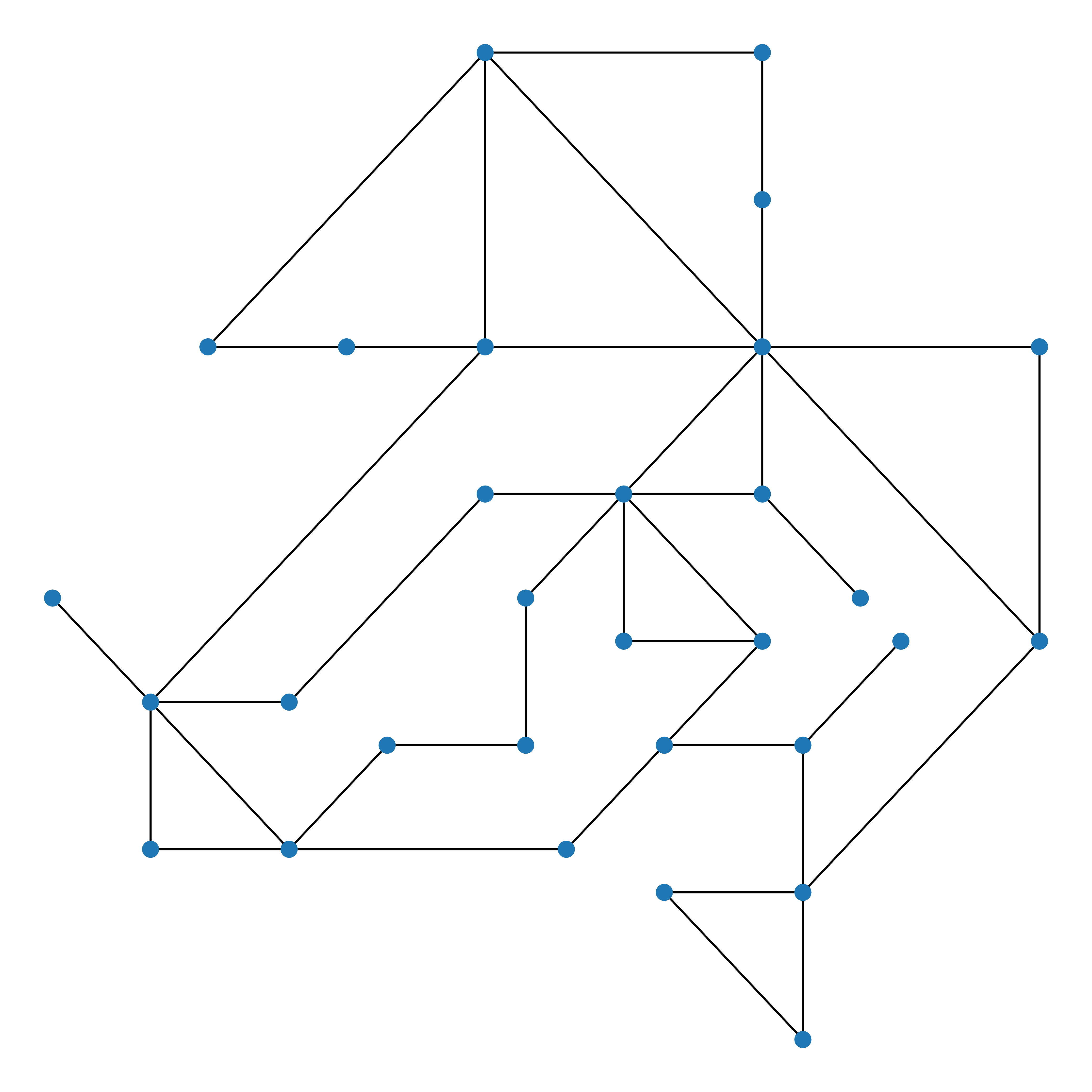}} & $K$ & 4 \\
    & & & & $s$ & 1 \\
    & & & & $\ell_{\min}$ & 2 \\
    & & & & $d_{\min}$ & 1 \\
    & & & & $w_{\text{RP}}$ & 0.1 \\
    & & & & $w_{\text{OR}}$ & 0.4 \\
    & & & & $w_{\text{EV}}$ & 0.5 \\ \midrule
    \multirow[c]{7}{*}{\shortstack{Grid A \\[1em] Nodes: 39 \\ Edges: 50 \\[1em] Crossings:\\ Initial: 1\\Final: 1}} & 
    \multirow[c]{7}{*}{\includegraphics[width=1.1in]{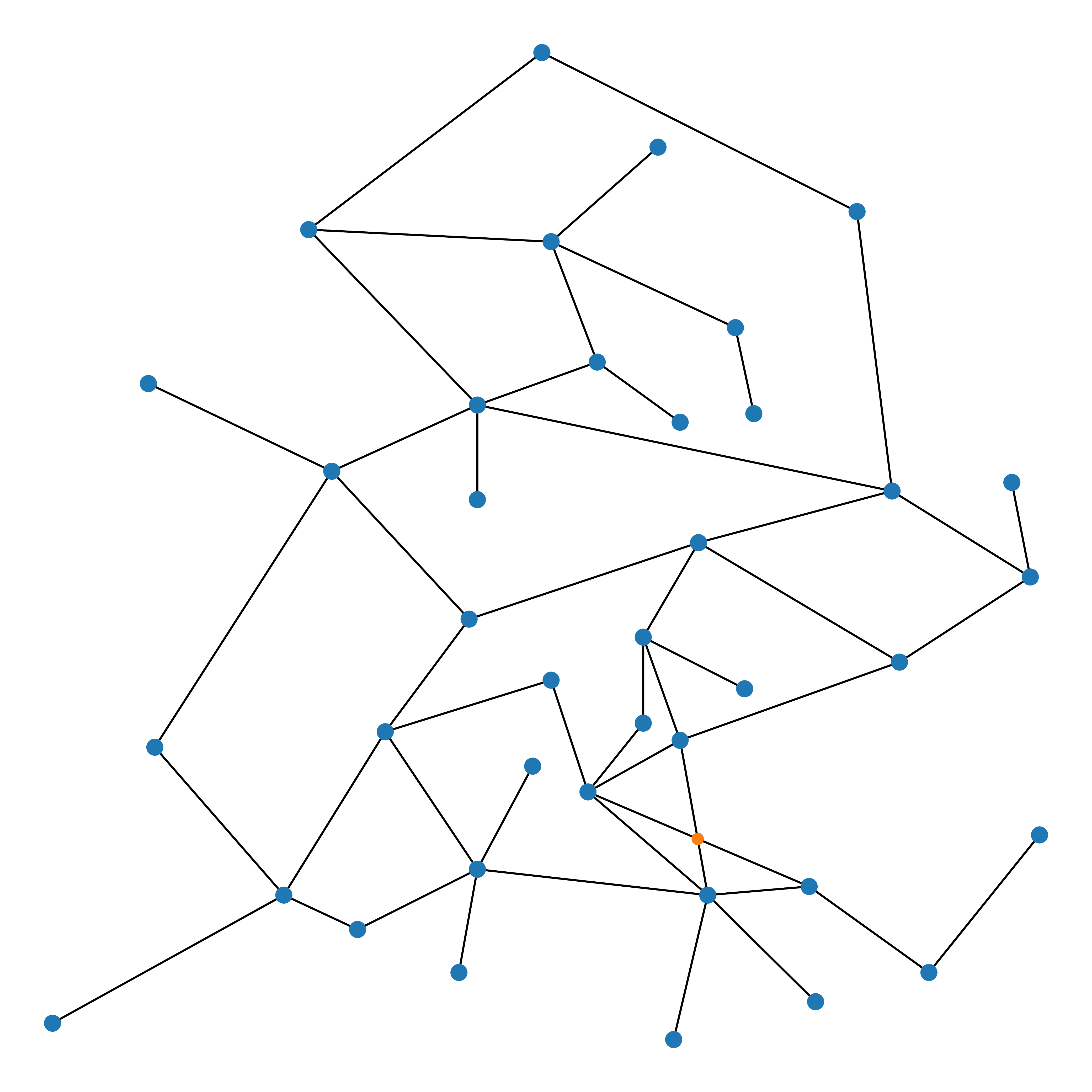}} & 
    \multirow[c]{7}{*}{\includegraphics[width=1.1in]{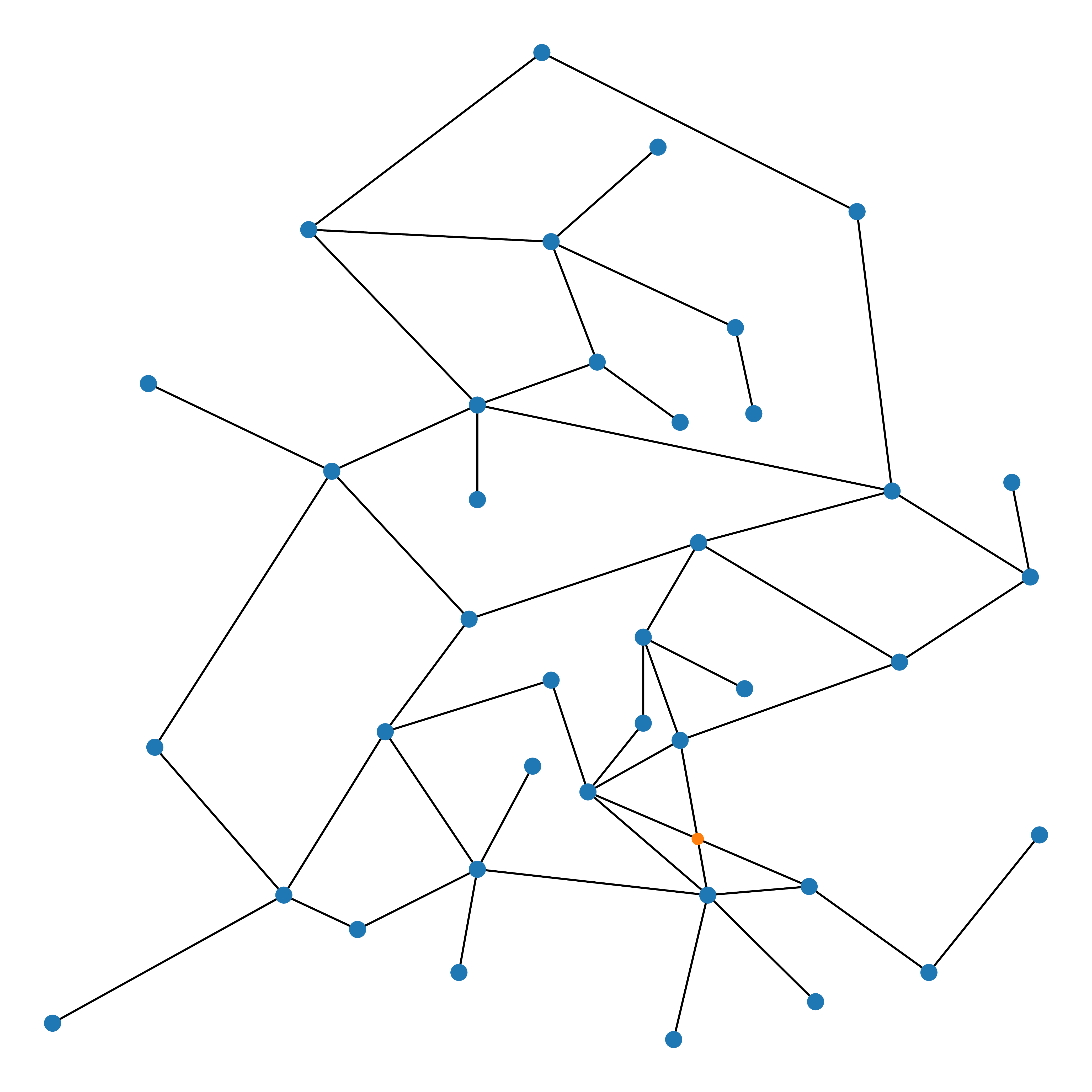}} & 
    \multirow[c]{7}{*}{\includegraphics[width=1.1in]{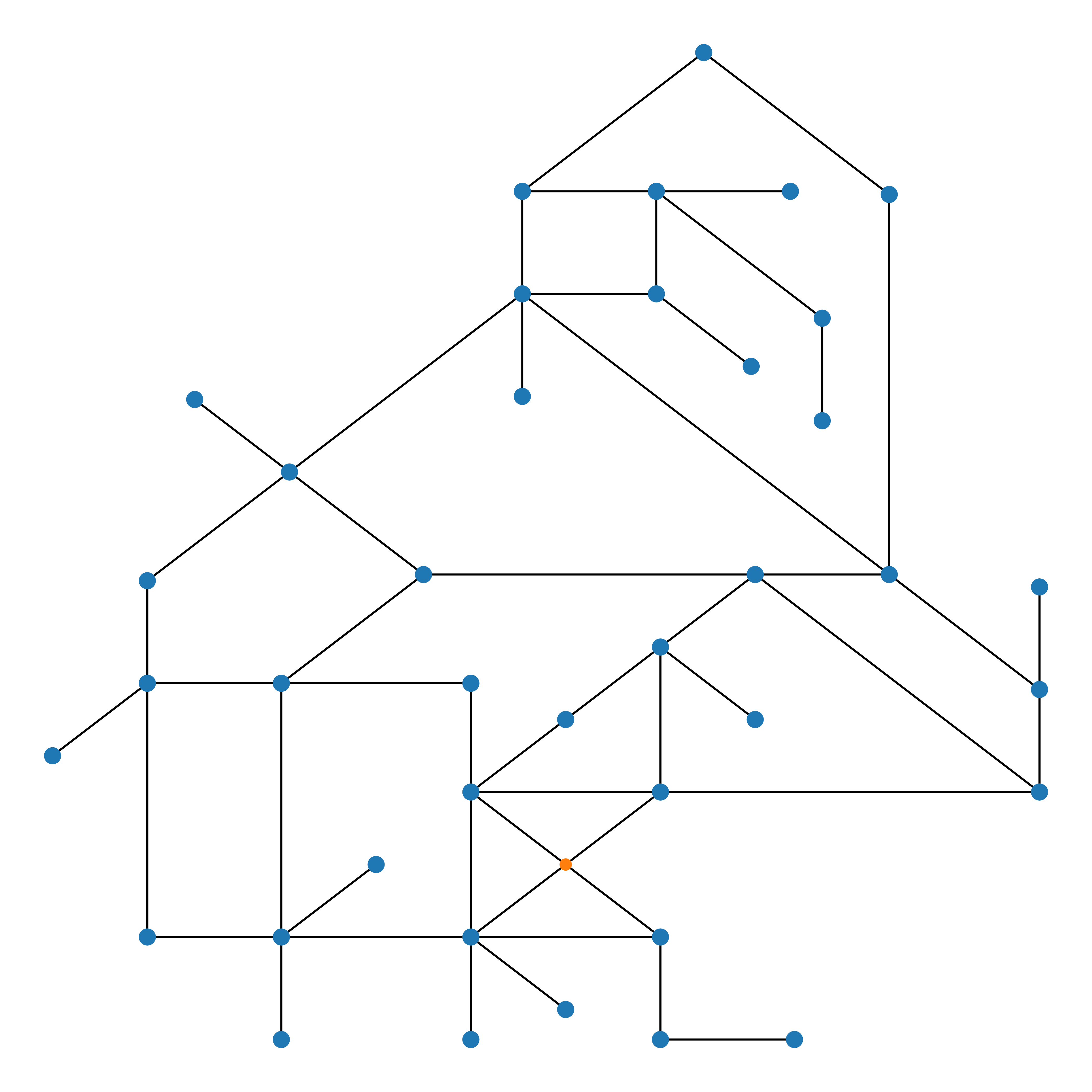}} & $K$ & 4 \\
    & & & & $s$ & 1 \\
    & & & & $\ell_{\min}$ & 2 \\
    & & & & $d_{\min}$ & 1 \\
    & & & & $w_{\text{RP}}$ & 0.2 \\
    & & & & $w_{\text{OR}}$ & 0.3 \\
    & & & & $w_{\text{EV}}$ & 0.5 \\ \midrule
    \multirow[c]{7}{*}{\shortstack{IEEE 57 \\[1em] Nodes: 57 \\ Edges: 80 \\[1em] Crossings:\\ Initial: 14\\Final: 5}} & 
    \multirow[c]{7}{*}{\includegraphics[width=1.1in]{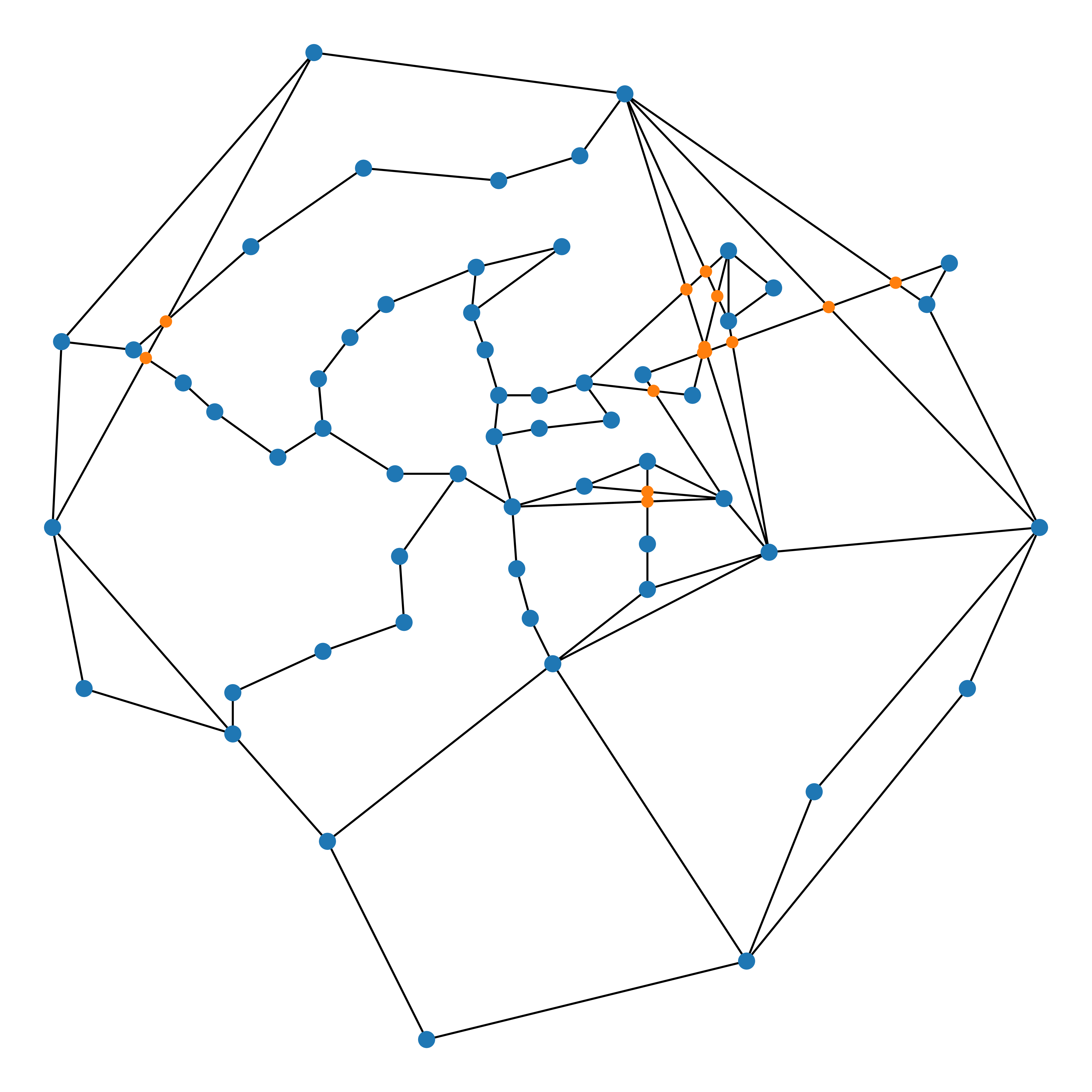}} & 
    \multirow[c]{7}{*}{\includegraphics[width=1.1in]{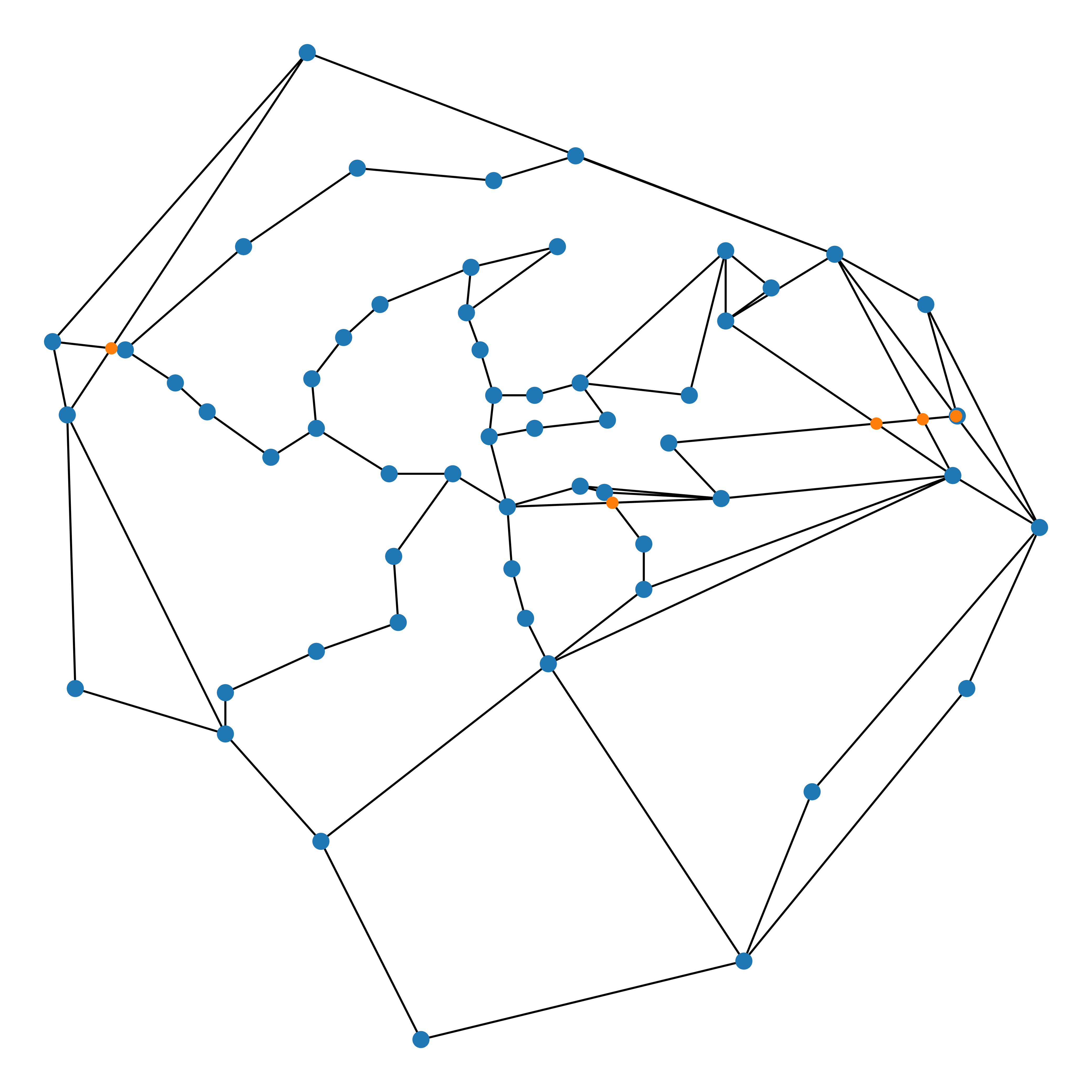}} & 
    \multirow[c]{7}{*}{\includegraphics[width=1.1in]{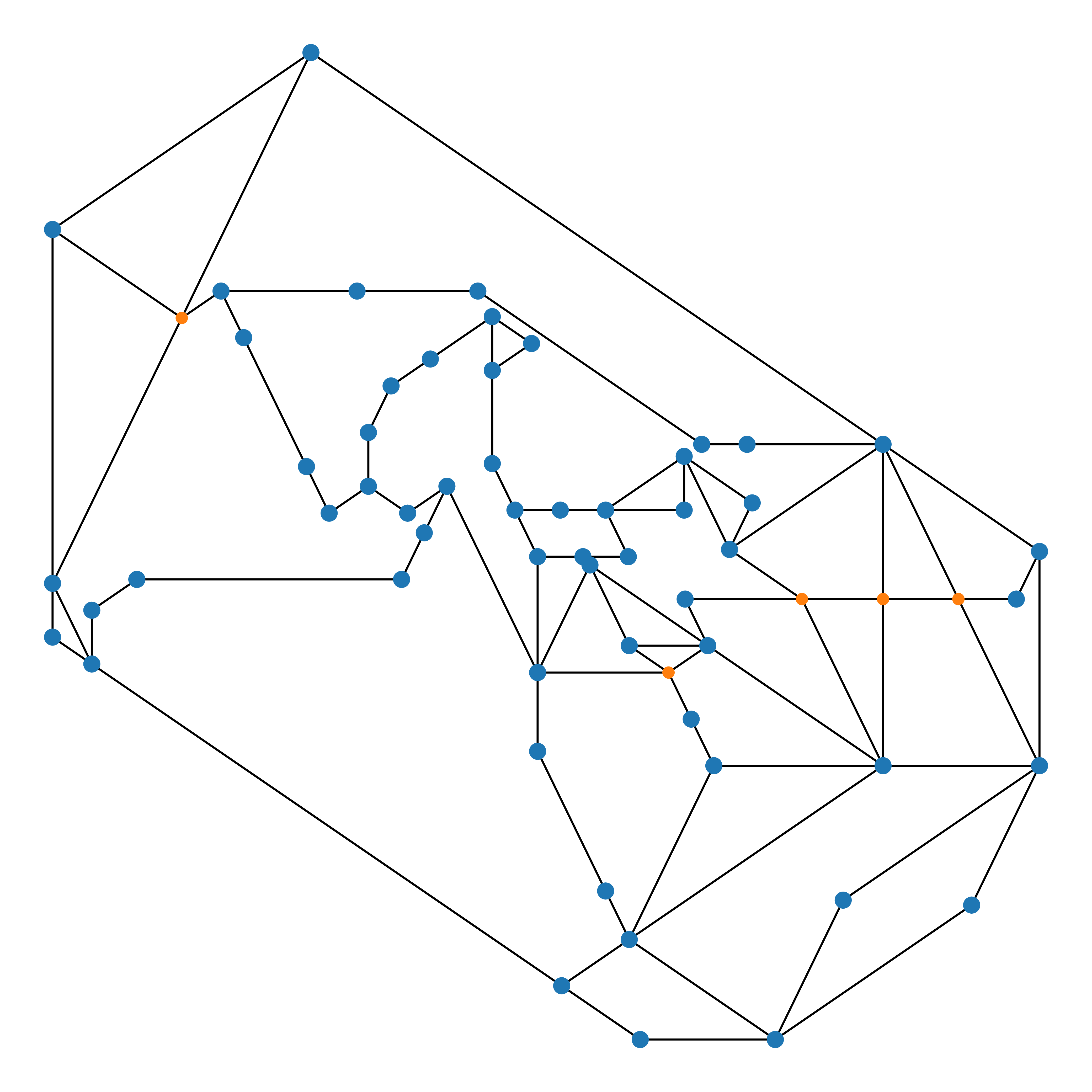}} & $K$ & 6 \\
    & & & & $s$ & 1 \\
    & & & & $\ell_{\min}$ & 1 \\
    & & & & $d_{\min}$ & 0.1 \\
    & & & & $w_{\text{RP}}$ & 0.2 \\
    & & & & $w_{\text{OR}}$ & 0.3 \\
    & & & & $w_{\text{EV}}$ & 0.5 \\ \midrule
    \multirow[c]{7}{*}{\shortstack{Grid B \\[1em] Nodes: 67 \\ Edges: 89 \\[1em] Crossings:\\ Initial: 5\\Final: 3}} & 
    \multirow[c]{7}{*}{\includegraphics[width=1.1in]{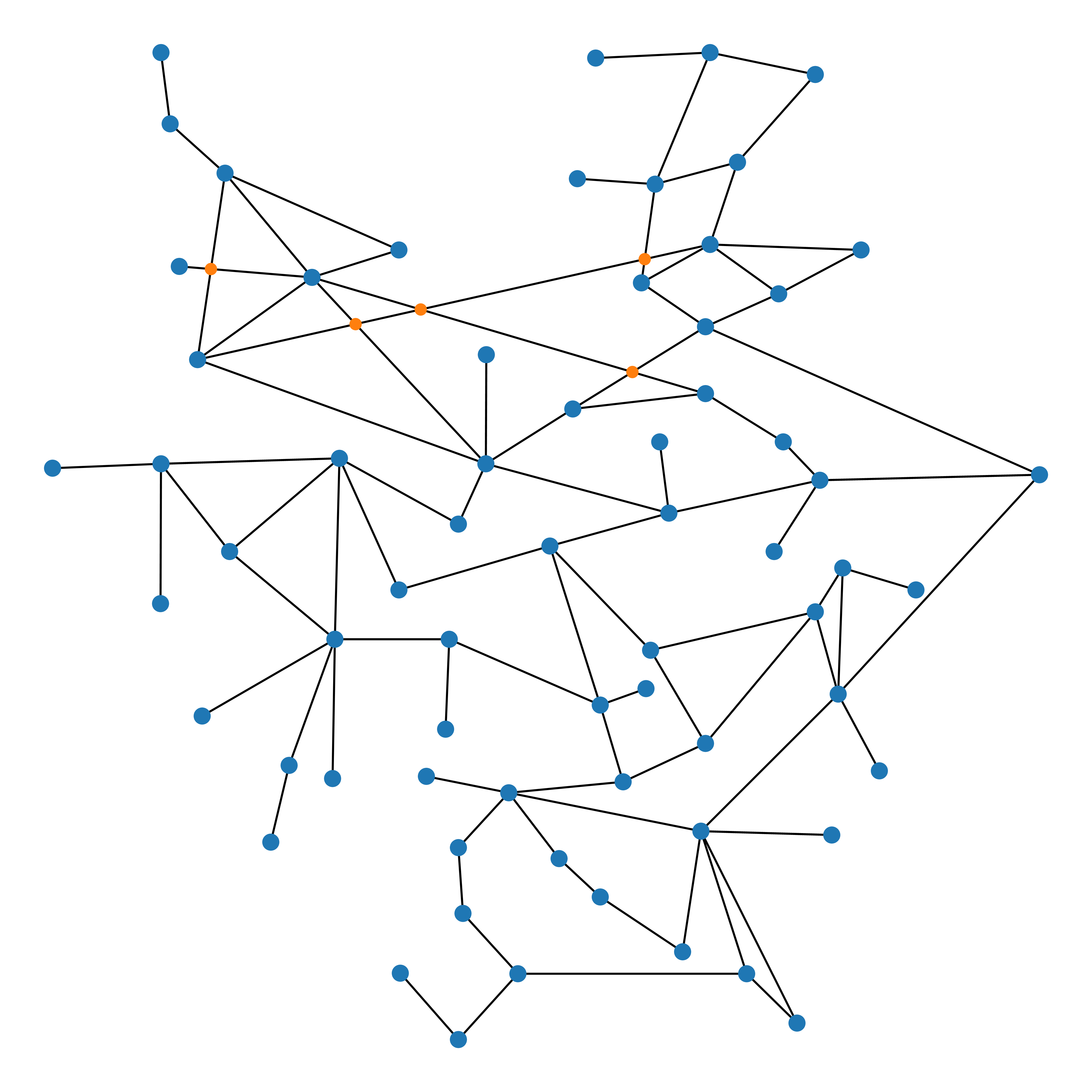}} & 
    \multirow[c]{7}{*}{\includegraphics[width=1.1in]{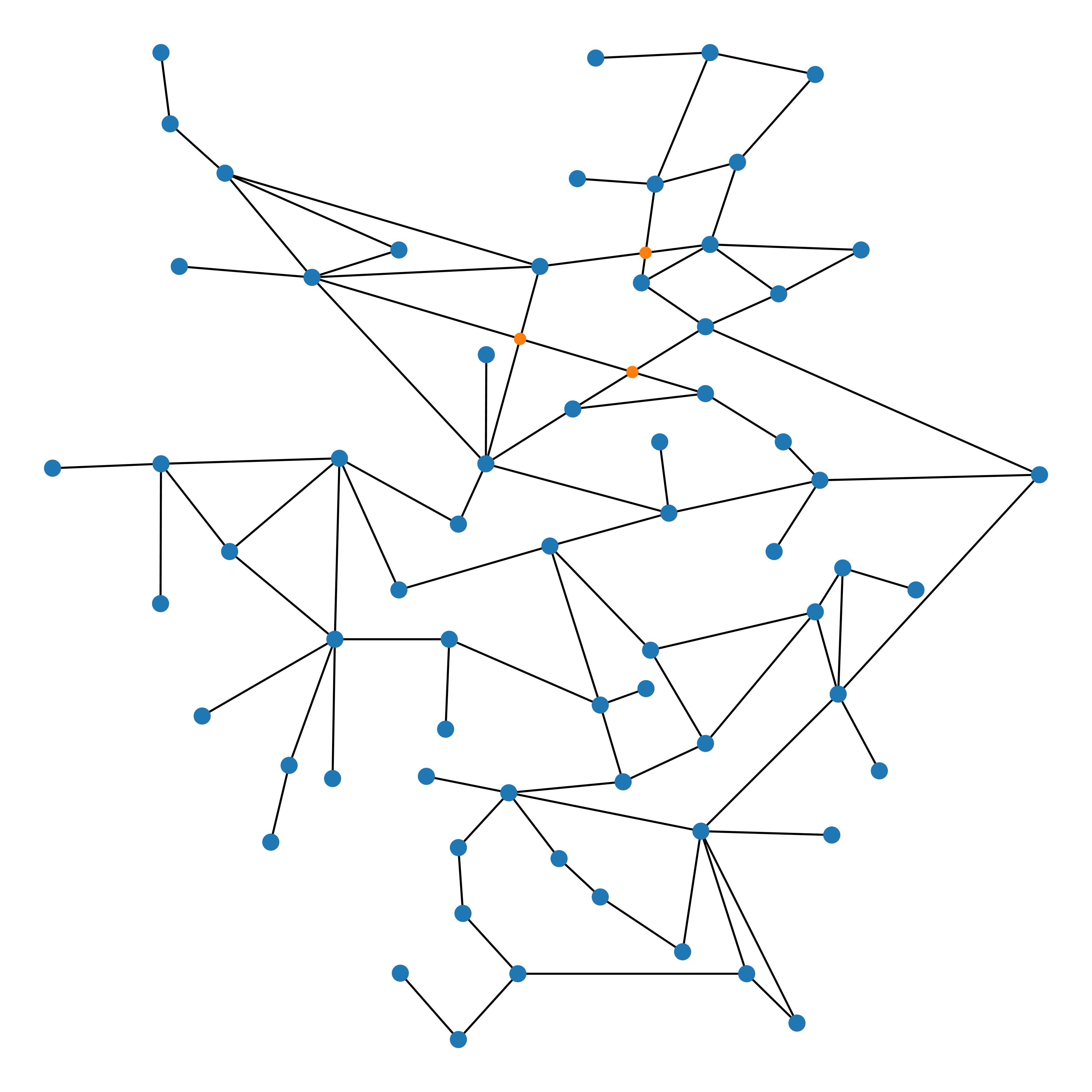}} & 
    \multirow[c]{7}{*}{\includegraphics[width=1.1in]{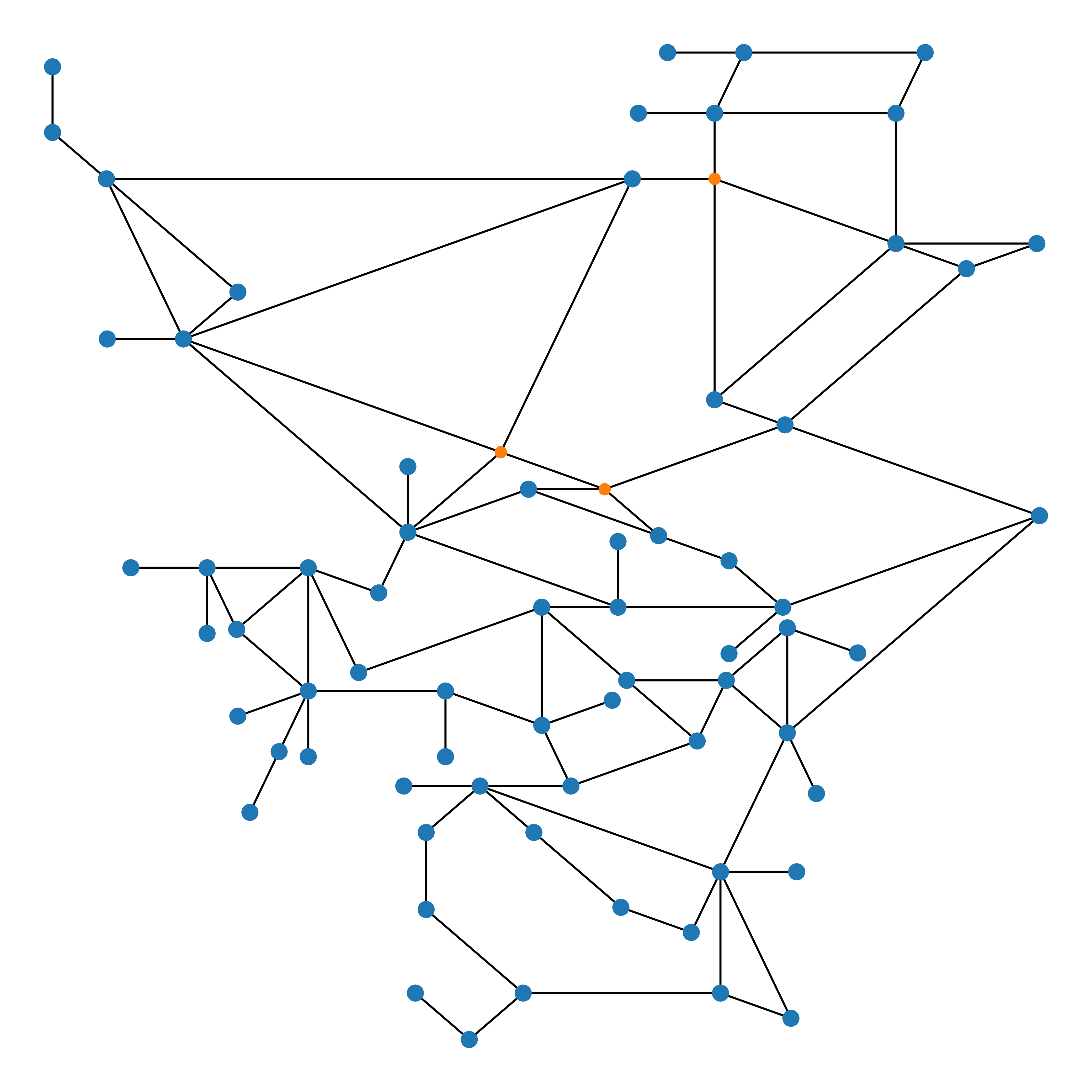}} & $K$ & 8 \\
    & & & & $s$ & 1 \\
    & & & & $\ell_{\min}$ & 5 \\
    & & & & $d_{\min}$ & 0.1 \\
    & & & & $w_{\text{RP}}$ & 0.5 \\
    & & & & $w_{\text{OR}}$ & 0.3 \\
    & & & & $w_{\text{EV}}$ & 0.2 \\ \midrule
    \multirow[c]{7}{*}{\shortstack{IEEE 118 \\[1em] Nodes: 118 \\ Edges: 179 \\[1em] Crossings:\\ Initial: 24\\Final: 13}} & 
    \multirow[c]{7}{*}{\includegraphics[width=1.1in]{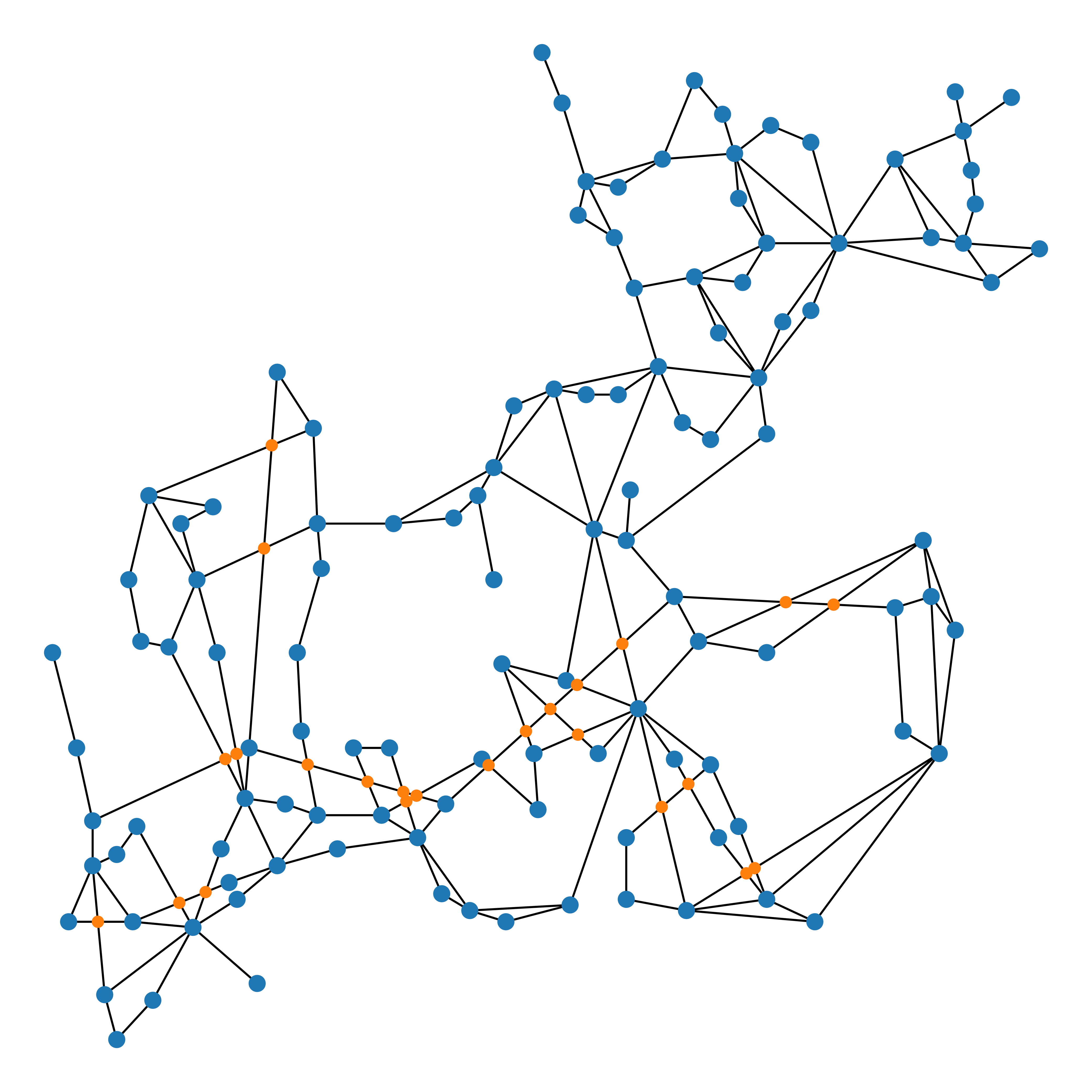}} & 
    \multirow[c]{7}{*}{\includegraphics[width=1.1in]{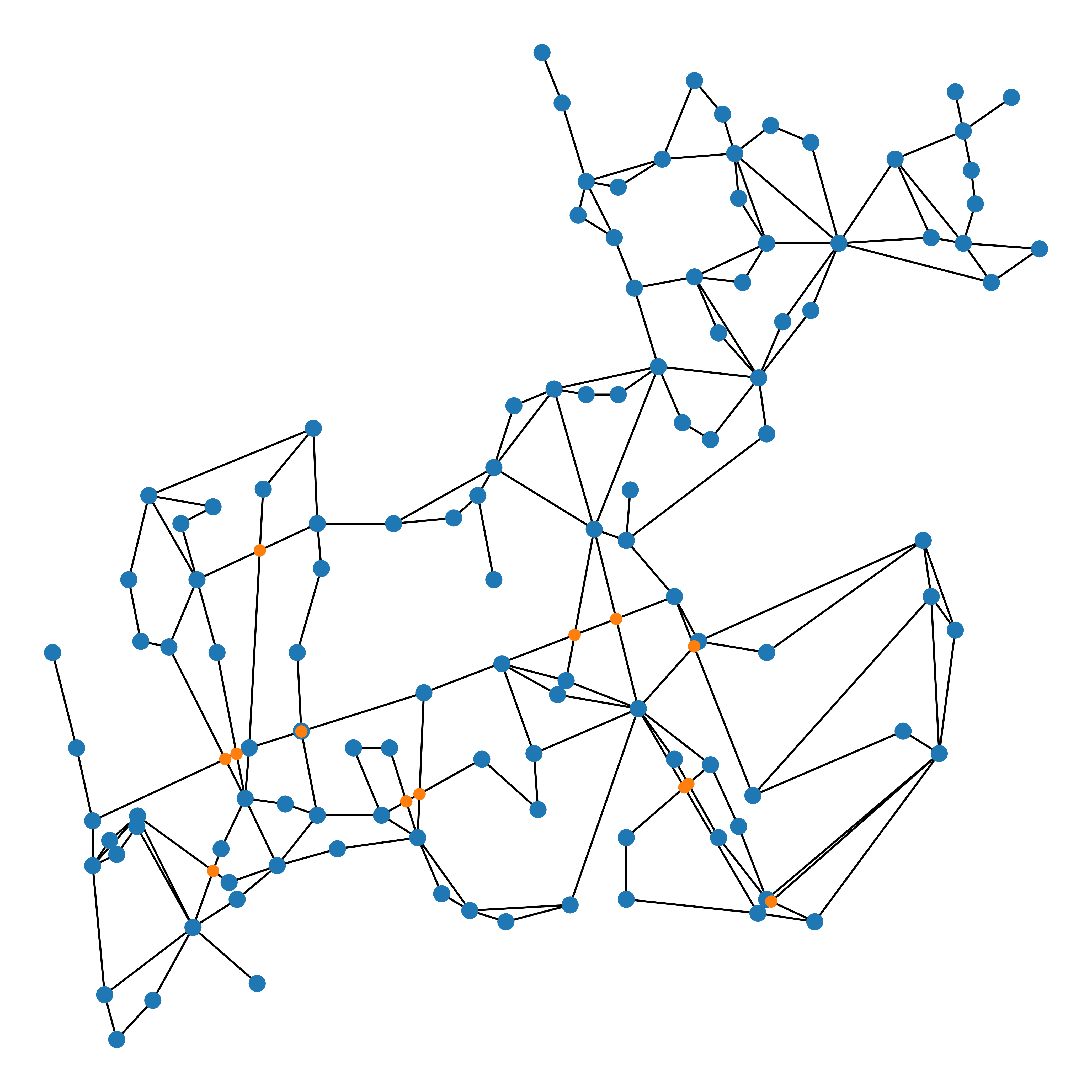}} & 
    \multirow[c]{7}{*}{\includegraphics[width=1.1in]{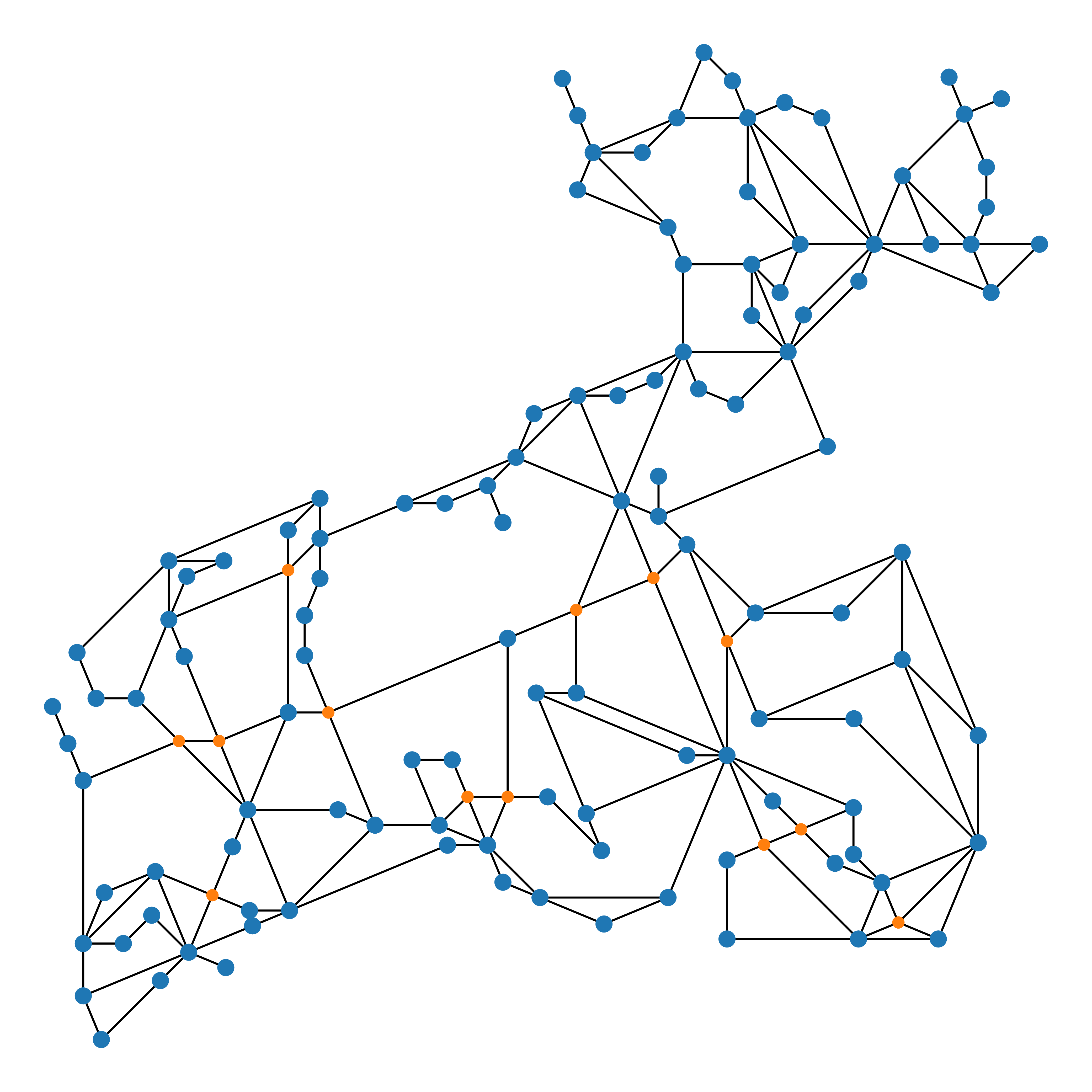}} & $K$ & 8 \\
    & & & & $s$ & -- \\
    & & & & $\ell_{\min}$ & 5 \\
    & & & & $d_{\min}$ & 0.1 \\
    & & & & $w_{\text{RP}}$ & 0.5 \\
    & & & & $w_{\text{OR}}$ & 0.3 \\
    & & & & $w_{\text{EV}}$ & 0.2 \\
    \bottomrule
    \end{tabular}%
  \label{tab:topology_diagrams_of_different_models}%
\end{table*}%
\setlength{\tabcolsep}{\oldtabcolsep}
\renewcommand{\arraystretch}{\oldarraystretch}

This section validates the proposed topology diagram generation framework through case studies on power transmission systems of different scales. Tests are conducted on a server with two Intel Xeon Gold 6248R CPUs and 128 GB RAM. The framework is implemented in Python scripts, with Gurobi Optimizer used to solve the MILP-based layout planning problem\footnote{The source code is available on \href{https://github.com/hust-psa/TopoGen}{https://github.com/hust-psa/TopoGen}.}. Metrics defined in Section \ref{sec:grid_topology_aesthetics} are used to compare the quality of topology diagrams generated by the proposed framework and other algorithms. Parameter selection strategies are discussed. Effects of simplification heuristics in reducing time overheads are also demonstrated.

\subsection{Effectiveness Analysis} 
\label{sub:comparison_of_topology_drawing_approaches}

The first group of case studies aims to demonstrate the capability of the proposed framework in generating topology diagrams that can meet the aesthetic convention of the power system community. Five power transmission system models are tested, including IEEE 30-bus, 57-bus, and 118-bus standard systems and another two models representing the 500 kV backbone structure of two provincial power grids in China. With the raw layouts $\Gamma^0(G)$ of the five grid models input into the proposed framework, the crossing-reduced layouts $\bar{\Gamma}(G)$ and the optimized layouts $\hat{\Gamma}(G)$ can be generated in succession. The yielded topology diagrams, along with the parameter values used for generating these diagrams, are displayed in Table \ref{tab:topology_diagrams_of_different_models}, with blue and orange dots representing ordinary nodes and line crossings, respectively. As can be seen from the results, the layouts $\hat{\Gamma}(G)$ of all power transmission system models typically maintain the spatial relationships of nodes as in the initial layouts $\Gamma^0(G)$ but reduce the number of line crossings and significantly improve the overall tidiness and readability. As the mental map of readers is respected, topology diagrams generated by the proposed framework are much easier to understand.

\subsection{Parameter Selection Strategies} 
\label{sub:parameter_selection_strategy}

\begin{figure}[!tb]
    \centering
    \subfloat[\label{fig:k_raw} Crossing-reduced layout]{\includegraphics[width=1.3in]{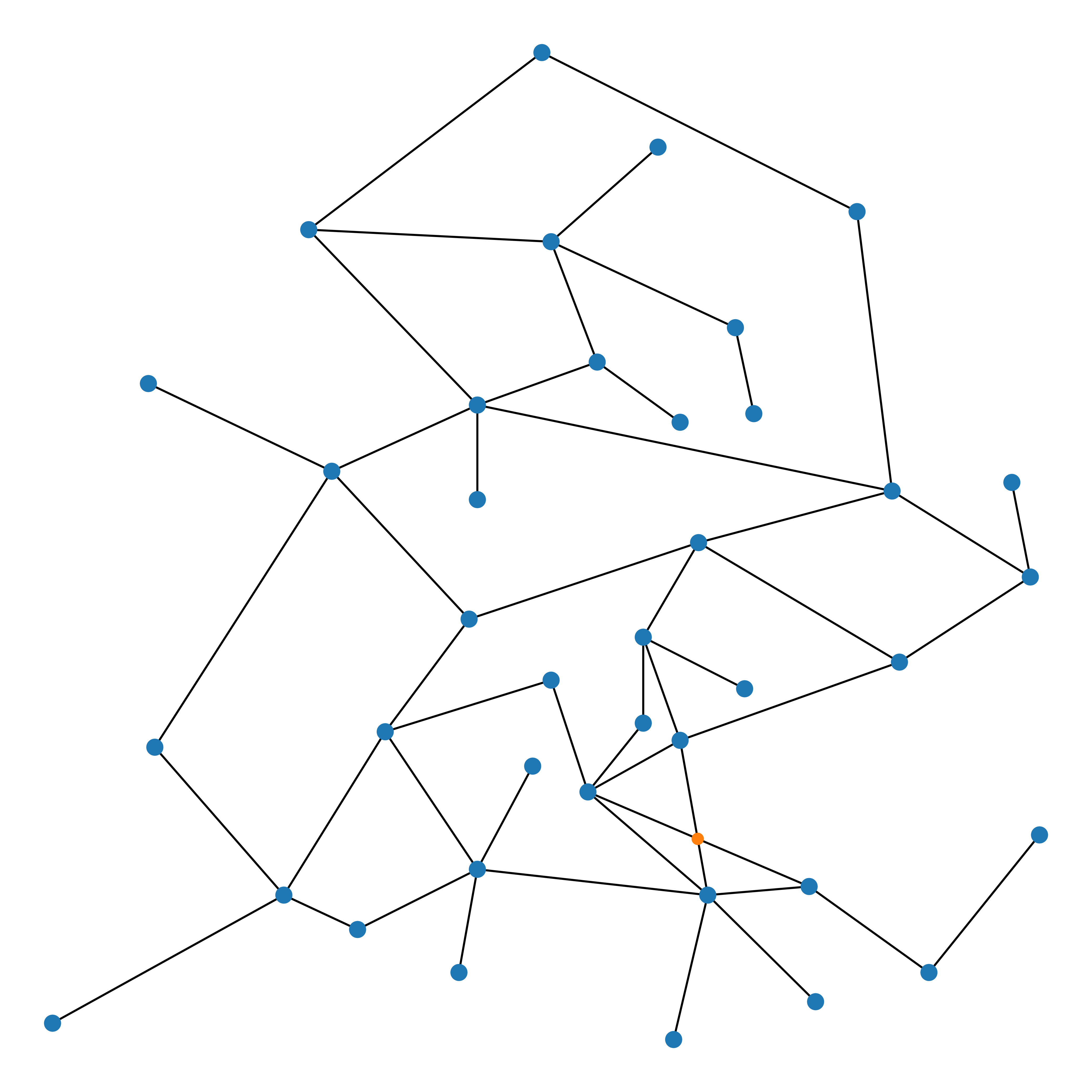}}
    \hfil
    \subfloat[\label{fig:k_4} Optimized on $K=4$]{\includegraphics[width=1.3in]{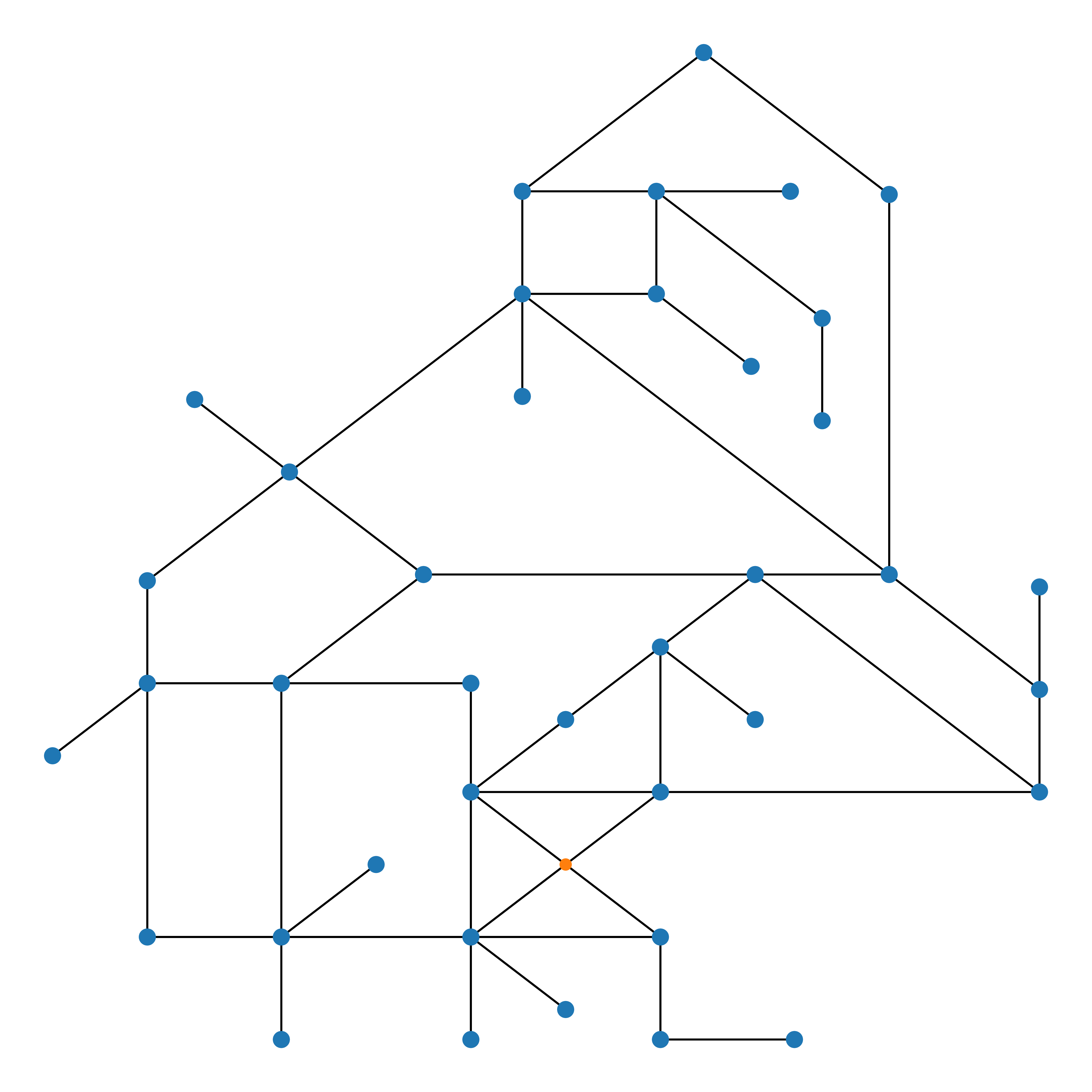}}\\[-0.5em]
    \subfloat[\label{fig:k_6} Optimized on $K=6$]{\includegraphics[width=1.3in]{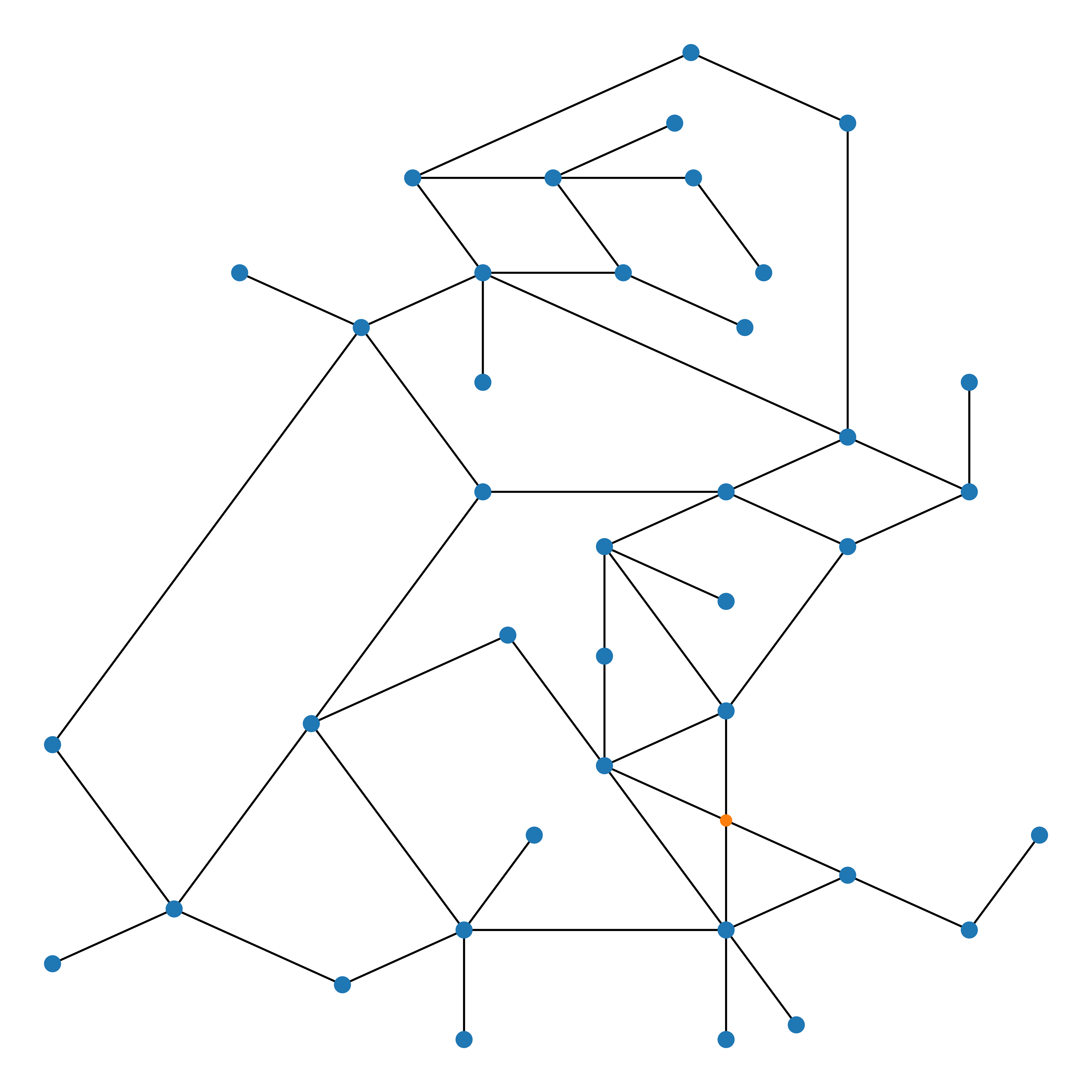}}
    \hfil
    \subfloat[\label{fig:k_8} Optimized on $K=8$]{\includegraphics[width=1.3in]{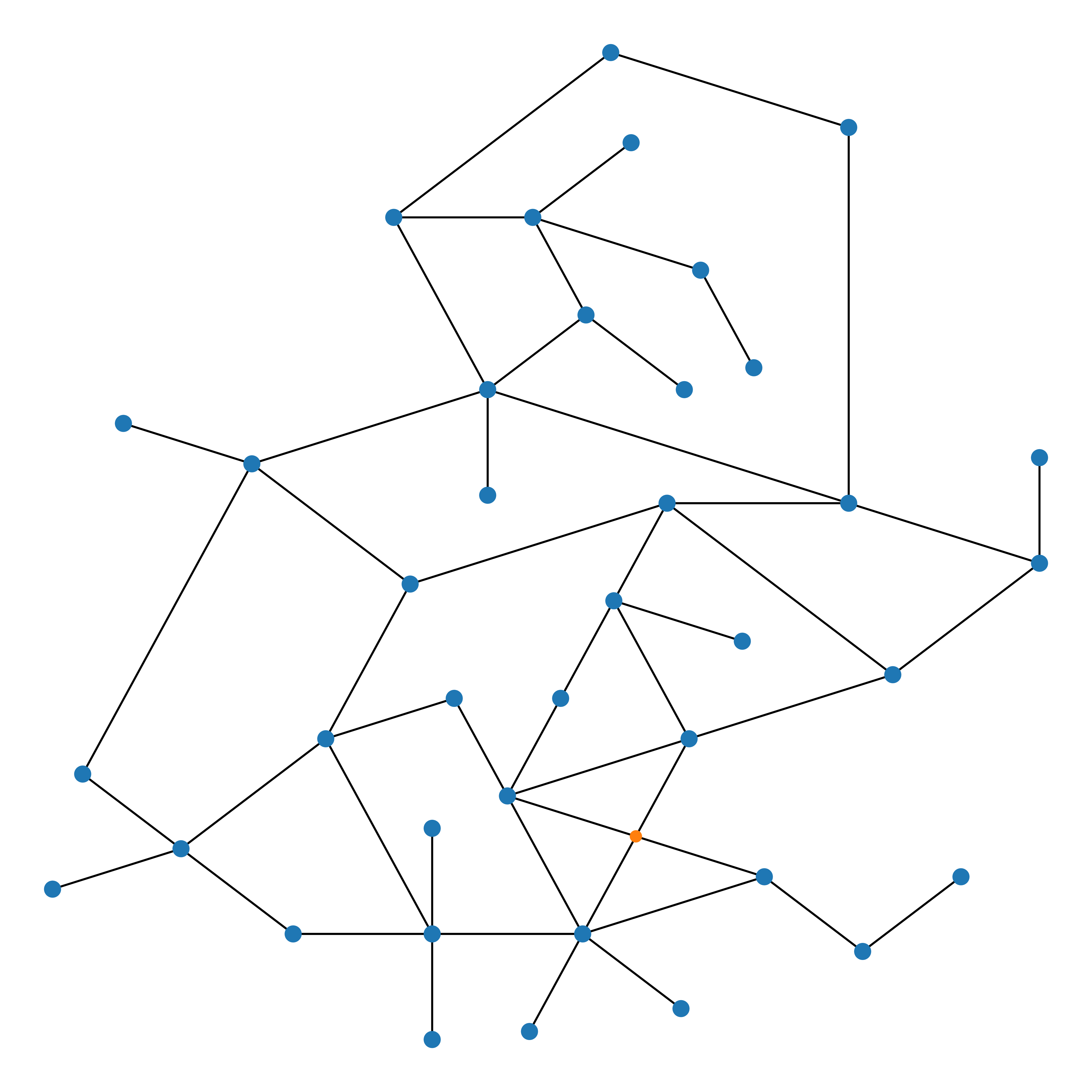}}
    \caption{Optimized layouts of the Grid A model on different values of $K$.}
    \label{fig:layouts_of_different_k}
\end{figure}

As listed in Table \ref{tab:topology_diagrams_of_different_models}, seven parameters are involved in the MILP-based layout planning process, including the cardinality $K$ of the coordinate system $\mathcal{C}_K$, the flexibility margin of edge direction $s$, the minimum edge length $\ell_{\min}$, the minimum distance between two edges $d_{\min}$, and weight coefficients $w_{\text{RP}}$, $w_{\text{OP}}$, $w_{\text{EV}}$ balancing the three aesthetic criteria RP, OP, and EV. Among them, $K$ is the most important parameter of the layout planning process. To make the MILP problem have feasible solutions, $K$ should be at least larger than half of the maximum degree of the power grid model. Note that this value is just a theoretical lower bound of $K$. If incident edges of each node are evenly distributed around the node, setting $K$ to half of the maximum degree may work well. However, such a value can hardly tackle cases where many incident edges squeeze within a small sector. A more practical selection is to let $K = \max(\{\pi / \bar{a}_v \mid v \in \mathcal{V}\})$. In the equation, $\bar{a}_v$ is the average included angle between neighboring edges incident to $v$, which can be calculated by dividing the angle from the first incident edge in Quadrant I to the last incident edge in Quadrant IV by the degree of $v$. As shown in Fig. \ref{fig:layouts_of_different_k}, a small value of $K$ often leads to a slightly larger distortion from the original layout $\Gamma^0(G)$. Empirically, when the model size is small, choosing a small $K$, say $K=4$, helps improve the regularity of the output diagram. Conversely, a relatively larger $K$ will make the output layout more resemble the input one, which is beneficial for the comprehension of readers. Besides, with the expansion of the model scale, selecting a larger $K$ will introduce more optimization variables and thus may facilitate the search for feasible solutions.

The flexibility margin $s$ is related to the similarity between the input and output topology diagrams one the one hand, and has an influence on finding feasible solutions of the optimization problem on the other hand. For simple grid models where the node degree is generally small, $s$ can be directly set to 1 to make the output layout more similar to the input layout. However, for large-scale grid models, $s$ should be assigned with a larger value. That is because it is possible to have more than three incident edges of a node crowded in one sector in large models. In this case, $s=1$ is no longer feasible because there is no way to arrange all the edges in three sectors without introducing overlap. As in the IEEE 118 case in Table \ref{tab:topology_diagrams_of_different_models}, to provide full flexibility for arranging the incident edges of a node $v$, $s$ can be set to $\lceil(n_v-1)/2\rceil$, where $n_v$ denotes the degree of $v$.

The selection of $\ell_{\min}$ and $d_{\min}$ is also relatively arbitrary for small-scale power grid models. However, just like $K$ and $s$, casual selection of these parameters for large power grids with complex structures may also obstruct the search for feasible solutions meeting the massive number of constraints. To leave more freedom for optimizing the layout, nodes should not be placed too closely together. Thus, it is suggested to increase the ratio $\ell_{\min} / d_{\min}$ for large power grid models. Other parameters, i.e., the weights of cost terms in the optimization objective \eqref{eq:objective}, are designated mainly out of the demand of users.

\renewcommand{\arraystretch}{1.3}
\setlength{\tabcolsep}{5pt}
\begin{table*}[!t]
  \centering
  \caption{Comparison of Topology Drawing Algorithms}
    \begin{tabular}{cc|cc|cc|cc|cc|cc|cc|cc}
    \toprule
    \multicolumn{4}{c}{\textbf{Original}} & \multicolumn{4}{c}{\textbf{Proposed}} & \multicolumn{4}{c}{\textbf{Birchfield \textit{et al.} \cite{Birchfield_Overbye_2018}}} & \multicolumn{4}{c}{\textbf{Classic Force-Directed}} \\\midrule
    \multicolumn{4}{c|}{\includegraphics[width=1.4in]{Figures/bus_118/raw.png}} & \multicolumn{4}{c|}{\includegraphics[width=1.4in]{Figures/bus_118/opt.png}} & \multicolumn{4}{c|}{\includegraphics[width=1.4in]{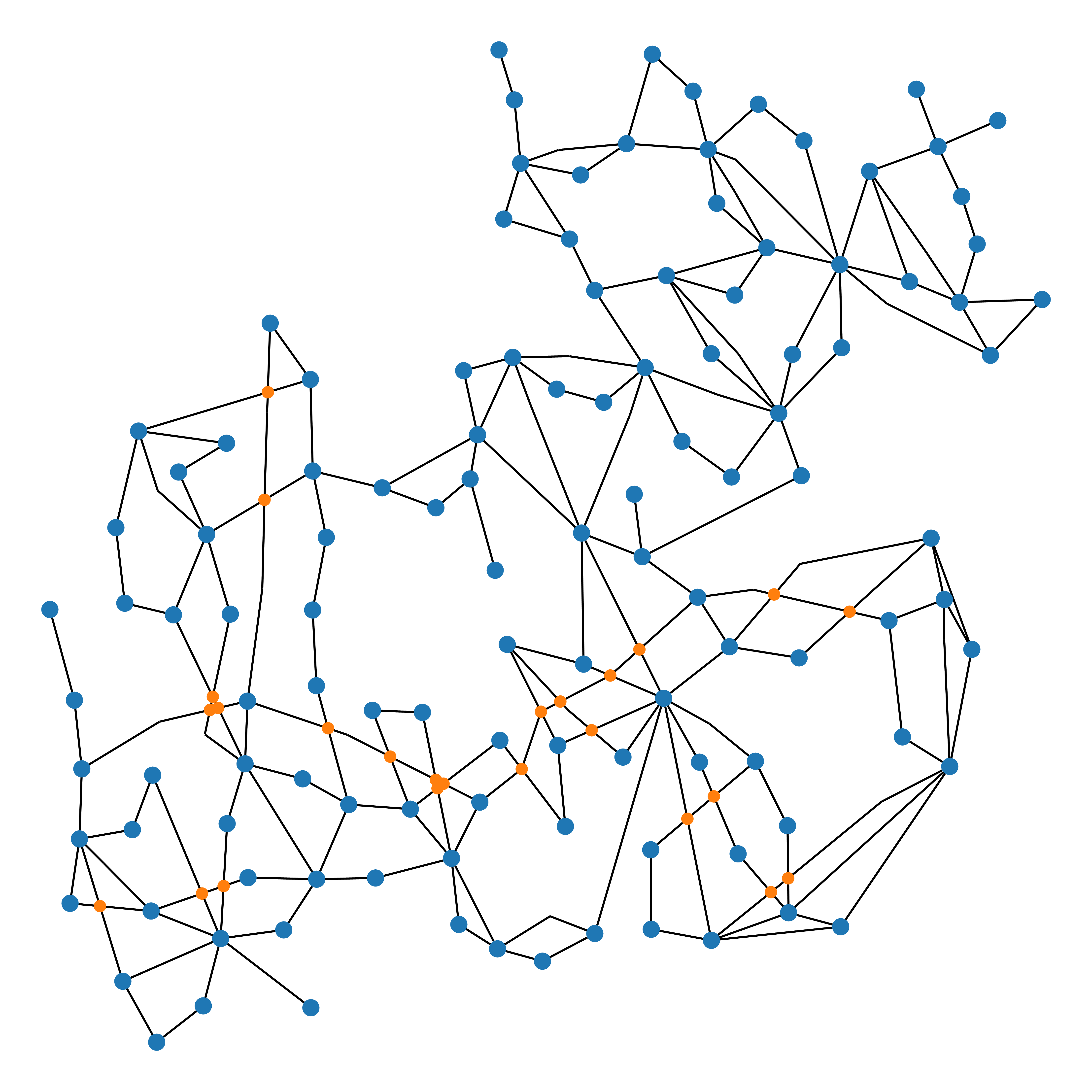}} & \multicolumn{4}{c}{\includegraphics[width=1.4in]{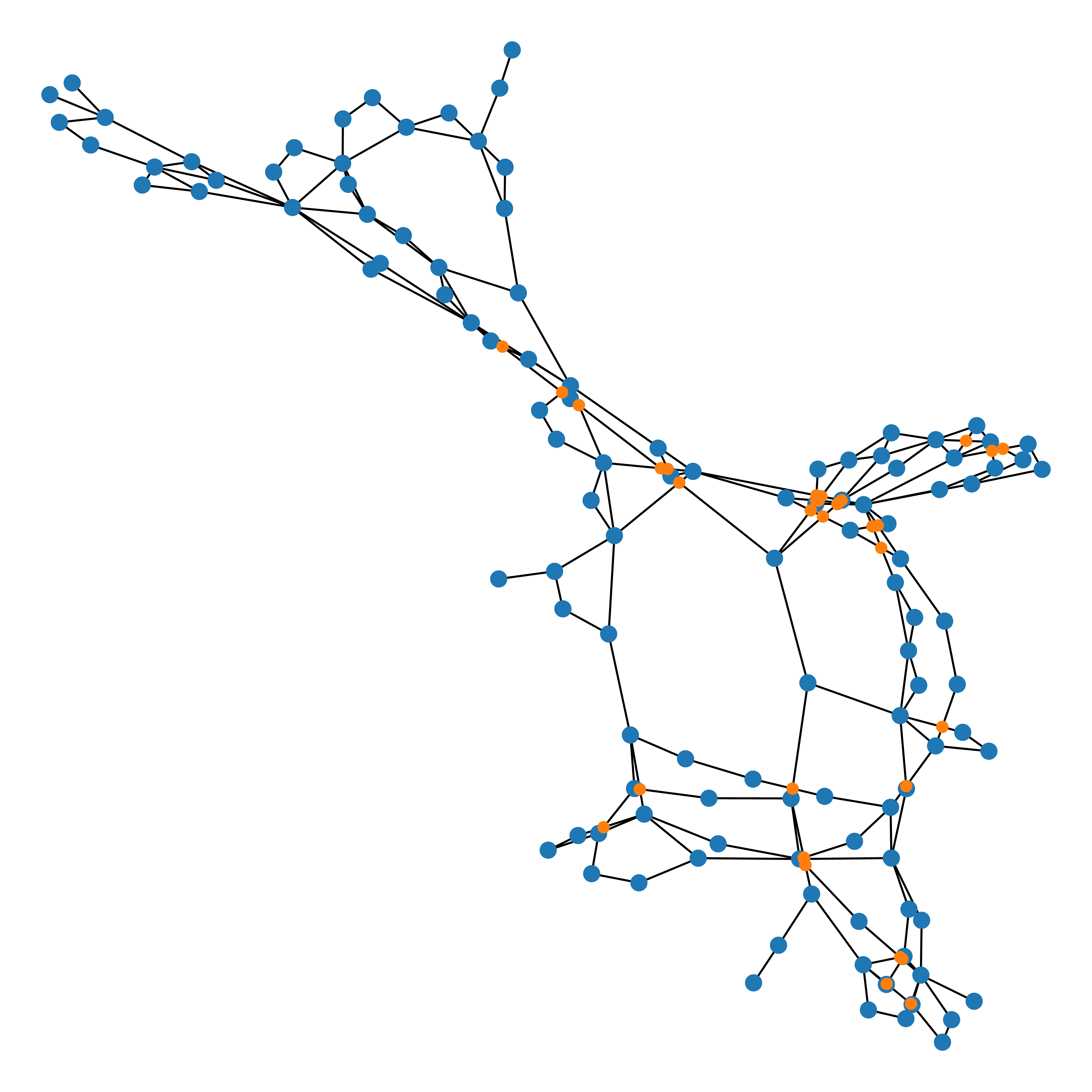}} \\\hline
    \multicolumn{4}{c|}{\rule{0pt}{2.6ex}IEEE 118} & $m_\text{EX}$ & \textbf{-13} & $m_\text{EL}$ & \textbf{0.523} & $m_\text{EX}$ & -25 & $m_\text{EL}$ & 0.517 & $m_\text{EX}$ & -33 & $m_\text{EL}$ & 0.300 \\
    \multicolumn{4}{c|}{Nodes: 118} & $m_\text{ND}$ & 0.436 & $m_\text{IA}$ & \textbf{0.214} & $m_\text{ND}$ & \textbf{0.623} & $m_\text{IA}$ & 0.131 & $m_\text{ND}$ & 0.416 & $m_\text{IA}$ & 0.009 \\
    \multicolumn{4}{c|}{Edges: 179} & $m_\text{RP}$ & 0.929 & $m_\text{OR}$ & 0.515 & $m_\text{RP}$ & \textbf{0.957} & $m_\text{OR}$ & 0.502 & $m_\text{RP}$ & 0.537 & $m_\text{OR}$ & 0.487 \\
    \multicolumn{4}{c|}{Crossings: 24} & $m_\text{EV}$ & -0.034 & & & $m_\text{EV}$ & -0.025 & & & $m_\text{EV}$ & -0.039 & & \\\midrule
	\multicolumn{4}{c}{\textbf{ForceAtlas2 \cite{Jacomy_Venturini_2014}}} & \multicolumn{4}{c}{\textbf{tsNET* \cite{Kruiger_Rauber_2017}}} & \multicolumn{4}{c}{\textbf{DeepDrawing \cite{Wang_Jin_2020}}} & \multicolumn{4}{c}{\textbf{SGD\textsuperscript{2} \cite{Ahmed_Luca_2022}}} \\\midrule
 	\multicolumn{4}{c|}{\includegraphics[width=1.4in]{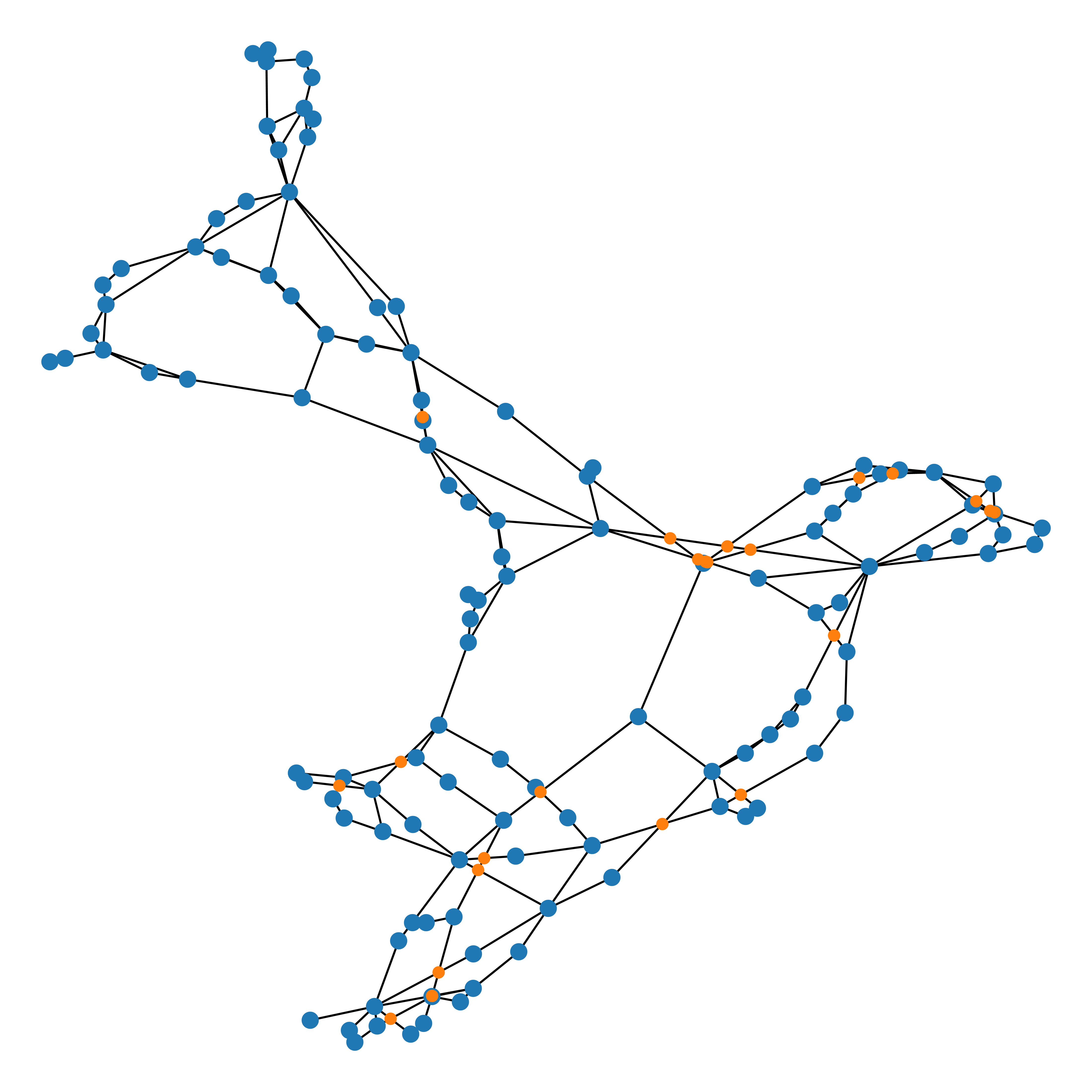}} & \multicolumn{4}{c|}{\includegraphics[width=1.4in]{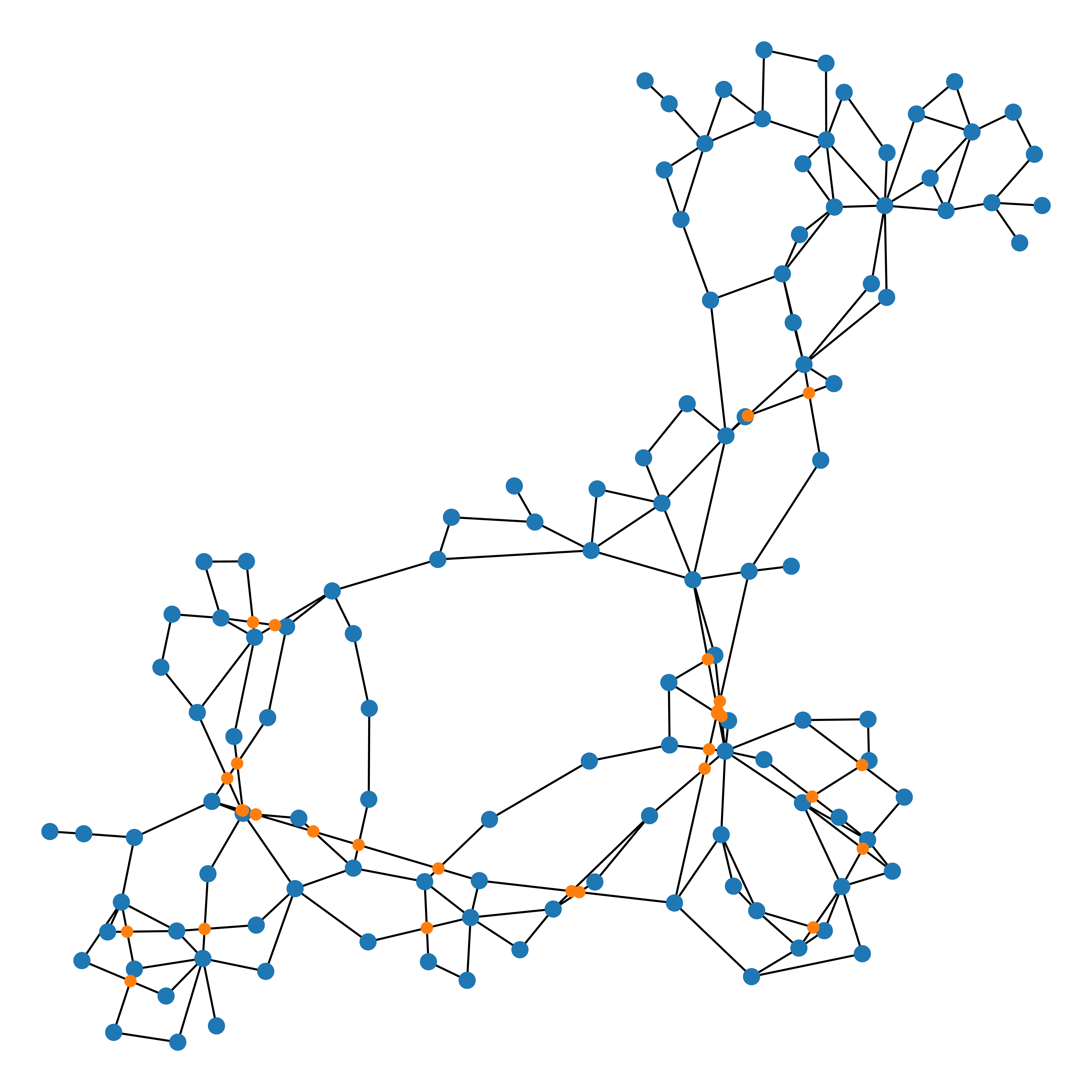}} & \multicolumn{4}{c|}{\includegraphics[width=1.4in]{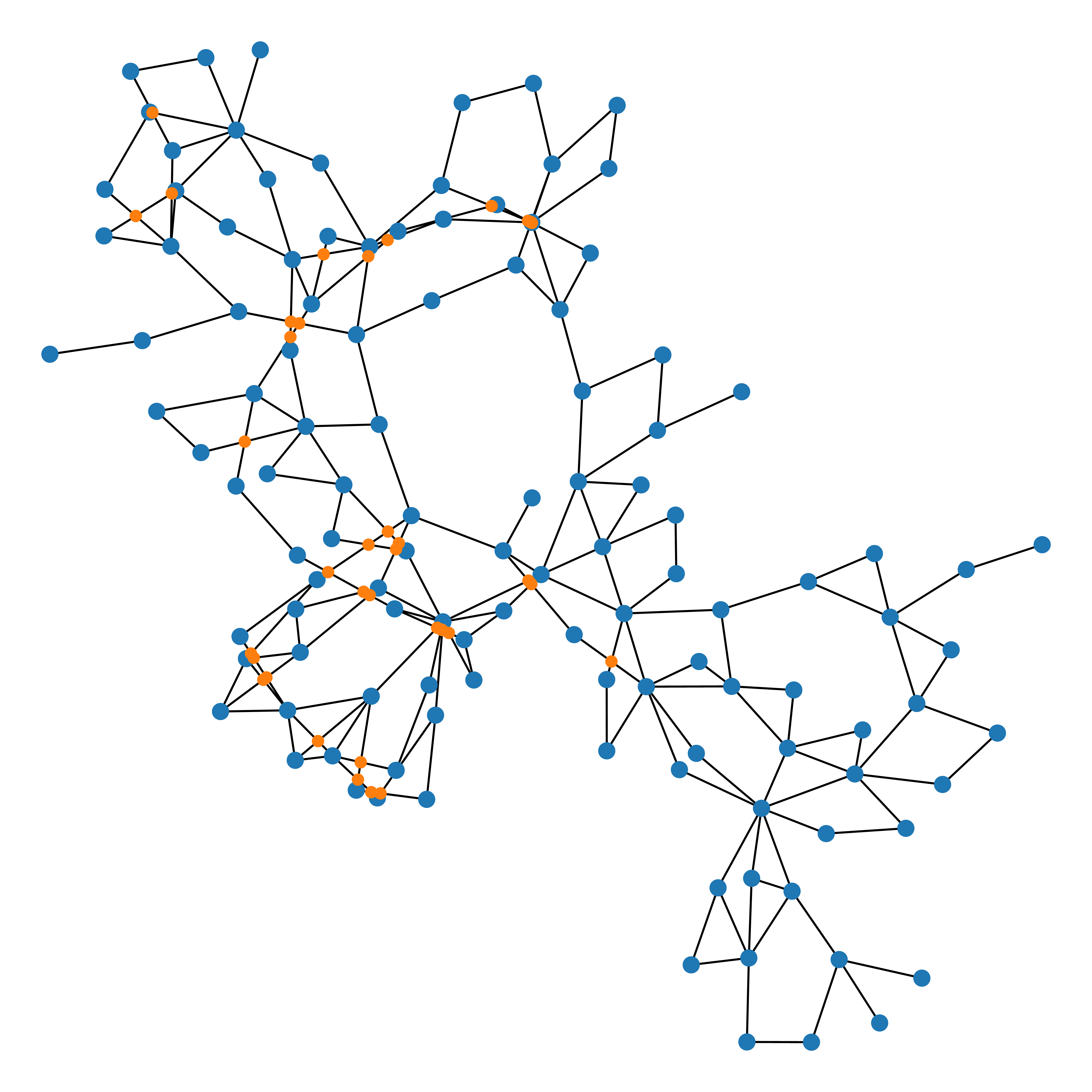}} & \multicolumn{4}{c}{\includegraphics[width=1.4in]{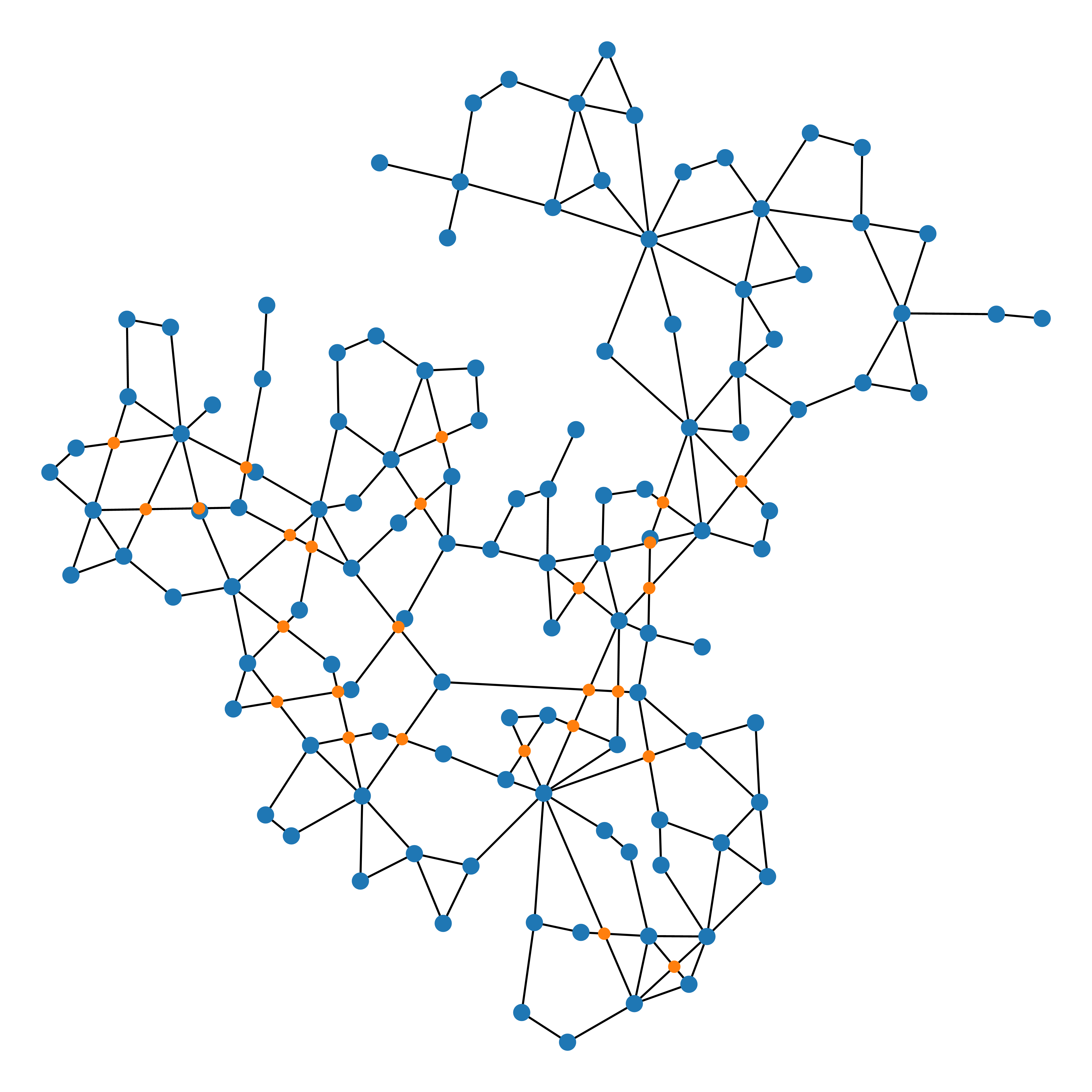}} \\\hline
    \rule{0pt}{2.6ex}$m_\text{EX}$ & -23 & $m_\text{EL}$ & 0.176 & $m_\text{EX}$ & -30 & $m_\text{EL}$ & 0.330 & $m_\text{EX}$ & -40 & $m_\text{EL}$ & 0.472 & $m_\text{EX}$ & -26 & $m_\text{EL}$ & 0.416 \\
    $m_\text{ND}$ & 0.317 & $m_\text{IA}$ & 0.007 & $m_\text{ND}$ & 0.492 & $m_\text{IA}$ & 0.034 & $m_\text{ND}$ & 0.559 & $m_\text{IA}$ & 0.011 & $m_\text{ND}$ & 0.572 & $m_\text{IA}$ & 0.390 \\
    $m_\text{RP}$ & 0.626 & $m_\text{OR}$ & 0.457 & $m_\text{RP}$ & 0.734 & $m_\text{OR}$ & 0.531 & $m_\text{RP}$ & 0.478 & $m_\text{OR}$ & 0.502 & $m_\text{RP}$ & 0.633 & $m_\text{OR}$ & \textbf{0.533} \\
    $m_\text{EV}$ & -0.038 & & & $m_\text{EV}$ & -0.030 & & & $m_\text{EV}$ & -0.028 & & & $m_\text{EV}$ & \textbf{-0.021} & & \\
    \bottomrule
    \end{tabular}%
  \label{tab:comparison_of_topology_drawing_algorithms}%
\end{table*}%
\setlength{\tabcolsep}{\oldtabcolsep}
\renewcommand{\arraystretch}{\oldarraystretch}

\subsection{Comparison with Existing Graph Drawing Algorithms} 
\label{sub:comparison_with_existing_graph_drawing_algorithms}

In this group of case studies, the proposed topology diagram generation framework is compared with existing graph drawing algorithms, including the topology layout algorithm proposed by Birthfield \textit{et al.} in \cite{Birchfield_Overbye_2018}, as well as general-purpose graph drawing algorithms such as the classic force-directed algorithm, ForceAtlas2 \cite{Jacomy_Venturini_2014}, tsNET* \cite{Kruiger_Rauber_2017}, DeepDrawing \cite{Wang_Jin_2020}, and SGD\textsuperscript{2} \cite{Ahmed_Luca_2022}. The output diagrams of tested algorithms, along with their aesthetic metrics as discussed in Section \ref{sec:grid_topology_aesthetics}, are listed in Table \ref{tab:comparison_of_topology_drawing_algorithms}. From the viewpoint of qualitative analysis, only the proposed framework and the algorithm in \cite{Birchfield_Overbye_2018} can maintain the relative positions of nodes in the initial layout. The output of tsNET* is somewhat similar to the input layout on a macroscopic level. However, when focusing on each locality, it can be found that the nodes are still rearranged to a great extent. The output diagrams of the two force-directed algorithms, i.e., the classic force-directed algorithm and ForceAtlas2, resemble each other in general. They apparently do not meet the aesthetics recognized by the power system community. The legibility of the output of SGD\textsuperscript{2} is much better than force-directed algorithms, as multiple aesthetic criteria are optimized as a whole. The DeepDrawing algorithm is trained with a set of layouts generated by the proposed framework. The numbers of nodes in those layouts vary from 20 to 80. By learning the layout style of the proposed framework, DeepDrawing can generate layouts with slightly better quality than force-directed algorithms. Nonetheless, as a data-driven approach, there is no guarantee that DeepDrawing can always maintain node positions and reduce edge crossings. As shown in Table \ref{tab:comparison_of_topology_drawing_algorithms}, it has the highest number of edge crossings among all tested algorithms.

Then, a quantitative comparison between the proposed framework and the Birthfield's algorithm is conducted. Birthfield's algorithm has better spatial evenness than the proposed algorithm, which is manifested by the non-negligible advantanges in $m_\text{ND}$ and $m_\text{EV}$. The differences of the two algorithms in $m_\text{RP}$, $m_\text{OR}$, and $m_\text{EL}$ are not big. In contrast, there are only 13 crossings in the output of the proposed framework, which is almost a half of Birchfield's algorithm. That is because Birchfield's algorithm only considers to prevent lines from overlapping with nodes, without an explicit reduction of line intersections. Besides, since Birchfield's algorithm mitigates line-node overlapping by introducing waypoints, it may be difficult for readers to track long transmission lines in complex grid models. In summary, among the seven aesthetic metrics defined in Section \ref{sec:grid_topology_aesthetics}, the proposed framework ranks first in three, second in one, and third in one; it has the best quantitative performance among all tested algorithms.

\subsection{Validation of Crossing Reduction Heuristics} 
\label{sub:validation_of_crossing_reduction_heuristics}

With the effectiveness of the proposed framework in reducing line crossings demonstrated in Table \ref{tab:comparison_of_topology_drawing_algorithms}, this subsection aims to validate the capability of the two simplification heuristics in Section \ref{sec:edge_crossing_reduction} in enhancing computation efficiency.

The performance of the crossing reduction module applying different combinations of simplification heuristics H1 and H2 to the IEEE 118-bus test system is shown in Table \ref{tab:performance_of_simplification_heuristics}. The case where H1 and H2 are both unapplied, i.e., the native approach in \cite{Radermacher_Reichard_2019}, is regarded as the benchmark. It can be seen from Table \ref{tab:performance_of_simplification_heuristics} that the efficiency of the crossing reduction process can be improved with either one of the two heuristics. H1 has more strength in decreasing time overheads than H2, yet H2 is more likely to produce layouts with fewer line intersections since the introduction of inflection points brings an additional degree of freedom. The combination of H1 and H2 retains their respective advantages and achieves a decent crossing reduction performance in a shorter time. Relative to the benchmark, applying H1 and H2 simultaneously reduces 27\% fewer line crossings but is nearly 35 times faster.

\begin{table}[!t]
  \centering
  \caption{Performance of Simplification Heuristics}
	\begin{tabular}{cccc}
	\toprule
	\textbf{Conditions} & \boldmath $C[\bar{\Gamma}(G)]$ \unboldmath & \textbf{Relative Rate} & \textbf{Time (s)} \\
	\midrule
	w/o H1 and H2 & 9 & -- & 8682.67 \\
	w/ H1 only & 14 & 67\% & 704.06 \\
	w/ H2 only & 10 & 93\% & 2299.81 \\
	w/ H1 and H2 & 13 & 73\% & 249.49 \\
	\bottomrule
	\end{tabular}%
  \label{tab:performance_of_simplification_heuristics}%
\end{table}%

\begin{table}[!t]
  \centering
  \caption{Performance of Different Values of \texorpdfstring{$d_v^{\mathcal{T}}$}{d_v^T}}
    \begin{tabular}{cccccc}
    \toprule
    \boldmath $d_v^{\mathcal{T}}$\unboldmath & \textbf{3} & \textbf{4} & \textbf{5} & \textbf{6} & \textbf{7} \\ \midrule
    $C[\bar{\Gamma}(G)]$ & 15 & 13 & 13 & 13 & 12 \\
    Time (s) & 119.41 & 249.49 & 480.84 & 831.27 & 1035.14 \\
    \bottomrule
    \end{tabular}%
  \label{tab:different_d_v_T}%
\end{table}%

It is noteworthy that the simplification heuristic H1 has two tunable parameters, namely the BFS-tree depth $d_v^{\mathcal{T}}$ and the expansion radius $r$. Both parameters jointly delineate the shape of the bounded canvas $\bar{H}_v^r$ in which the optimal placement of the node $v$ is searched. Typically, $d_v^{\mathcal{T}}$ has a more decisive influence on $\bar{H}_v^r$ than $r$; the former determines the region of the convex hull $H_v$, while the latter (usually with a small value) is used to slightly expand the coverage of $H_v$ to obtain $\bar{H}_v^r$. A too small $d_v^{\mathcal{T}}$ will cause the crossing reduction module makes little sense, as the search of the optimal position is confined to a narrow area. For small power grid models where the efficiency is not a predominant issue, $d_v^{\mathcal{T}}$ can be set to a relative large value. In contrast, such a strategy may not suit large-scale power grids. It can be seen from Table \ref{tab:different_d_v_T} that it is cost-inefficient to apply a large $d_v^{\mathcal{T}}$ to the IEEE 118-bus test system, as the number of reduced crossings does not grow at a rate commensurate with the time consumption. As a rule of thumb, $d_v^{\mathcal{T}}=4$ is suggested for power grid models with relatively large scales, while for smaller models, its value could be slightly larger.

\subsection{Validation of Planarity Relaxation Strategy} 
\label{sub:validation_of_planarity_relaxation}

\begin{figure}[!tb]
    \centering
    \subfloat[\label{fig:staged_cr} Crossing-reduced layout]{\includegraphics[width=1.4in]{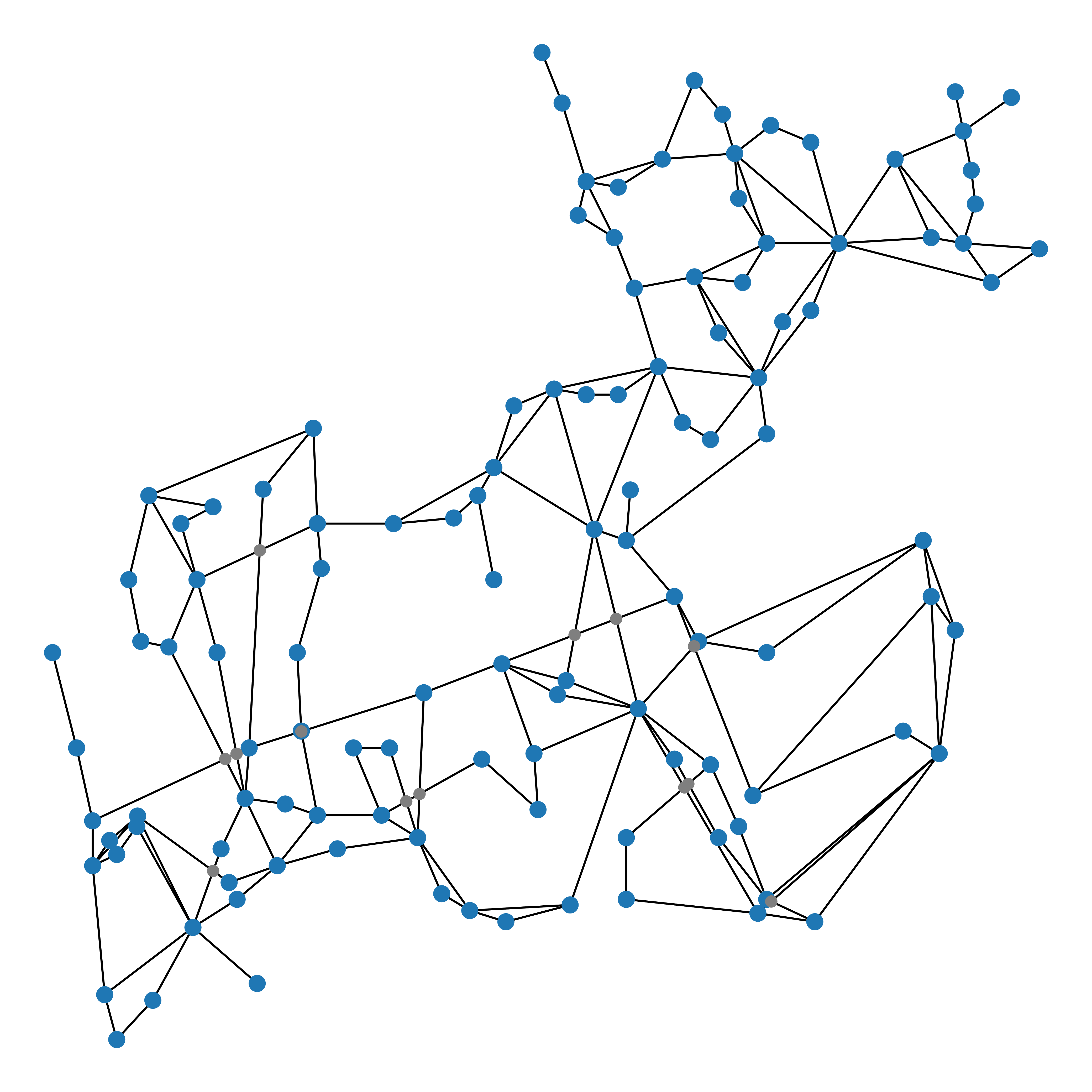}}
    \hfil
    \subfloat[\label{fig:staged_1} Layout of Round 1]{\includegraphics[width=1.4in]{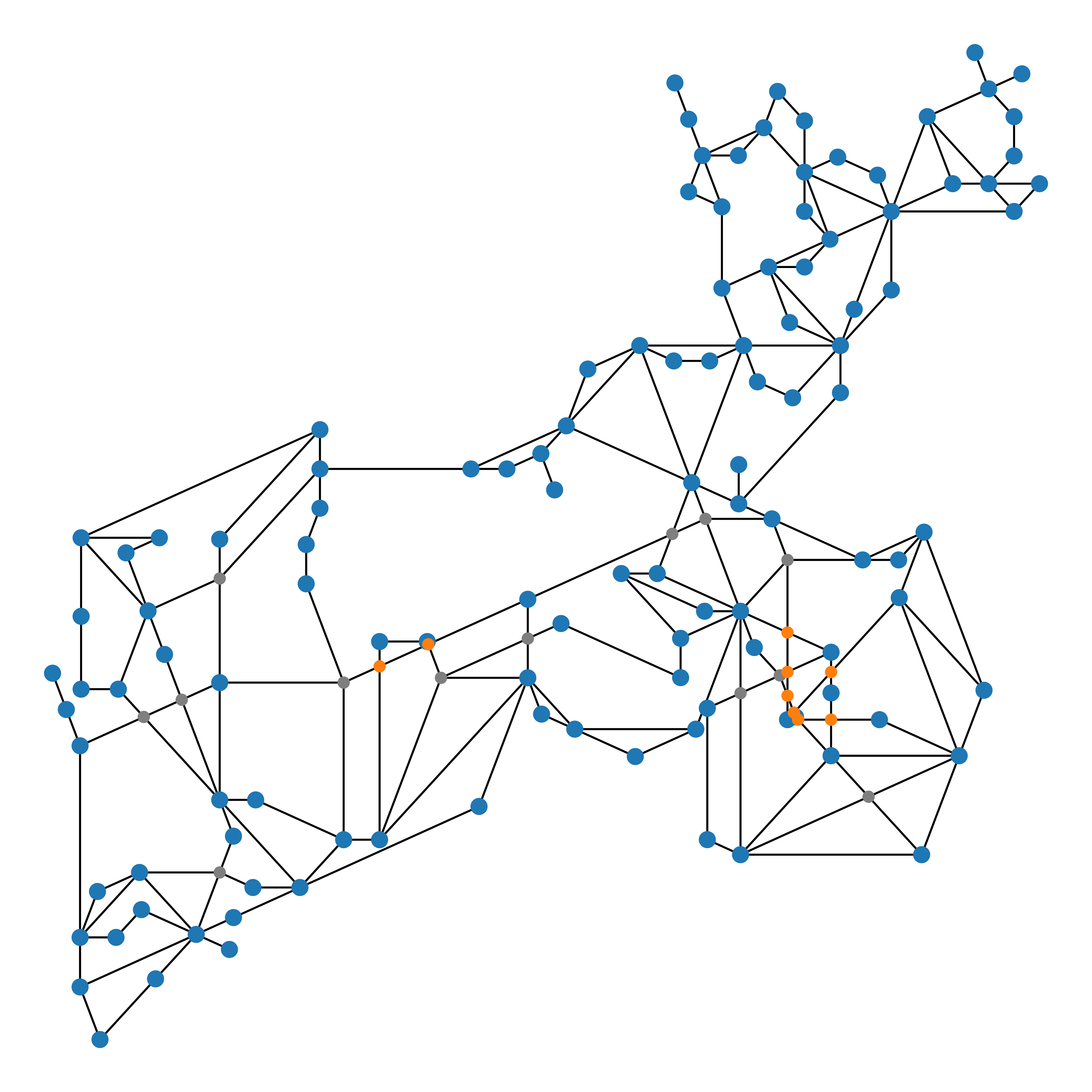}}\\[-0.5em]
    \subfloat[\label{fig:staged_2} Layout of Round 2]{\includegraphics[width=1.4in]{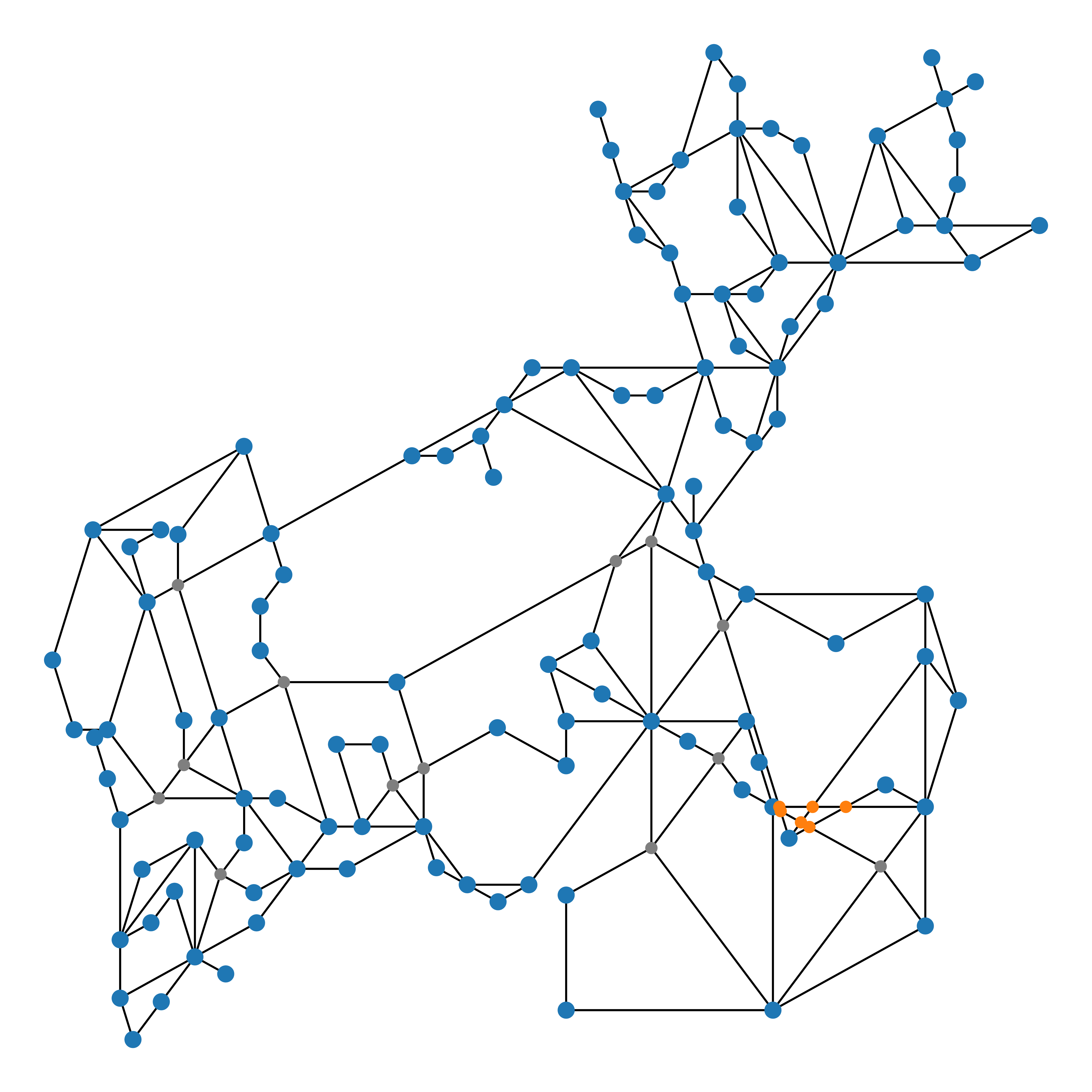}}
    \hfil
    \subfloat[\label{fig:staged_3} Layout of Round 3]{\includegraphics[width=1.4in]{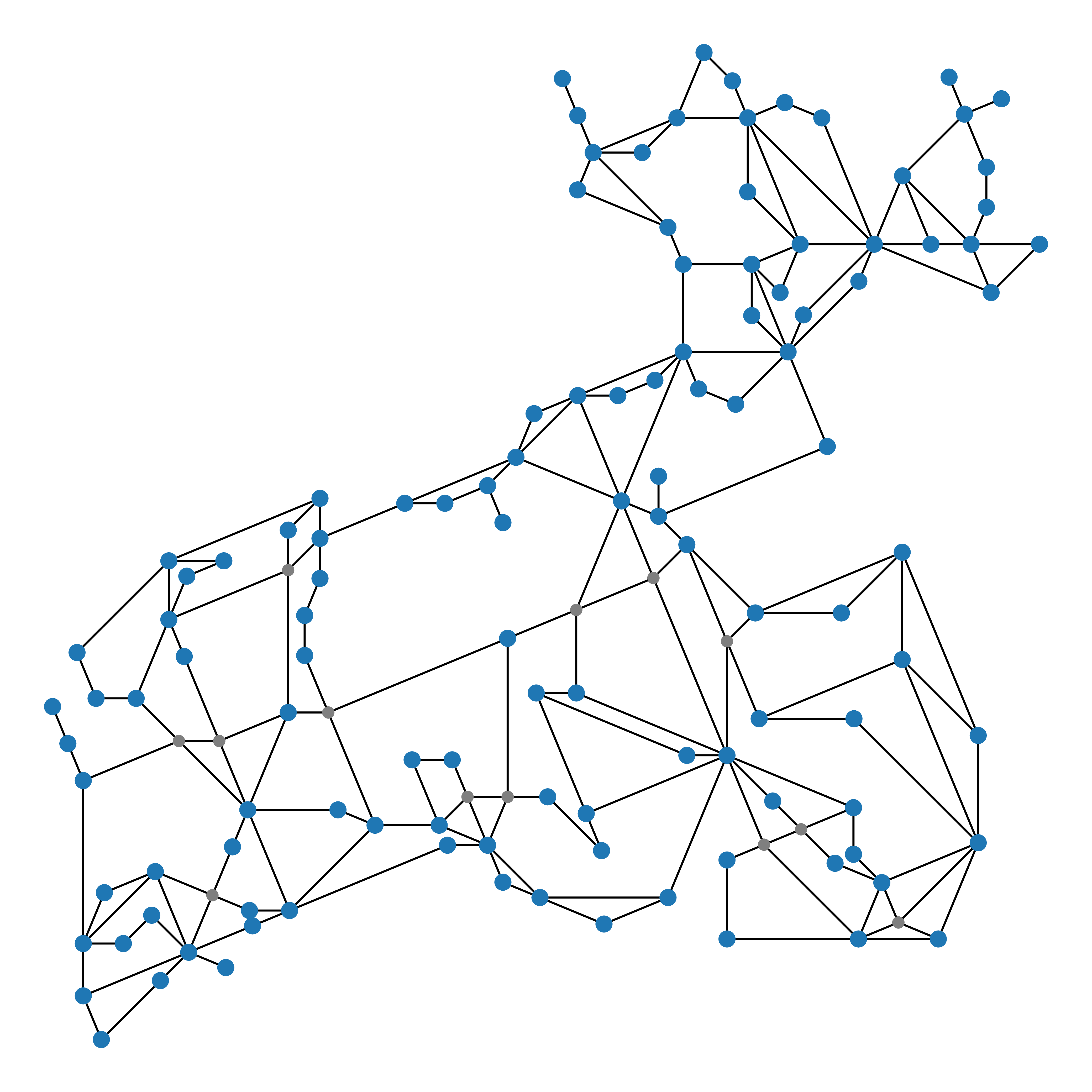}}
    \caption{The layout planning process of the IEEE 118-bus test system with the planarity relaxation adopted. Line crossings existing in the crossing-reduced layout are marked in gray, with newly appearing ones marked in orange.}
    \label{fig:staged_drawing_of_118}
\end{figure}

As discussed in Section \ref{sub:modeling_of_constraints}, the number of constraints related to planarity is of quadratic complexity to the total number of edges, which is considerable for large-scale power grids. A relaxation strategy, i.e., staged addition of such constraints, is thus proposed to enhance MILP solving efficiency. The relaxation strategy is validated in this subsection through case studies on the IEEE 118-bus test system. 

In the first round, planarity-related constraints are not included in the MILP model. It takes about 4 s for the optimizer to find the first feasible solution. As shown in Fig. \subref*{fig:staged_1}, the yielded topology diagram contains only nine pairs of intersecting edges. By adding planarity constraints related to those intersecting edge pairs to the MILP problem, a feasible solution of the second round appears at about 36.7 s. As shown in Fig. \subref*{fig:staged_2}, the output layout of this round contains six new line crossings. By integrating planarity constraints relevant to those edges, the first feasible solution of the third round is obtained at about 53.3 s. As shown in Fig. \subref*{fig:staged_3}, no new intersecting edges occur in the output of the third round of optimization. In comparison, the optimizer spends over 60 hours exploring the solution space of the MILP problem containing a complete set of planarity constraints for all edge pairs but still cannot find a single feasible solution. Therefore, it can be concluded that the proposed relaxation strategy can significantly improve the efficiency of generating topology diagrams, especially for large-scale power transmission systems.

\subsection{Efficiency Analysis} 
\label{sub:topology_diagrams_generation_efficiency}

This subsection analyzes the efficiency of the proposed framework. The first group of case studies is conducted on the five moderate-scale power transmission systems in Table \ref{tab:topology_diagrams_of_different_models}. Simplification heuristics H1 and H2 are both applied to the crossing reduction module, with the tunable BFS-tree depth $d_v^\mathcal{T} = 3$ and the expansion radius $r=0.1\times \bar{\ell}$, where $\bar{\ell}$ is the average edge length of the considered subgraph $H_v$. A round of MILP optimization is regarded to be converged if the gap between the lower and upper objective bounds is less than 30\% of the absolute value of the incumbent objective value. Measured time consumptions of crossing reduction and layout planning processes are displayed in Table \ref{tab:time_consumption}. 

It can be seen that with the scale growth of power transmission systems, the time overhead of generating topology diagrams rises rapidly. Even with the simplification heuristics and the planarity relaxation strategy both applied to the crossing reduction and layout planning modules, respectively, it still takes over 20 minutes to produce a satisfactory diagram for the IEEE 118-bus test system. This high time overhead stems from the theoretical complexity of computational geometry algorithms and the speed bottleneck of present MILP solvers, hindering the straightforward application of the proposed framework to generate topology diagrams for large-scale power transmission systems with hundreds or even thousands of nodes.

\begin{table}[!t]
  \centering
  \caption{Time Consumption of Topology Diagram Generation \textup{(s)}}
	\begin{tabular}{cm{5.5em}<{\centering}m{6em}<{\centering}c}
	\toprule
	\textbf{Model} & \textbf{Crossing Reduction} & \textbf{Layout Optimization} & \textbf{Total} \\
	\midrule
	IEEE 30 & 13.79 & 1.63 & 15.42 \\
	Grid A & 4.13 & 4.92 & 9.05 \\
	IEEE 57 & 60.82 & 252.62 & 313.44 \\
	Grid B & 17.26 & 227.75 & 245.01 \\
	IEEE 118 & 217.27 & 1024.89 & 1242.16\\
	\bottomrule
	\end{tabular}%
  \label{tab:time_consumption}%
\end{table}%

\begin{figure*}[!tb]
    \centering
    \subfloat[\label{fig:partition_2046} Original diagram and the graph partitioning scheme]{\includegraphics[width=0.9\textwidth]{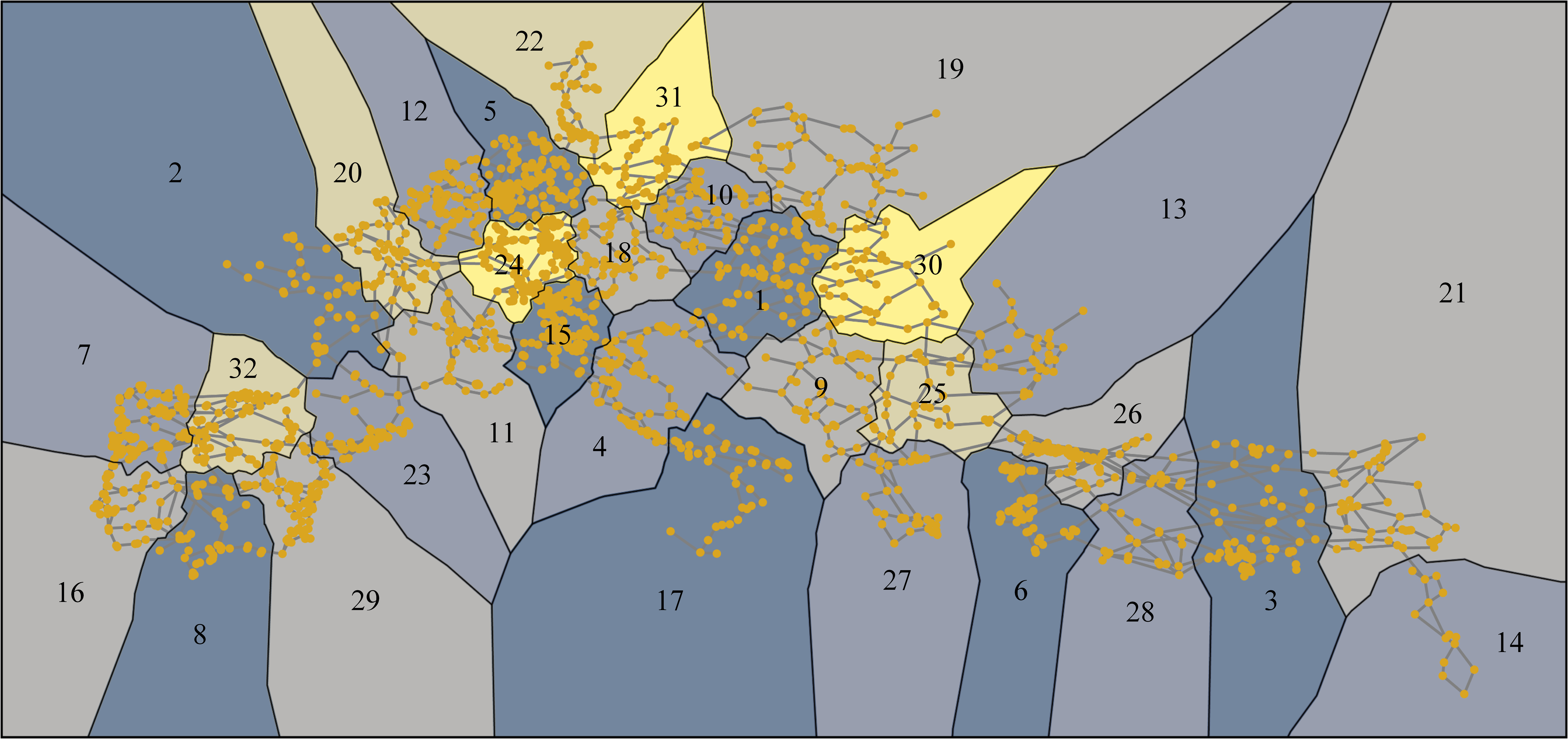}}\\
    \subfloat[\label{fig:final_2046} Optimized diagram generated by the proposed framework]{\includegraphics[width=0.9\textwidth]{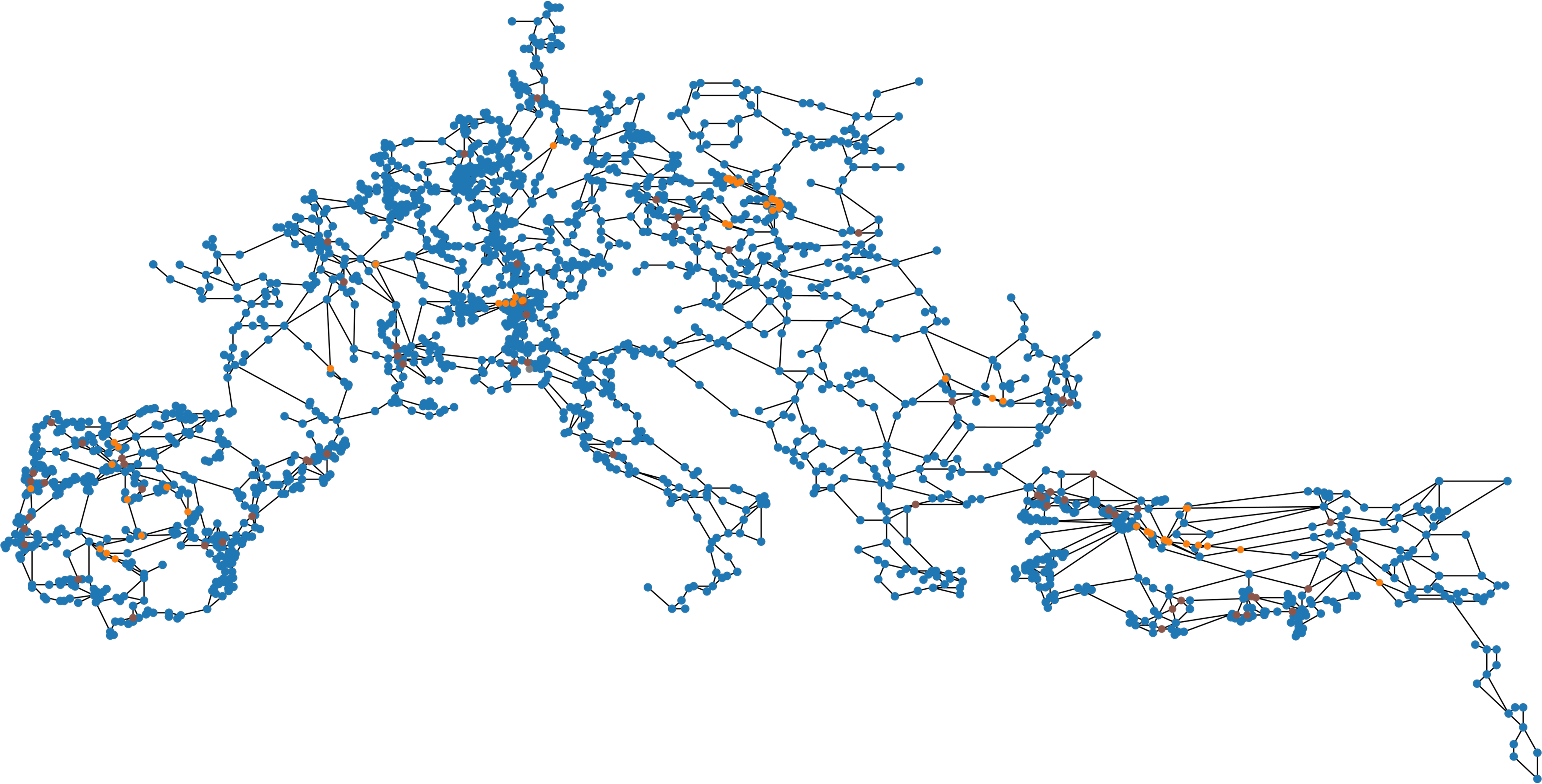}}
    \caption{The (a) original and (b) optimized topology diagrams of the 380 kV European continental power transmission system.}
\end{figure*}

An intuitive idea to circumvent this problem is to partition large power systems into several regions and generate topology diagrams for each region in parallel. The final diagram can be obtained by splicing all partial diagrams together. Such a divide-and-conquer strategy is promising in decreasing the time overhead of the proposed framework, not only because because problem dimensions of crossing reduction and layout optimization can be considerably narrowed, but also because it enables parallel computation. The following case study generates a topology diagram for the major part of the 380 kV European continental power transmission system\footnote{Data source: \href{https://github.com/PyPSA/pypsa-eur/tree/master/data/entsoegridkit}{https://github.com/PyPSA/pypsa-eur/tree/master/data}}. The power transmission system contains 2,046 nodes, covering over 20 countries from Western Europe to the Middle East. During the topology diagram generation process, nodes are divided into 32 clusters based on the K-Means algorithm. By doing so, the whole grid model can be partitioned into 32 submodels and a set of tie-lines, as illustrated in Fig. \subref*{fig:partition_2046}. Crossing reduction and layout planning are conducted on submodels, with 16 CPU threads working simultaneously. After subgraphs are rearranged, tie-lines are added to the optimized layout by connecting the terminal nodes straightforwardly. The final layout is presented in Fig. \subref*{fig:final_2046}, with the comparison of aesthetic metrics displayed in Table \ref{tab:comparison_2046}. 

It can be seen from Table \ref{tab:comparison_2046} that the proposed framework competes the traditional force-directed algorithm in all aspects discussed in Section \ref{sec:grid_topology_aesthetics} and improves the aesthetic quality of the original diagram in terms of EX, EL, ND, IA, and OR. Moreover, with parallel computation, the topology diagram generation process only takes 2748.35 seconds. That is to say, the time overhead of the 2,046-bus case is only 2.21 times of the IEEE 118-bus case, although the former has 17.34 times more nodes than the latter. Hence, it can be concluded that the proposed framework is available for generating topology diagrams for power transmission systems of very large scales with the aid of graph partitioning and parallel computation.

\begin{table}[!t]
  \centering
  \caption{Comparison of Aesthetic Metrics}
	\begin{tabular}{cccccccc}
	\toprule
	\textbf{Layout} & $m_{\text{EX}}$ & $m_{\text{EL}}$ & $m_{\text{ND}}$ & $m_{\text{IA}}$ & $m_{\text{RP}}$ & $m_{\text{OR}}$ & $m_{\text{EV}}$ \\
	\midrule
	$\hat{\Gamma}(G)$ & \textbf{-121} & \textbf{0.112} & \textbf{0.221} & \textbf{0.023} & \textbf{0.926} & \textbf{0.549} & \textbf{-0.010} \\
	$\Gamma^0(G)$ & -179 & 0.001 & 0.002 & 0.001 & -- & 0.435 & -0.010 \\
	$\Gamma^\text{FD}(G)$ & -876 & 0.089 & 0.129 & 0.001 & 0.205 & 0.399 & -0.018 \\
	\bottomrule
	\end{tabular}%
  \label{tab:comparison_2046}%
\end{table}%


\subsection{Future Work} 

In Fig. \subref*{fig:final_2046}, both orange and brown dots represent line crossings. The difference is that brown dots are those contained in subgraphs and participating in layout planning, while orange dots are those introduced when linking up tie-lines. Among the 121 crossings, 60 are formed when drawing tie-lines, which implies that graph partitioning and subgraph assembly processes still have vast room for improvement. For the graph partitioning process, the boundary of each subgraph needs to be strategically determined for the convenience of subsequent splicing. For the subgraph assembly process, graphic elements near the boundary of different subgraphs should be coordinated to prevent overlapping or producing new intersections. These designs are out of this paper's scope and left for future work.

\section{Conclusion}
\label{sec:conclusion}

This paper proposes an automated framework to generate topology diagrams for power transmission systems. Input with an initial layout, the framework first reduces line crossings by applying computational geometry heuristics based on visibility region analysis and then rearranges node positions by modeling the layout planning problem into a mixed-integer linear programming optimization problem. Compared with the classical force-directed algorithm, the proposed framework can produce topology diagrams that better conform to the aesthetic consensus of the power system community for transmission grids with complex meshed structures. Particularly, spatial relationships among nodes are largely preserved, significantly fostering the recognition and understanding of output diagrams. With the aid of graph partitioning and parallel computation, the framework is also available for large-scale power systems. The proposed framework can be used as a fundamental visualization module in various power system applications to relieve the burden of drawing power grid topology diagrams manually. 

\bibliographystyle{IEEEtran}
\bibliography{IEEEabrv,bibi}

\begin{thebibliography}{10}
\providecommand{\url}[1]{#1}
\csname url@samestyle\endcsname
\providecommand{\newblock}{\relax}
\providecommand{\bibinfo}[2]{#2}
\providecommand{\BIBentrySTDinterwordspacing}{\spaceskip=0pt\relax}
\providecommand{\BIBentryALTinterwordstretchfactor}{4}
\providecommand{\BIBentryALTinterwordspacing}{\spaceskip=\fontdimen2\font plus
\BIBentryALTinterwordstretchfactor\fontdimen3\font minus
  \fontdimen4\font\relax}
\providecommand{\BIBforeignlanguage}[2]{{%
\expandafter\ifx\csname l@#1\endcsname\relax
\typeout{** WARNING: IEEEtran.bst: No hyphenation pattern has been}%
\typeout{** loaded for the language `#1'. Using the pattern for}%
\typeout{** the default language instead.}%
\else
\language=\csname l@#1\endcsname
\fi
#2}}
\providecommand{\BIBdecl}{\relax}
\BIBdecl

\bibitem{Ten_Wuergler_2008}
C.-W. Ten, E.~Wuergler, H.-J. Diehl, and H.~B. Gooi, ``Extraction of geospatial
  topology and graphics for distribution automation framework,'' \emph{{IEEE}
  Trans. Power Syst.}, vol.~23, no.~4, pp. 1776--1782, 2008.

\bibitem{Wong_Schneider_2009}
P.~C. Wong, K.~Schneider, P.~Mackey, H.~Foote, G.~Chin~Jr., R.~Guttromson, and
  J.~Thomas, ``A novel visualization technique for electric power grid
  analytics,'' \emph{{IEEE} Trans. Vis. Comput. Graphics}, vol.~15, no.~3, pp.
  410--423, 2009.

\bibitem{Overbye_Wert_2021}
T.~J. Overbye, J.~L. Wert, K.~S. Shetye, F.~Safdarian, and A.~B. Birchfield,
  ``The use of geographic data views to help with wide-area electric grid
  situational awareness,'' in \emph{Proc. IEEE TPEC}, College Station, TX, USA,
  2021, pp. 1--6.

\bibitem{Li_Feng_2008}
X.~Li, X.~Feng, Z.~Zeng, X.~Xu, and Y.~Zhang, ``Distribution feeder one-line
  diagrams automatic generation from geographic diagrams based on gis,'' in
  \emph{Proc. 3rd DRPT}, Nanjing, China, 2008, pp. 2228--2232.

\bibitem{Wu_Lin_2018}
L.~Wu, Y.~Lin, and W.~Pang, ``Distribution network topology modelling and
  automatic mapping based on cim and gis,'' in \emph{Proc. IEEE 4th ITOEC},
  Chongqing, China, 2018, pp. 1--5.

\bibitem{Shang_Hu_2019}
L.~Shang, R.~Hu, H.~Ci, W.~Zhang, and G.~Ouyang, ``Automatic generation
  algorithm of distribution network topology map based on {GIS} drawing,''
  \emph{{IOP} Conf. Series: Earth and Environ. Sci.}, vol. 384, no. 012231, pp.
  1--7, 2019.

\bibitem{Birchfield_Overbye_2018}
A.~B. Birchfield. and T.~J. Overbye, ``Techniques for drawing geographic
  one-line diagrams: Substation spacing and line routing,'' \emph{{IEEE} Trans.
  Power Syst.}, vol.~33, no.~6, pp. 7269--7276, 2018.

\bibitem{Canales_Garibay_1979}
R.~Canales-Ruiz, D.~T. Garibay, and A.~Alonso-Concheiro, ``Optimal automatic
  drawing of one-line diagrams,'' \emph{{IEEE} Trans. Power App. Syst.}, vol.
  PAS-98, no.~2, pp. 387--392, 1979.

\bibitem{Raman_Khincha_1986}
N.~Raman, H.~Khincha, and K.~Parthasarathy, ``Automatic generation of power
  system one-line diagrams,'' \emph{IFAC Proc. Vol.}, vol.~19, no.~16, pp.
  225--229, 1986.

\bibitem{Nagendra_Deekshit_2004}
P.~{Nagendra Rao} and R.~Deekshit, ``Distribution feeder one-line diagram
  generation: a visibility representation,'' \emph{Elect. Power Syst. Res.},
  vol.~70, no.~3, pp. 173--178, 2004.

\bibitem{Peng_Wang_2016}
W.~Peng and J.~Wang, ``A novel method for automatic generation of one-line
  diagram of distribution network with multiple feeders,'' in \emph{Proc. 5th
  ICCSNT}, Changchun, China, 2016, pp. 59--64.

\bibitem{Ding_Meng_2016}
D.~Wei, M.~Zhaoyong, C.~Yaomin, and Z.~Boxi, ``Automatic generation for
  single-line diagram of distribution network,'' in \emph{Proc. 11th IEEE
  ICIEA}, Hefei, China, 2016, pp. 348--353.

\bibitem{Hussain_Aslam_2018}
A.~Hussain, M.~Aslam, and S.~M. Arif, ``A standards-based approach for
  auto-drawing single line diagram of multivendor smart distribution systems,''
  \emph{Int. J. Elect. Power Energy Syst.}, vol.~96, pp. 357--367, 2018.

\bibitem{Cuffe_Keane_2017}
P.~Cuffe and A.~Keane, ``Visualizing the electrical structure of power
  systems,'' \emph{{IEEE} Syst. J.}, vol.~11, no.~3, pp. 1810--1821, 2017.

\bibitem{KOVACEV2023118733}
N.~Kovačev, M.~Gavrić, and I.~Lendák, ``Algorithm for visualizing substation
  areas in electric power systems,'' \emph{Expert Syst. With Appl.}, vol. 212,
  p. 118733, 2023.

\bibitem{Eades_1984}
P.~Eades, ``A heuristic for graph drawing,'' \emph{Congressus Numerantium},
  vol.~42, pp. 149--160, 1984.

\bibitem{Walshaw_2000}
C.~Walshaw, ``A multilevel algorithm for force-directed graph drawing,'' in
  \emph{Proc. 8th Int. Symp. Graph Drawing}, Colonial Williamsburg, VA, USA,
  2000, p. 171–182.

\bibitem{Teja_Yemula_2014}
S.~C. Teja and P.~K. Yemula, ``Power network layout generation using force
  directed graph technique,'' in \emph{Proc. 18th NPSC}, Guwahati, India, 2014,
  pp. 1--6.

\bibitem{Mota_Mota_2011}
A.~de~Assis~Mota and L.~T.~M. Mota, ``Drawing meshed one-line diagrams of
  electric power systems using a modified controlled spring embedder algorithm
  enhanced with geospatial data,'' \emph{J. Comput. Sci.}, vol.~7, no.~2, pp.
  234--241, 2011.

\bibitem{Kovacev_Lendak_2013}
N.~Kovačev, I.~Lendák, D.~Čapko, and A.~Erdeljan, ``Electric power
  distribution system visualization with graph partitioning,'' in \emph{Proc.
  IEEE AFRICON}, Pointe aux Piments, Mauritius, 2013, pp. 1--6.

\bibitem{Zhou_Sun_2016}
B.~Zhou, L.~Sun, H.~Zhang, Y.~Yin, W.~Ding, and W.~Huang, ``Automatic
  single-line diagram generation of distribution network with rings based on
  ga,'' in \emph{Proc. 12th WCICA}, Guilin, China, 2016, pp. 442--445.

\bibitem{Lendak_Erdeljan_2010}
I.~Lendak, A.~Erdeljan, D.~Čapko, and S.~Vukmirović, ``Algorithms in electric
  power system one-line diagram creation: The soft computing approach,'' in
  \emph{Proc. IEEE SMC}, Istanbul, Turkey, 2010, pp. 2867--2873.

\bibitem{Lin_Xing_2010}
R.~Lin, J.~Xing, H.~Yang, and W.~Chen, ``Intelligent automatic layout of
  one-line diagrams for district electrical distribution network,'' in
  \emph{Proc. IEEE APPEEC}, Chengdu, China, 2010, pp. 1--4.

\bibitem{Jacomy_Venturini_2014}
M.~Jacomy, T.~Venturini, S.~Heymann, and M.~Bastian, ``{ForceAtlas2}, a
  continuous graph layout algorithm for handy network visualization designed
  for the gephi software,'' \emph{PLOS ONE}, vol.~9, no.~6, pp. 1--12, 06 2014.

\bibitem{Zheng_Pawar_2019}
J.~X. Zheng, S.~Pawar, and D.~F.~M. Goodman, ``Graph drawing by stochastic
  gradient descent,'' \emph{{IEEE} Trans. Vis. Comput. Graphics}, vol.~25,
  no.~9, pp. 2738--2748, 2019.

\bibitem{Kruiger_Rauber_2017}
J.~F. Kruiger, P.~E. Rauber, R.~M. Martins, A.~Kerren, S.~Kobourov, and A.~C.
  Telea, ``Graph layouts by {t-SNE},'' \emph{Comput. Graph. Forum}, vol.~36,
  no.~3, pp. 283--294, 2017.

\bibitem{Wang_Jin_2020}
Y.~Wang, Z.~Jin, Q.~Wang, W.~Cui, T.~Ma, and H.~Qu, ``Deepdrawing: A deep
  learning approach to graph drawing,'' \emph{{IEEE} Trans. Vis. Comput.
  Graphics}, vol.~26, no.~1, pp. 676--686, 2020.

\bibitem{Ahmed_Luca_2022}
R.~Ahmed, F.~De~Luca, S.~Devkota, S.~Kobourov, and M.~Li, ``Multicriteria
  scalable graph drawing via stochastic gradient descent, ({SGD}$^{2}$),''
  \emph{{IEEE} Trans. Vis. Comput. Graphics}, vol.~28, no.~6, pp. 2388--2399,
  2022.

\bibitem{Radermacher_Reichard_2019}
M.~Radermacher, K.~Reichard, I.~Rutter, and D.~Wagner, ``Geometric heuristics
  for rectilinear crossing minimization,'' \emph{ACM J. Exp. Algorithmics},
  vol.~24, no. 1.12, pp. 1--21, 2019.

\bibitem{Nickel_Nollenburg_2020}
S.~Nickel and M.~N{\"o}llenburg, ``Towards data-driven multilinear metro
  maps,'' in \emph{Proc. 11th Int. Conf. Theory and Appl. Diagrams}, Tallinn,
  Estonia, 2020, pp. 153--161.

\bibitem{Purchase_2002}
H.~C. Purchase, ``Metrics for graph drawing aesthetics,'' \emph{J. Vis. Lang.
  Comput.}, vol.~13, no.~5, pp. 501--516, 2002.

\bibitem{Bennett_Ryall_2007}
C.~Bennett, J.~Ryall, L.~Spalteholz, and A.~Gooch, ``The aesthetics of graph
  visualization,'' in \emph{Proc. 3rd Eurographics Conf. Comput. Aesthetics
  Graph. Vis. Imag.}, Alberta, Canada, 2007, p. 57–64.

\bibitem{Cuffe_Keane_2016}
P.~Cuffe and A.~Keane, ``Novel quality metrics for power system diagrams,'' in
  \emph{Proc. IEEE ENERGYCON}, Leuven, Belgium, 2016, pp. 1--5.

\end{thebibliography}

\end{document}